\documentclass[12pt]{article}
\usepackage{amsfonts}
\usepackage[dvips]{color}
\usepackage{xcolor}

\usepackage{amssymb}
\usepackage{graphicx}
\usepackage{amsmath}
\usepackage{natbib}
\usepackage{rotating}
\usepackage{afterpage}
\usepackage{cancel}

\usepackage{setspace}
\newcommand{\bc}{\begin{center}}
\newcommand{\ec}{\end{center}}
\newcommand{\be}{\begin{equation}}
\newcommand{\ee}{\end{equation}}
\newcommand{\bea}{\begin{eqnarray}}
\newcommand{\eea}{\end{eqnarray}}
\newcommand{\bean}{\begin{eqnarray*}}
\newcommand{\eean}{\end{eqnarray*}}
\newcommand{\bt}{\begin{tabular}}
\newcommand{\et}{\end{tabular}}

\usepackage{html}   %
\usepackage{url}       %
\usepackage{xcolor}
\usepackage[margin=1in]{geometry}
\usepackage{graphicx}

\newtheorem{theorem}{Theorem}

\newtheorem{assumption}[theorem]{Assumption}

\newtheorem{definition}[theorem]{Definition}

\newtheorem{lemma}[theorem]{Lemma}

\numberwithin{theorem}{section}

\newcommand{\argmin}{\operatorname*{argmin}}

\newcommand{\sgn}{\operatorname*{sgn}}
\newcommand{\diag}{\operatorname*{diag}}
\newcounter{saveeqn}

\begin{document}

\title{\bf Forecasting with Dynamic Panel Data Models}

\author{
        Laura Liu\\ {\em University of Pennsylvania}
        \and
        Hyungsik Roger Moon\\ {\em University of Southern California} \\
        {\em USC Dornsife INET, and Yonsei}
        \and
        Frank Schorfheide\thanks{Correspondence:
                L. Liu and F. Schorfheide: Department of Economics, 3718
                Locust Walk, University of Pennsylvania, Philadelphia, PA 19104-6297. Email:
                yuliu4@sas.upenn.edu (Liu) and schorf@ssc.upenn.edu (Schorfheide).
                H.R. Moon: Department of Economics, University of Southern California, KAP 300, Los Angeles, CA
                90089. E-mail: moonr@usc.edu. We thank Xu Cheng, Frank Diebold, Peter Phillips, Akhtar Siddique, and participants at various seminars and conferences for helpful comments and suggestions. Moon and Schorfheide gratefully acknowledge financial support from the National Science Foundation under Grants SES 1625586 and SES 1424843, respectively. } \\
        {\em University of Pennsylvania} \\ {\em CEPR, NBER, and PIER} }

\date{\today}
\maketitle

\thispagestyle{empty}

%\author{Yu Laura Liu}
%\address{University of Pennsylvania}
%\author{Hyungsik Roger Moon}
%\address{University of Southern California, USC Dornsife INET, and Yonsei}
%\author{Frank Schorfheide}
%\address{University of Pennsylavia, CEPR, and NBER}
%\email{moonr@usc.edu}
%\urladdr{https://www-bcf.usc.edu/~moonr/}

\newpage

\begin{abstract}
This paper considers the problem of forecasting  a collection of short time series using cross sectional information in panel data. We construct point predictors using Tweedie's formula for the posterior mean of heterogeneous coefficients under a correlated random effects distribution. This formula utilizes cross-sectional information to transform the unit-specific (quasi) maximum likelihood estimator into an approximation of the posterior mean under a prior distribution that equals the population distribution of the random coefficients. We show that the risk of a predictor based on a non-parametric estimate of the Tweedie correction is asymptotically equivalent to the risk of a predictor that treats the correlated-random-effects distribution as known (ratio-optimality). Our empirical Bayes predictor performs well compared to various competitors in a Monte Carlo study. In an empirical application we use the predictor to forecast revenues for a large panel of bank holding companies and compare forecasts that condition on actual and severely adverse macroeconomic conditions.
\end{abstract}
%\date{\today}

\noindent JEL CLASSIFICATION: C11, C14, C23, C53, G21

\noindent KEY\ WORDS: Bank Stress Tests, Empirical Bayes, Forecasting, Panel Data, Ratio Optimality, Tweedies Formula

\thispagestyle{empty}
\setcounter{page}{0}
\newpage

\section{Introduction}

The main goal of this paper is to forecast a collection of short time series. Examples are the performance of start-up companies, developmental skills of small children, and revenues and leverage of banks after significant regulatory changes. In these applications the key difficulty lies in the efficient implementation of the forecast. Due to the short time span, each time series taken by itself provides insufficient sample information to precisely estimate unit-specific parameters. We will use the cross-sectional information in the sample to make inference about the distribution of heterogeneous parameters. This distribution can then serve as a prior for the unit-specific coefficients to sharpen posterior inference based on the short time series.

More specifically, we consider a linear dynamic panel model in which the unobserved individual heterogeneity, which we denote by the vector $\lambda_i$, interacts with some observed predictors: 
\begin{equation}
   Y_{it}= \lambda_i^{\prime}W_{it-1} + \rho^{\prime} X_{it-1} + \alpha^{\prime}Z_{it-1}  + U_{it}, \quad i=1,\ldots,N, \quad t=1,\ldots,T. \label{eq.yit} 
\end{equation}% 
Here, $(W_{it-1},X_{it-1},Z_{it-1})$ are predictors and $U_{it}$ is an unpredictable shock. Throughout this paper we adopt a correlated random effects approach in which the $\lambda_i$s are treated as random variables that are possibly correlated with some of the predictors. An important special case is the linear dynamic panel data model in which $W_{it-1}=1$, $\lambda_i$ is a heterogeneous intercept, and the sole predictor is the lagged dependent variable: $X_{it-1}=Y_{it-1}$. 

We develop methods to generate point forecasts of $Y_{iT+1}$, assuming  that the time dimension $T$ is short relative to the number of predictors $(W_{iT},X_{iT},Z_{iT})$. The forecasts are evaluated under a quadratic loss function. In this setting an accurate forecasts not only requires a precise estimate of the common parameters $(\alpha,\rho)$, but also of the parameters $\lambda_i$ that are specific to the cross-sectional units $i$. The existing literature on dynamic panel data models almost exclusively studied the estimation of the common parameters, treating the unit-specific parameters as a nuisance. Our paper builds on the insights of the dynamic panel literature and focuses on the estimation of $\lambda_i$, which is essential for the prediction of $Y_{it}$.   

The benchmark for our prediction methods is the so-called oracle forecast. The oracle is assumed to know the common coefficients $(\alpha,\rho)$ as well as the distribution of the heterogeneous coefficients $\lambda_i$, denoted by $\pi(\lambda_i|\cdot)$. Note that this distribution could be conditional on some observable characteristics of unit $i$. Because we are interested in forecasts for the entire cross section of $N$ units, a natural notion of risk is that of compound risk, which is a (possibly weighted) cross-sectional average of expected losses. In a correlated random-effects setting, this averaging is done under the distribution $\pi(\lambda_i|\cdot)$, which means that the compound risk associated with the forecasts of the $N$ units is the same as the integrated risk for the forecast of a particular unit $i$. It is well known, that the integrated risk is minimized by the Bayes predictor that minimizes the posterior expected loss conditional on time $T$ information for unit $i$. Thus, the oracle replaces $\lambda_i$ by its posterior mean.

The implementation of the oracle forecast is infeasible because in practice neither the common coefficients $(\rho,\alpha)$ nor the distribution of the unit-specific coefficients $\pi(\lambda_i|\cdot)$ is known. To obtain a feasible predictor, we extend the classical posterior mean formula attributed to separate works of Arthur Eddington and Maurice Tweedie to our dynamic panel data setup. According to this formula, the posterior mean of $\lambda_i$ can be expressed as a function of the cross-sectional density of certain sufficient statistics. Conditional on the common parameters, this distribution can then be estimated either parametrically or non-parametrically from the panel data set. The unknown common parameters can be replaced by a generalized method of moments (GMM) estimator, a likelihood-based correlated random effects estimator, or a Bayes estimator. 

Our paper makes three contributions. First, we show in the context of the linear dynamic panel data model that a feasible predictor based on a consistent estimator of $(\rho,\alpha)$ and a non-parametric estimator of the cross-sectional density of the relevant sufficient statistics can achieve the same compound risk as the oracle predictor asymptotically. Our main theorem extends a result from \cite{BrownGreenshtein2009} for a vector of means to a panel data model with estimated common coefficients. Importantly, this result also covers the case in which the distribution $\pi(\lambda_i|\cdot)$ degenerates to a point mass. As in \cite{BrownGreenshtein2009}, we are able to show that the rate of convergence to the oracle risk accelerates in the case of homogeneous $\lambda$ coefficients. Second, we provide a detailed Monte Carlo study that compares the performance of various implementations, both non-parametric and parametric, of our predictor. Third, we use our techniques to forecast pre-provision net-revenues of a panel of banks.

If the time series dimension is small, our feasible predictor performs much better than a naive predictor of $Y_{iT+1}$ that is based on within-group estimates of $\lambda_i$. A small $T$ leads to a noisy estimate of $\lambda_i$. Moreover, from a compound risk perspective, there will be a selection bias. Consider the special case of $\alpha=\rho=0$ and $W_{it}=1$. Here, $\lambda_i$ is simply a heterogeneous intercept. Very large (small) realizations of $Y_{it}$ will be attributed to large (small) values of $\lambda_i$, which means that the within-group mean will be upward (downward) biased for those units. The use of a prior distribution estimated from the cross-sectional information essentially corrects this bias, which facilitates the reduction of the prediction risk if it is averaged over the entire cross section. Alternatively, one could ignore the cross-sectional heterogeneity and estimate a (misspecified) model with a homogeneous coefficient $\lambda$. If the heterogeneity is small, this procedure is likely to perform well in a mean-squared-error sense. However, as the heterogeneity increases, the performance of a predictor that is based on a pooled estimation quickly deteriorates.
We illustrate the performance of various implementations of the feasible predictor in a Monte Carlo study and provide comparisons with other predictors, including one that is based on quasi maximum likelihood estimation of the unit-specific coefficients and one that is constructed from a pooled OLS estimator that ignores parameter heterogeneity.

In an empirical application we forecast pre-provision net revenues of bank holding companies. The stress tests that have become mandatory under the Dodd-Frank Act require banks to establish how revenues vary in stressed macroeconomic and financial scenarios. We capture the effect of macroeconomic conditions on bank performance by including the unemployment rate, an interest rate, and an interest rate spread in the vector $W_{it-1}$ in (\ref{eq.yit}). Our analysis consists of two steps. We first document the one-year-ahead forecast accuracy of the posterior mean predictor developed in this paper under the actual economic conditions, meaning that we set the aggregate covariates to their observed values. In a second step, we replace the observed values of the macroeconomic covariates by counterfactual values that reflect severely adverse macroeconomic conditions. We find that our proposed posterior mean predictor is considerably more accurate than a predictor that does not utilize any prior distribution. The posterior mean predictor shrinks the estimates of the unit-specific coefficients toward a common prior mean, which reduces its sampling variability. According to our estimates, the effect of stressed macroeconomic conditions on bank revenues is very small relative to the cross-sectional dispersion of revenues across holding companies.

Our paper is related to several strands of the literature. For $\alpha=\rho=0$ and $W_{it}=1$ the problem analyzed in this paper reduces to the problem of estimating a vector of means, which is a classic problem in the statistic literature. In this context, Tweedie's formula has been used, for instance, by \citet{Robbins1951} and more recently by \citet{BrownGreenshtein2009} and \citet{Efron2011} in a ``big data'' application. Throughout this paper we are adopting an empirical Bayes approach, that uses cross-sectional information to estimate aspects of the prior distribution of the correlated random effects and then conditions on these estimates. Empirical Bayes methods also have a long history in the statistics literature going back to \cite{Robbins1955} (see \citet{Robert1994} for a textbook treatment). 

We use compound decision theory as in \citet{Robbins1964}, \citet{BrownGreenshtein2009}, \citet{JiangZhang2009} to state our optimality result. Because our setup nests the linear dynamic panel data model, we utilize results on the consistent estimation of $\rho$ in dynamic panel data models with fixed effects when $T$ is small, e.g., \citet{AndersonHsiao1981}, \citet{ArellanoBond1991}, \citet{ArellanoBover1995}, \citet{BlundellBond1998}, \citet{AlvarezArellano2003}. Fully Bayesian approaches to the analysis of dynamic panel data models have been developed in 
\citet{ChamberlainHirano1999}, \citet{Hirano2002}, \citet{Lancaster2002}.

The papers that are most closely related to ours are \citet{GuKoenkerJAE2016,GuKoenker2014}. They also consider a linear panel data model and use Tweedie's formula to construct an approximation to the posterior mean of the heterogeneous regression coefficients. However, their papers focus on the use of the Kiefer-Wolfowitz estimator for the cross-sectional distribution of the sufficient statistics, whereas our paper explores various plug-in estimators for the homogeneous coefficients in combination with both parametric and nonparametric estimates of the cross-sectional distribution. Moreover, our paper establishes the ratio-optimality of the forecast and presents a different application.
Finally, \cite{Liu2016} develops a fully Bayesian (as opposed to empirical Bayes) approach to construct density forecast. She uses a Dirichlet process mixture to construct a prior for the distribution of the heterogeneous coefficients, which then is updated in view of the observed panel data.

There is an earlier panel forecast literature (e.g., see the survey article by \cite{Baltagi2008} and its references) that is based on the best linear unbiased prediction (BLUP) proposed by \cite{Goldberger1962}. Compared to the BLUP-based forecasts, our forecasts based on Tweedie's formula have several advantages. First, it is known that the estimator of the unobserved individual heterogeneity parameter based on the BLUP method corresponds to the Bayes estimator based on a Gaussian prior (see, for example, \cite{Robinson1991}), while our estimator based on Tweedie's formula is consistent with much more general prior distributions. Second, the BLUP method finds the forecast that minimizes the expected quadratic loss in the class of linear (in $(Y_{i0},...,Y_{iT})'$) and unbiased forecasts. Therefore, it is not necessarily optimal in our framework that constructs the optimal forecast without restricting the class of forecasts. Third, the existing panel forecasts based on the BLUP were developed for panel regressions with random effects and do not apply to correlated random effects settings.

%The related existing literature with our theoretical part of the paper includes the work of consistent estimation of $\rho$ in dynamic panel data models with fixed effects when $T$ is small (e.g., \citet{AndersonHsiao1981}, \citet{ArellanoBond1991}, \citet{ArellanoBover1995}, \citet{BlundellBond1998}, \citet{AlvarezArellano2003}), the literature of Bayesian analysis of panel data models (e.g., \citet{ChamberlainHirano1999}, \citet{Hirano2002}, \citet{Lancaster2002}), 
%the work on empirical Bayes (e.g., \cite{Robbins1955} and the survey in Chapter 10 of \citet{Robert2007} and the references in the chapter), the work studied Tweedie's formula (e.g., \citet{Robbins1951},  \citet{Efron2011}, \citet{KnoxStockWatson2001}, \citet{BrownGreenshtein2009}, and \citet{GuKoenker2014},) and the compound decision theory and optimality (e.g., \citet{Robbins1964}, \citet{BrownGreenshtein2009}, \citet{JiangZhang2009}).
%{\color{blue} Discuss \citet{GuKoenker2014} in more details.} 

There is a small academic literature on econometric techniques for stress test. Most papers analyze revenue and balance sheet data for the relatively small set of bank holding companies with consolidated assets of more than 50 billion dollars. There are slightly more than 30 of these companies and they are subject to the Comprehensive Capital Analysis and Review conducted by the Federal Reserve Board of Governors. An important paper in this literature is \cite{CovasRumpZakrajsek2014}, which uses quantile autoregressive models to forecast bank balance sheet and revenue components. We work with a much larger panel of bank holding companies that comprises, depending on the sample period, between 460 and 725 institutions. 

The remainder of the paper is organized as follows. Section~\ref{sec:model} introduces the panel data model considered in this paper, derives the likelihood function, and provides an important identification result. Decision theoretic foundations for the proposed predictor and a derivation of the oracle forecast are provided in Section~\ref{sec:decisiontheory}. Section~\ref{sec:implementation} discusses feasible implementation strategies for the predictor and we show in Section~\ref{sec:ratio.optimality} in the context of a basic dynamic panel data model that our proposed predictor asymptotically has the same risk as the oracle forecast. A simulation study is provided in Section~\ref{sec:monte-carlo-simulations}. The empirical application is presented in Section~\ref{sec:empiricalapplication} and Section~\ref{sec:conclusion} concludes. Technical derivations, proofs, the description of the data set used in the empirical analysis, and further empirical results are relegated to the Appendix.

%========================================================================================================

\section{A Dynamic Panel Forecasting Model} 
\label{sec:model}

We consider a panel with observations for cross-sectional units $i=1,\ldots,N$ in periods $t=1,\ldots,T$. Observation $Y_{it}$ is assumed to be generated by (\ref{eq.yit}). We distinguish three types of regressors. First, the $k_w\times 1$ vector $W_{it}$ interacts with the heterogeneous coefficients $\lambda_i$. In many panel data applications $W_{it} = 1$, meaning that $\lambda_i$ is simply a heterogenous intercept. We allow $W_{it}$ to also include deterministic time effects such as seasonality, time trends and/or strictly exogenous variables observed at time $t$. To distinguish deterministic time effects $w_{1,t+1}$ from cross-sectionally varying and strictly exogenous variables $W_{2,it}$, we partition the vector into $W_{it} = (w_{1,t+1},W_{2,it})$.\footnote{Because $W_{it}$ is a predictor for $Y_{it+1}$ we use a $t+1$ subscript for the deterministic trend component $w_{1}$.} The dimensions of the two components are $k_{w_1}$ and $k_{w_2}$, respectively.
Second, $X_{it}$ is a $k_x \times 1$ vector of sequentially exogenous predictors with homogeneous coefficients.
The predictors $X_{it}$ may include lags of $Y_{it+1}$ and we collect all the predetermined variables other than the lagged dependent variable into the subvector $X_{2,it}$.
Third, $Z_{it}$ is a $k_z$-vector of strictly exogenous regressors, also with common coefficients.

Our main goal is to construct optimal forecasts of $(Y_{1T+1},...,Y_{NT+1})$ conditional on the entire panel observations $\{(Y_{it},W_{it-1},X_{it-1},Z_{it-1})$, $i = 1,\ldots,N$ and $t=1,...,T$ using the forecasting model (\ref{eq.yit}).
An important special case of model (\ref{eq.yit}) is the basic dynamic panel data model
\begin{equation}
  Y_{it} = \lambda_i + \rho Y_{it-1} + U_{it} \label{m.restricted.linear.panel.regression},
\end{equation}
which is obtained by setting $W_{it} = 1$, $X_{it} = Y_{it}$ and $\alpha=0$.
The restricted model (\ref{m.restricted.linear.panel.regression}) has been widely studied in the literature. However, most studies  focus on consistently estimating the common parameter $\rho$ in the presence of an increasing (with the cross-sectional dimension $N$) number of $\lambda_i$s.
%\footnote{The incidental parameter problem dates back to  \cite{NeymanScott1948}. \cite{Nickell1981} showed that the standard within-group estimator for dynamic models with fixed effects generates an inconsistent estimate of $\rho$.} 
In forecasting applications, we also need to estimate the $\lambda_i$s.  
In Section~\ref{subsec:model.likelihood} we specify the likelihood function for model (\ref{eq.yit}) and in Section~\ref{subsec:model.identification} we establish the identifiability of the model parameters, including the distribution of the heterogeneous coefficients $\lambda_i$.

\subsection{The Likelihood Function}
\label{subsec:model.likelihood}

Let $Y_i^{t_1:t_2} =(Y_{it_1},...,Y_{it_2})$ and use a similar notation to collect $W_{it}$s, $X_{it}$s, and $Z_{it}s$.
We begin by making some assumptions on the joint distribution of $\{Y_i^{1:T+1},X_i^{0:T},W_{2,i}^{0:T},Z_i^{0:T},\lambda_i\}_{i=1}^N$
conditional on the regression coefficients $\rho$ and $\alpha$ and the vector of volatility parameters $\gamma$ (to be introduced below). We drop the deterministic trend regressors $w_{1,t}$ from the notation for now. We use $\mathbb{E}[\cdot]$ to denote expectations and $\mathbb{V}[\cdot]$ to denote variances.

\begin{assumption} \label{as.likelihood} \hspace*{1cm}\\[-5ex]
	\setstretch{1}
        \begin{itemize}
                \item [(i)]  $(Y_i^{1:T+1},\lambda_i,X_i^{0:T},W_{2i}^{0:T},Z_i^{0:T})$ are independent across $i$.
                \item [(ii)] $(\lambda_i,X_{i0},W_{2,i}^{0:T},Z_i^{0:T})$ are iid with joint density 
                  \[\pi(\lambda,x_0,w_2^{0:T},z^{0:T})= \pi(\lambda|x_0,w_2^{0:T},z^{0:T}) \pi(x_0,w_2^{0:T},z^{0:T}).\] 
                \item[(iii)]  For $t=1,\ldots,T$, the distribution of $X_{2,it}$ conditional on $(Y_i^{1:t},X_i^{0:t-1},W_{2,i}^{0:T}, Z_i^{0:T})$ does not depend on the heterogeneous parameters $\lambda_i$ and parameters $(\rho,\alpha,\gamma_1,...\gamma_{T})$. 
                \item[(iv)]  The distribution of $(W_{2,i}^{0:T},Z_i^{0:T})$ does not depend on $\lambda_i$ and $(\rho,\alpha,\gamma_1,...,\gamma_{T})$.
                \item[(v)] $U_{it} = \sigma_{t}(X_{i0},W_{2,i}^{0:T},Z_i^{0:T},\gamma_t) V_{it}$, where $V_{it}$  is $iid$ across $i=1,...,N$ and independent over $t=1,...,T+1$ with $\mathbb{E}[V_{it}] = 0$ and $\mathbb{V}[V_{it}] = 1$
                for $t=1,\ldots,T+1$ and $(V_{i1},\ldots,V_{iT})$ are independent of $X_{i0},W_{2,i}^{0:T},Z_i^{0:T}$. We assume $\sigma_{t}(X_{i0},W_{2,i}^{0:T},Z_i^{0:T},\gamma_t)$ is a function that depends on the unknown finite-dimensional parameter vector $\gamma_t$. 
        \end{itemize}
\end{assumption}

Assumption~\ref{as.likelihood}(i) states that conditionally on the predictors, the $Y_{it+1}$s are cross-sectionally independent. Thus, we assume that all the spatial correlation in the dependent variables is due to the observed predictors. 
Assumption~\ref{as.likelihood}(ii) formalizes the correlated random effects assumption.
%stating that the distribution of $\lambda_i$ may depend on the initial observation of the sequentially exogenous predictors and the entire sequence of strictly exogenous and deterministic predictors. 
The subsequent Assumptions~\ref{as.likelihood}(iii) and (iv) imply that $\lambda_i$ may affect $X_{it}$ only indirectly through $Y_i^{1:t}$ -- an assumption that is clearly satisfied in the dynamic panel data model (\ref{m.restricted.linear.panel.regression}) -- and that the strictly exogenous predictors do not depend on $\lambda_i$. 
In Assumption~\ref{as.likelihood}(v), we allow the unpredictable shocks $U_{it}$ to be conditionally heteroskedastic in both the cross section and over time. We allow $\sigma_t(\cdot)$ to be dependent on the initial condition of the sequentially exogenous predictors, $X_{i0}$, and other exogenous variables. Because throughout the paper we assume that the time dimension $T$ is small, the dependence through $X_{i0}$ can generate a persistent ARCH effect.

%However, the assumption does not cover an autoregressive conditional heteroskedasticity (ARCH) in which $\sigma_t(\cdot)$ depends on $X_{it}$. Instead,  We could relax the assumption of independence and homogeneity of the $V_{it}$s by conditioning on $(Y_i^{1:t-1},X_i^{0:t-1}, W_{2,i}^{0:T}, Z_i^{0:T}, \lambda_i)$. However, we decided to proceed under the stronger assumption because it simplifies the notation and it is sufficient to cover the applications considered in this paper. 

We now turn to the likelihood function. We use lower case $(y_{it}, w_{it}, x_{it}, z_{it})$ to denote the realizations of 
the random variables $(Y_{it},X_{it},W_{it},Z_{it})$. The parameters that control the volatilities $\sigma_t(\cdot)$ are stacked into the vector $\gamma=[\gamma_1',...,\gamma_T']'$ and we collect the homogeneous parameters into the vector $\theta = [\alpha',\rho',\gamma']'$. We use $H_i = (X_{i0},W_{2,i}^{0:T},Z_i^{0:T})$
for the exogenous conditioning variables and $h_i = (x_{i0},w_{2,i}^{0:T},z_i^{0:T})$ for their realization. Finally, we denote the density of $V_i$ by $\varphi(v)$. 
Recall that we used $x_{2,it}$ to denote predetermined predictors other than the lagged dependent variable. According to Assumption~\ref{as.likelihood}(iii) the 
density $q_{t}(x_{2,it}|y_i^{1:t}, x_i^{0:t-1},w_{2i},z_i)$ 
does not provide any information about $\lambda_i$ and
will subsequently be absorbed into a constant of proportionality. 
Combining the likelihood function for the observables 
with the conditional distribution of the heterogeneous coefficients leads to
\be
p(y_i,x_{2,i},\lambda_{i}|h_i,\theta) \label{eq.likelihoodprior} 
\propto
\left( \prod_{t=1}^{T} \frac{1}{\sigma_t(h_i,\gamma_t)}\varphi \left( \frac{y_{it}-\lambda_i^{\prime}w_{it-1} - \rho^{\prime} x_{it-1} - \alpha^{\prime} z_{it-1}}{\sigma_t(h_i,\gamma_t)} \right) \right) \pi(\lambda_i|h_i).
\ee
Because conditional on the predictors the observations are cross-sectionally independent, the joint densities for observations $i=1,\ldots,N$ can be obtained by taking the product across $i$ of (\ref{eq.likelihoodprior}).

%Recall that we used $X_{2,it}$ to denote predetermined predictors other than the lagged dependent variable. The contribution of observation $i$ to the
%likelihood function is given by 
%\be
%p(y_i,x_{2,i}|h_i,\lambda_{i},\theta)   \label{eq.likelihood} 
%= \left( \prod_{t=1}^{T} \frac{1}{\sigma_t(h_i,\gamma_t)}\varphi \left( \frac{y_{it}-\lambda_i^{\prime}w_{it-1} - \rho^{\prime} x_{it-1} - \alpha^{\prime} z_{it-1}}{\sigma_t(h_i,\gamma_t)} \right) \right) Q_{i}, 
%\ee
%where
%\[
%Q_{i} = \prod_{t=1}^{T} q_{t}(x_{2,it}|y_i^{1:t}, x_i^{0:t-1},w_{2i},z_i)               
%\]
%and $q_t(x_{2,it}|\cdot)$ is the conditional density of the pre-determined regressors.
%By Assumption~\ref{as.likelihood}(iii) the term $Q_{i}$ does not provide any information about $(\lambda_i)$ and
%we will subsequently absorb it into a constant of proportionality. 
%The likelihood function~(\ref{eq.likelihood}) can be combined with the conditional distribution
%of the heterogeneous coefficients
%$\pi(\lambda_i|h_i)$, which 
%leads to
%\be
%    p(y_i,x_{2,i},\lambda_{i}|h_i,\theta) \label{eq.likelihoodprior} 
%        \propto
%\left( \prod_{t=1}^{T} \frac{1}{\sigma_t(h_i,\gamma_t)}\varphi \left( \frac{y_{it}-\lambda_i^{\prime}w_{it-1} - \rho^{\prime} x_{it-1} - \alpha^{\prime} z_{it-1}}{\sigma_t(h_i,\gamma_t)} \right) \right) \pi(\lambda_i|h_i).
%\ee
%Because conditional on the predictors the observations are cross-sectionally independent, the joint densities for observations $i=1,\ldots,N$ can be obtained by taking products across $i$ of (\ref{eq.likelihood}) and (\ref{eq.likelihoodprior}), respectively.

\subsection{Identification}
\label{subsec:model.identification}

We now provide conditions under which the forecasting model (\ref{eq.yit}) is identifiable. While the identification
of the finite-dimensional parameter vector $\theta$ is fairly straightforward, the empirical Bayes approach pursued in this paper also requires the identification of the correlated random effects distribution $\pi(\lambda_i|h_i)$ from the cross-sectional information in the panel. Before presenting a general result which is formally proved in the Online Appendix, we sketch the identification argument in the context of the restricted dynamic model (\ref{m.restricted.linear.panel.regression}) with heterogeneous intercept and heteroskedastic innovations.

The identification can be established in three steps. First, the identification of the homogeneous regression coefficient $\rho$ follows from a standard argument used in the instrumental variable (IV) estimation of dynamic panel data models. To eliminate the dependence on $\lambda_i$ define $Y_{it}^* = Y_{it} - \frac{1}{T-t} \sum_{s=t+1}^T Y_{is}$ and $X_{it-1}^* = Y_{it-1} - \frac{1}{T-t} \sum_{s=t+1}^T Y_{is-1}$.
Then, because $\mathbb{E}[U_{it}|Y_i^{0:t-1},\lambda_i] = 0$, the orthogonality conditions
$\mathbb{E}\big[ (Y_{it}^* - \rho X_{it-1}^*) Y_{it-1} \big] = 0$ for 
$t=1,\ldots,T-1$ in combination with a relevant rank condition can be used to identify $\rho$ (see, e.g., \cite{ArellanoBover1995}). Second, to identify the variance parameters $\gamma$, 
let $Y_i$, $X_{i}$, and $U_i$ denote the $T \times 1$ vectors that stack $Y_{it}$, $Y_{it-1}$, and $U_{it}$, respectively, for $t=1,\ldots,T$. Moreover, let $\iota$ be a $T\times 1$ vector of ones and define
$\Sigma_i^{1/2}(\tilde{\gamma}) = \mbox{diag}\big( \sigma_1(h_i, \tilde{\gamma}_1), \ldots, \sigma_T(h_i, \tilde{\gamma}_T) \big)$, 
$S_i(\tilde{\gamma}) = \Sigma_i^{-1/2}(\tilde{\gamma}) \iota$, and $M_i(\tilde{\gamma}) = I - S_i(S_i'S_i)^{-1} S_i'$.
%\begin{eqnarray*}
%	\Sigma_i^{1/2}(\tilde{\gamma}) &=& \mbox{diag}\big( \sigma_1(h_i, \tilde{\gamma}_1), \ldots, \sigma_T(h_i, \tilde{\gamma}_T) \big), \\
%	S_i(\tilde{\gamma}) &=& \Sigma_i^{-1/2}(\tilde{\gamma}) \iota, \quad M_i(\tilde{\gamma}) = I - S_i(S_i'S_i)^{-1} S_i'.
%\end{eqnarray*}
Using this notation, we obtain
\be
    M_i(\tilde{\gamma}) \Sigma_i^{-1/2}(\tilde{\gamma}) \big( Y_i - X_{i} \rho \big) \label{eq:identifyMV} 
     = M_i(\tilde{\gamma}) S_i(\tilde{\gamma}) \lambda_i + M_i(\tilde{\gamma}) \Sigma_i^{-1/2}(\tilde{\gamma}) U_i 
     = M_i(\tilde{\gamma}) V_i. \nonumber 
\ee
This leads to the conditional moment condition
\be
   \mathbb{E} \big[ M_i(\tilde{\gamma}) \Sigma_i^{-1/2}(\tilde{\gamma}) \big( Y_i - X_{i} \rho \big)\big( Y_i - X_{i} \rho \big)'\Sigma_i^{-1/2}(\tilde{\gamma})M_i'(\tilde{\gamma}) - M_i(\tilde{\gamma}) \big| H_i \big] = 0 \label{eq.identification.momcondgamma} 
\ee
if and only if $\tilde{\gamma} = \gamma$, which identifies $\gamma$.
Third, let 
\be
   \tilde{Y}_i = \Sigma_i^{-1/2}(\gamma) \big( Y_i - X_{i} \rho \big) = S_i(\gamma) \lambda_i + V_i.
\ee
The identification of $\pi(\lambda_i|h_i)$ can be established using a characteristic function argument similar to that in \citet{ArellanoBonhomme2012}. For the general model (\ref{eq.yit}) we make the following assumptions:

\begin{assumption} \label{as.identification} \hspace*{1cm}\\[-5ex]
	\setstretch{1}
        \begin{itemize}
                \item [(i)]  The parameter vectors $\alpha$ and $\rho$ are identifiable.
                \item[(ii)] For each $t=1,\ldots,T$ and almost all $h_i$ $\sigma^2_t(h_i,\tilde{\gamma}_t) = \sigma^2_t(h_i,\gamma_t) $ implies $\tilde{\gamma}_t = \gamma_t$. Moreover, $\sigma^2_t(h_i,\gamma_t) > 0.$
                \item[(iii)] The characteristic functions for $\lambda_i|(H_i=h_i)$ and $V_i$ are non-vanishing almost everywhere.
                \item[(iv)] $W_i = [W_{i0},...,W_{iT-1}]'$ has full rank $k_w$. 
        \end{itemize}
\end{assumption}

Because the identification of $\alpha$ and $\rho$ in panel data models with fixed or random effects is well	established, we make the high-level Assumption~\ref{as.identification}(i) that the homogeneous parameters are identifiable.\footnote{Textbook / handbook chapter treatments can be found in, for instance, \cite{Baltagi1995}, \cite{ArellanoHonore2001},  \cite{Arellano2003} and \cite{Hsiao2014}.}
We discuss in the appendix how the  identification argument for $\rho$ in the basic dynamic panel data model can be extended to a more general specification as in (\ref{eq.yit}). Assumption~\ref{as.identification}(ii) enables us to identify the volatility parameters $\gamma$, and (iii) and (iv) deliver the identifiability of the distribution of heterogeneous coefficients. The following theorem summarizes the identification result and is proved in the Appendix.

\begin{theorem} \label{thm.identification} \setstretch{1}
Suppose that Assumptions~\ref{as.likelihood} and~\ref{as.identification} are satisfied. Then the parameters $\alpha$, $\rho$, and $\gamma$ as well as the correlated random effects distribution $\pi(\lambda_i|h_i)$ and the distribution of $V_{it}$ in model (\ref{eq.yit}) are identified.
\end{theorem}

%==============================================================================================================

\section{Decision-Theoretic Foundation} 
\label{sec:decisiontheory}

We adopt a decision-theoretic framework in which forecasts are evaluated based on cross-sectional sums of mean-squared error losses. Such losses are called compound loss functions. Section~\ref{subsec:decisiontheory.risk} provides a formal definition of the compound risk (expected loss). In Section~\ref{subsec:decisiontheory.optimal} we derive the optimal forecasts under the assumption that the cross-sectional distribution of the $\lambda_i$s is known (oracle forecast). While it is infeasible to implement this forecast in practice, the oracle forecast provides a natural benchmark for the evaluation of feasible predictors. Finally, in Section~\ref{subsec:decisiontheory.ratiooptimality} we introduce the concept of ratio optimality, which describes forecasts that asymptotically (as $N \longrightarrow \infty$) attain the same risk as the oracle forecast.

\subsection{Compound Risk}
\label{subsec:decisiontheory.risk}

Let $L( \widehat{Y}_{iT+1},Y_{iT+1}) $ denote the loss associated with
forecast $\hat{Y}_{i,T+1}$ of individual $i^{\prime }s$ time $T+1$
observation, $Y_{iT+1}$. In this paper we consider the conventional quadratic loss function,  
\begin{equation*}
L( \widehat{Y}_{iT+1},Y_{iT+1}) =(\widehat{Y}_{iT+1}-Y_{iT+1})^{2}.
\end{equation*}
The main goal of the paper is to construct optimal forecasts for groups
of individuals selected by a known selection rule in terms of observed
data. We express the selection rule as
\be
D_i=D_i({\cal Y}^N) \in \{0,1\}, \quad i=1,\ldots,N,
\ee
where $D_i({\cal Y}^N)$ is a measurable function of the observations ${\cal Y}^N$, 
${\cal Y}^N = ({\cal Y}_1,\ldots,{\cal Y}_N)$, and ${\cal Y}_i = (Y_i^{0:T},X_i^{1:T},H_i )$.
For instance, suppose that 
$D_i({\cal Y}^N) = \mathbb{I}\{Y_{iT} \in A \}$ for $A \subset \mathbb{R}$. In this case, the selection is homogeneous across $i$ and, for individual $i$, depends only on its own sample. Alternatively, suppose that units are selected based on the ranking of an index, e.g., the empirical quantile of  $Y_{iT}$. In this case, the selection dummy $D_i$ depends on $(Y_{1T},...,Y_{NT})$ and thereby also on the data for the other $N-1$ individuals.

The compound loss of interest is the average of the individual losses
weighted by the selection dummies: 
\begin{equation*}
L_{N}( \widehat{Y}_{T+1}^N,Y_{T+1}^N) 
=\sum_{i=1}^{N} D_i({\cal Y}^N) L( \widehat{Y}_{iT+1}, Y_{iT+1}),
\end{equation*}
where $Y^N_{T+1} = (Y_{1T+1},\ldots,Y_{NT+1})$.
The compound risk is the expected compound loss
\begin{equation}
R_{N}( \widehat{Y}_{T+1}^N) = 
\mathbb{E}_{\theta}^{{\cal Y}^N,\lambda^N,U^N_{T+1}} \left[ L_{N} (\widehat{Y}_{T+1}^N,Y_{T+1}^N) \right]. 
\label{def:compound.risk}
\end{equation}%
We use the $\theta$ subscript for the expectation operator to indicate
that the expectation is conditional on $\theta$.\footnote{Strictly speaking, the expectation also conditions on the deterministic trend terms $W_1$}. The superscript $({\cal Y}^N,\lambda^N,U^N_{T+1})$
indicates that we are integrating with respect to the observed data ${\cal Y}^N$ and the unobserved heterogeneous coefficients $\lambda^N = (\lambda_1,\ldots,\lambda_N)$ and $U^N_{T+1}=(U_{1T+1},\ldots,U_{NT+1})$.

\subsection{Optimal Forecast and Oracle Risk} 
\label{subsec:decisiontheory.optimal}

We now derive the optimal forecast that minimizes the compound risk. 
The risk achieved by the optimal forecast will be called the oracle risk, which is the target risk to achieve.
In the compound decision theory it is assumed that the oracle knows the vector $\theta$
as well as the distribution of the heterogeneous coefficients $\pi(\lambda_i,h_i)$
and observes ${\cal Y}^N$.
However, the oracle does not know the specific $\lambda_i$ for unit $i$.
In order to find the optimal forecast, note that conditional on $\theta$ the compound risk
takes the form of an integrated risk that can be expressed as
\be
  R_{N}( \widehat{Y}_{T+1}^N) = 
  \mathbb{E}_{\theta}^{{\cal Y}^N} \left[ \mathbb{E}_{\theta,{\cal Y}^N}^{\lambda^N,U^N_{T+1}}[ L_{N} (\widehat{Y}_{T+1}^N,Y_{T+1}^N) ] \right]. 
\ee
The inner expectation can be interpreted as posterior risk, which is obtained by conditioning
on the observations ${\cal Y}^N$ and integrating over the heterogeneous parameter $\lambda^N$ and the shocks $U^N_{T+1}$. The outer
expectation averages over the possible trajectories ${\cal Y}^N$. 

It is well known that the integrated risk is minimized by choosing the forecast that minimizes the posterior risk for each realization ${\cal Y}^N$. Using the independence across $i$, the posterior risk can be written as follows:
\begin{eqnarray}
  \lefteqn{\mathbb{E}_{\theta,{\cal Y}^N}^{\lambda^N,U^N_{T+1}} [ L_{N} (\widehat{Y}_{T+1}^N,Y_{T+1}^N) ]}\\    \label{eq.posteriorrisk.decomposition} 
%  &=& \sum_{i=1}^{N} D_i({\cal Y}^N) \mathbb{E}_{\theta,{\cal Y}_i}^{\lambda_i,U_{iT+1}} 
%        [ (\widehat{Y}_{iT+1} - Y_{iT+1})^2]  \nonumber \\
  &=& \sum_{i=1}^{N} D_i ({\cal Y}^N)
  \left\{
  \left(
  \widehat{Y}_{iT+1} - \mathbb{E}_{\theta,{\cal Y}_i}^{\lambda_i,U_{iT+1}}[Y_{iT+1}]
    \right)^2
  +  
   \mathbb{V}_{\theta,{\cal Y}_i}^{\lambda_i,U_{iT+1}}[Y_{iT+1}]
  \right\} \nonumber
\end{eqnarray}
where $\mathbb{V}_{\theta,{\cal Y}_i}^{\lambda_i,U_{iT+1}}[\cdot]$ is the posterior variance.
The decomposition of the risk into a squared bias term and the posterior variance of $Y_{iT+1}$ implies that $\mathbb{E}_{\theta,{\cal Y}_i}^{\lambda_i,U_{iT+1}} [Y_{iT+1}]$ is the optimal predictor. Because $U_{iT+1}$ is mean-independent of $\lambda_i$ and ${\cal Y}_i$, we obtain
\be
\widehat{Y}_{iT+1}^{opt} 
= \mathbb{E}_{\theta,{\cal Y}_i}^{\lambda_i,U_{iT+1}} [Y_{iT+1}]
= \mathbb{E}_{\theta,{\cal Y}_i}^{\lambda_i} [\lambda_i]^{\prime} W_{iT} + \rho'X_{iT} + \alpha' Z_{iT}.
\label{eq.oracle.forecast}
\ee
Note that the posterior expectation
of $\lambda_i$ only depends on observations for unit $i$, even if the selection rule $D_i({\cal Y}^N)$ also depends on the data from other units $j \not=i$. The result is summarized in the following theorem: 

\begin{theorem}[Optimal Forecast]\label{thm:optimal.forecast.corr.random} \setstretch{1}
        Suppose Assumptions \ref{as.likelihood} are satisfied.          
        The optimal forecast that minimizes the composite risk in (\ref{def:compound.risk}) is given by $\widehat{Y}_{iT+1}^{opt}$ in~(\ref{eq.oracle.forecast}). 
        The compound risk of the optimal forecast is 
        \be
        R_{N}^{\text{opt}} 
%         &=& \mathbb{E}_\theta^{{\cal Y}^N}
%                \left[  \sum_{i=1}^{N} 
%                D_i({\cal Y}^N) \mathbb{V}_{\theta,{\cal Y}_i}^{\lambda_i} 
%                \left[ Y_{iT+1} \right]
%                  \right]  \\
         = \mathbb{E}_\theta^{{\cal Y}^N} 
                \left[  \sum_{i=1}^{N} D_i({\cal Y}^N) \left( 
                W_{iT}'\mathbb{V}_{\theta,{\cal Y}_i}^{\lambda_i} 
                \left[ \lambda_i \right] W_{iT}
             + \sigma_{T+1}^{2}(H_i,\gamma_{T+1}) \right) \right]. \label{eq.rnopt}
        \ee
\end{theorem}

According to (\ref{eq.rnopt}), the compound oracle risk has two components. The first component reflects uncertainty with respect to the heterogeneous coefficient $\lambda_i$ and the second component captures uncertainty about the error term $U_{iT+1}$. Unfortunately, the direct implementation of the optimal forecast is infeasible because neither the parameter vector $\theta$ nor the correlated random effect distribution (or prior) $\pi(\cdot)$ are known. Thus, the oracle risk $R_{N}^{\text{opt}}$ provides a lower bound for the risk that is attainable in practice.

%Direct implementation of the optimal forecast is infeasible because neither the parameter vector $\theta$ nor the correlated random effect distribution (or prior) $\pi(\cdot)$ are known. From Section~\ref{sec:implementation} onwards we will pursue an empirical Bayes strategy that replaces the unknown objects by estimates that are based on the cross-sectional information. In this regard the identification result of Section~\ref{subsec:model.identification} will become important. 

\subsection{Ratio Optimality}
\label{subsec:decisiontheory.ratiooptimality}

The identification result presented in Section~\ref{subsec:model.identification} implies that as the cross-sectional dimension $N \longrightarrow \infty$, it might be possible to learn the unknown parameter $\theta$ and random-effects distribution $\pi(\cdot)$ and construct a feasible estimator that asymptotically attains the oracle risk. Following \cite{BrownGreenshtein2009}, we say that a predictor achieves ratio optimality if the regret $R_{N}(\widehat{Y}^N_{T+1}) - R_N^{\text{opt}}$ of the forecast $\widehat{Y}^N_{T+1}$ is negligible relative to the part of the optimal risk that is due
to uncertainty about $\lambda_i$:

\begin{definition}
        \label{def:ratio.optimality} \setstretch{1}
        For a given $\epsilon_0>0$, we say that forecast $\widehat{Y}^N_{T+1}$ achieves  $\epsilon_0$-ratio optimality, if 
        \begin{equation}
        \limsup_{N \rightarrow \infty}
        \dfrac{ R_{N}(\widehat{Y}^N_{T+1}) - R_N^{\text{opt}}}
        {\mathbb{E}_\theta^{{\cal Y}^N} \left[ \sum_{i=1}^N D_i({\cal Y}^N) W_{iT}'\mathbb{V}_{\theta,{\cal Y}_i}^{\lambda_i}[\lambda_i] W_{iT} \right] +  N^{\epsilon_0}} \leq  0.      
        \end{equation}
\end{definition}

Using (\ref{eq.posteriorrisk.decomposition}), the risk differential in the numerator (called regret) can be written as 
\be
 R_{N}(\widehat{Y}^N_{T+1}) - R_N^{\text{opt}} 
 = \mathbb{E}_\theta^{{\cal Y}^N} \left[ \sum_{i=1}^{N} D_i ({\cal Y}^N)
   \left( \widehat{Y}_{iT+1} - \mathbb{E}_{\theta,{\cal Y}_i}^{\lambda_i,U_{iT+1}}[Y_{iT+1}] \right)^2 \right].
\ee
For illustrative purposes, Consider the basic dynamic panel data model~(\ref{m.restricted.linear.panel.regression}).
For this model $\mathbb{E}_{\theta,{\cal Y}_i}^{\lambda_i,U_{iT+1}}[Y_{iT+1}] = \mathbb{E}_{{\cal Y}_i}^{\lambda_i}[\lambda_i] + \rho Y_{iT}$.
A natural class of predictors is given by
$\widehat{Y}_{iT+1} = \widehat{\mathbb{E}}_{{\cal Y}_i}^{\lambda_i}[\lambda_i] + \hat{\rho} Y_{iT}$,
where $\widehat{\mathbb{E}}_{{\cal Y}_i}^{\lambda_i}[\lambda_i]$ is an approximation of the posterior mean of $\lambda_i$
that replaces the unknown $\rho$ and distribution $\pi(\cdot)$ by suitable estimates. The autoregressive coefficient in this model can be $\sqrt{N}$-consistently estimated, which suggests that $\sum_{i=1}^N (\hat{\rho}-\rho)^2Y_{iT}^2 = O_p(1)$. 
Thus, whether a predictor attains ratio optimality crucially depends on the rate at which the discrepancy between
$\mathbb{E}_{{\cal Y}_i}^{\lambda_i}[\lambda_i]$ and $\widehat{\mathbb{E}}_{{\cal Y}_i}^{\lambda_i}[\lambda_i]$ vanishes.

The denominator of the ratio in Definition~\ref{def:ratio.optimality} is divergent. The rate of divergence depends 
on the posterior variance of $\lambda_i$. If the posterior variance is strictly greater than zero, then the denominator is of order $O(N)$. Note that for each unit $i$, the posterior variance is based on a finite number of observations $T$. Thus, for the posterior variance to be equal to zero, it must be the case that the prior density $\pi(\lambda)$ is a pointmass, meaning that
there is a homogeneous intercept $\lambda$. In this case the definition of ratio optimality requires that the regret vanishes at a faster rate, because the rate of the numerator drops from $O(N)$ to $N^{\epsilon_0}$. 
Subsequently, we will pursue an empirical Bayes strategy to construct an approximation $\widehat{\mathbb{E}}_{{\cal Y}_i}^{\lambda_i}[\lambda_i]$ based on the cross-sectional information and show that it attains ratio-optimality. 

In the linear panel literature, researchers often use the first difference to eliminate $\lambda_i$. In this case, the natural forecast of $Y_{iT+1}$ in the basic dynamic panel data model~(\ref{m.restricted.linear.panel.regression}) would be 
$\widehat{Y}_{iT+1}^{FD}(\rho) = Y_{iT} + \rho (Y_{iT} - Y_{iT-1})$,
which is different from $\widehat{Y}_{iT+1}^{opt}$ in (\ref{eq.oracle.forecast}).
Thus, we can immediately deduce from Theorem~\ref{thm:optimal.forecast.corr.random} that $\widehat{Y}_{iT+1}^{FD}(\rho)$ is not an optimal forecast.
The quasi-differencing of $Y_{it}$ introduces a predictable moving-average error term that is
ignored by the predictor $\widehat{Y}_{iT+1}^{FD}(\rho)$.

\section{Implementation of the Optimal Forecast}
\label{sec:implementation}

We will construct a consistent approximation of the posterior mean $\mathbb{E}_{\theta,{\cal Y}^i}^{\lambda_i,U_{iT+1}}[\lambda_i]$ using a convenient formula which is named after the statistician Maurice Tweedie (though it had been previously derived by the astronomer Arthur Eddington). This formula is presented in Section~\ref{subsec:implementation.tweedie}. In Section~\ref{subsec:implementation.tweedie.parametric} we discuss the parametric estimation of the correction term and in Section~\ref{subsec:implementation.tweedie.nonparametric} we consider a nonparametric kernel-based estimation. The QMLE and Generalized Method-of-Moments (GMM) estimation of the parameter $\theta$ are discussed in Sections~\ref{subsec:implementation.theta.qmle} and~\ref{subsec:implementation.theta.gmm}.

\subsection{Tweedie's Formula}
\label{subsec:implementation.tweedie}

When the innovations $U_{it}$ are conditionally normally distributed, we can derive 
a convenient formula for the posterior expectation 
$\mathbb{E}_{\theta,{\cal Y}_i}^{\lambda_i} [\lambda_i]$ of the individual heterogeneous parameter $\lambda_i$.

\begin{assumption} \label{as:error.normal}  The unpredictable shock $V_{it}$ has a standard normal distribution:
        \[ 
        V_{it} \: | \: (Y_{i}^{1:t-1}, X_{i}^{0:t-1},W_{2i}, Z_{i},\lambda_i) 
        \sim N(0,1), \quad t=1,...,T.
        \] 
\end{assumption}

The assumption of normally distributed $V_{it}$'s is not as restrictive as it may seem. Recall that the shocks $U_{it}$ are defined as $V_{it} \sigma_t(X_{i0},W_{2,i}^{0:T},Z_i^{0:T},\gamma_t)$. Thus, due to the potential heteroskedasticity, the distribution of shocks is a mixture of normals. The only restriction is that the random variables characterizing the scale of the mixture component are observed. Moreover, even in the homoskedastic case $\sigma_t = \sigma$, the distribution of $Y_{it}$ given the regressors is non-normal because the distribution of the $\lambda_i$ parameters is fully flexible.
Using Assumption \ref{as:error.normal} we will now further manipulate the density $p(y_i,x_{2,i},\lambda_i|h_i,\theta)$ in~(\ref{eq.likelihoodprior}).\footnote{In principle, the normality assumption could be generalized to the assumption that the distribution of $V_{it}$ belongs to the exponential family.} To simplify the notation we will drop the $i$ subscript. Define 
\be
\tilde{y}_t(\theta) = y_t - \rho'x_{t-1} - \alpha'z_{t-1}, \quad
\Sigma(\theta) = \mbox{diag}(\sigma_1^2,\ldots,\sigma^2_T),
\label{eq.likelihood.deftildey}
\ee
and let $\tilde{y}(\theta)$ and $w$ be matrices with rows $\tilde{y}_t(\theta)$ and $w_{t-1}'$, $t=1,...,T$. Because the subsequent calculations condition on $\theta$ we will omit the $\theta$-argument from $\tilde{y}$, $\Sigma$, and functions thereof. Replacing 
$\varphi(v)$ in~(\ref{eq.likelihoodprior}) with a Gaussian density function we obtain:
\begin{eqnarray*}
\lefteqn{ p(y,x_2,\lambda|h,\theta) } \label{eq.likelihoodprior.Gaussian}\\
&\propto& \exp \left\{ - \frac{1}{2} (\hat{\lambda} - \lambda)'w'\Sigma^{-1}w (\hat{\lambda} - \lambda) \right\}       
\exp \left\{ - \frac{1}{2} (\tilde{y} - w \hat \lambda)'\Sigma^{-1} 
(\tilde{y} - w \hat \lambda ) \right\} \pi(\lambda|h). \nonumber 
\end{eqnarray*}
The factorization of $p(y,x_2,\lambda|h,\theta)$ implies that 
\begin{equation}
\hat{\lambda} = (w'\Sigma^{-1}w)^{-1}w'\Sigma^{-1}\tilde{y} \label{def.sufficient.stat}
\end{equation}
is a sufficient statistic and that we can express the posterior distribution of $\lambda$ as
\[
p(\lambda|y,x_2,h,\theta) 
= p(\lambda|\hat{\lambda},h,\theta)
= \frac{ p(\hat{\lambda}|\lambda,h,\theta) \pi(\lambda|h) }
{ p(\hat{\lambda}| h,\theta) },    
\]
where 
\be
p(\hat{\lambda}|\lambda,h,\theta) 
= (2 \pi)^{-k_w/2} |w'\Sigma^{-1}w|^{1/2} 
\exp \left\{ - \frac{1}{2} (\hat{\lambda} - \lambda)'w'\Sigma^{-1}w (\hat{\lambda} - \lambda) \right\}.
\label{eq.lambdahat.likelihood}
\ee

To obtain a representation for the posterior mean, we now 
differentiate the equation $\int  p(\lambda|\hat{\lambda}, h,\theta) d\lambda = 1$
with respect to $\hat{\lambda}$. Exchanging the order of integration and differentiation and using the properties of the exponential function, we obtain
\begin{eqnarray*}
        0 
        &=& w'\Sigma^{-1}w \int (\lambda - \hat{\lambda}) 
        p(\lambda|\hat{\lambda}, h,\theta) d\lambda
        - \frac{\partial}{\partial \hat{\lambda}} 
        \ln p(\hat{\lambda}|h,\theta) \\
        &=& w'\Sigma^{-1} w \big(  \mathbb{E}_{\theta,{\cal Y}}^\lambda[\lambda]  - \hat{\lambda} \big)
        - \frac{\partial}{\partial \hat{\lambda}} 
        \ln p(\hat{\lambda}|h,\theta) .
\end{eqnarray*}
Solving this equation for the posterior mean yields Tweedie's formula, which is summarized in the following theorem.

\begin{theorem}\label{thm:tweedie} \setstretch{1}
        Suppose that Assumptions~\ref{as.likelihood} and \ref{as:error.normal} hold. The posterior mean
        of $\lambda_i$ has the representation
        \begin{equation}        
        \mathbb{E}_{\theta,{\cal Y}_i}^{\lambda_i}[\lambda_i]
        = \hat{\lambda}_i(\theta) + \bigg( W_i^{0:T-1'}\Sigma^{-1}(\theta) W_i^{0:T-1} \bigg)^{-1}  
        \frac{\partial}{\partial \hat{\lambda}_i(\theta)} \ln p(\hat{\lambda}_i(\theta)|H_i,\theta).
        \label{eq.tweedie}
        \end{equation} 
        The optimal forecast is given by
        \begin{eqnarray}
        \widehat{Y}_{iT+1}^{opt}(\theta) &=& 
        \left( \hat{\lambda}_i(\theta) + \bigg( W_i^{0:T-1'}\Sigma^{-1}(\theta) W_i^{0:T-1} \bigg)^{-1}  
        \frac{\partial}{\partial \hat{\lambda}_i(\theta)} \ln p(\hat{\lambda}_i(\theta)|H_i,\theta)
        \right)'W_{T+1} \nonumber
        \\
        &&+ \rho'X_{iT} + \alpha'Z_{iT}. \label{eq.tweedie.opt.forecast}
        \end{eqnarray}
\end{theorem}

Tweedie's formula was used by \cite{Robbins1951} to estimate a vector of means $\lambda^N$ for the model $Y_i|\lambda_i \sim N(\lambda_i,1)$, $\lambda_i \sim \pi(\cdot)$, $i=1,\ldots,N$. Recently, it was extended by \cite{Efron2011} to the family of exponential distribution, allowing for a unknown finite-dimensional parameter $\theta$. Theorem \ref{thm:tweedie} extends Tweedie's formula to the estimation of correlated random effect parameters in a dynamic panel regression setup. 

The posterior mean takes the form of the sum of the sufficient statistic $\hat{\lambda}_i(\theta)$ and a correction term that reflects the prior distribution of $\lambda_i$. The correction term is expresses as a function of the marginal density of the sufficient statistic $\hat{\lambda}_i(\theta)$ conditional on $H_i$ and $\theta$. Thus, it is not necessary to solve a deconvolution problem that separates the prior density $\pi(\lambda_i|h_i)$ from the distribution of the error terms $V_{it}$.
We expressed Tweedie's formula in (\ref{eq.tweedie}) in terms of the conditional density $p(\hat{\lambda}_i(\theta)|H_i,\theta)$. However, because the posterior mean is a function of the log density differentiated with respect to $\hat{\lambda}_i(\theta)$, the conditional density can be replaced by a joint density:
\[
\frac{\partial}{\partial \hat{\lambda}_i(\theta)} \ln p(\hat{\lambda}_i(\theta)|H_i,\theta) = \frac{\partial}{\partial \hat{\lambda}_i(\theta)} \ln p(\hat{\lambda}_i(\theta),H_i|\theta).
\]
The construction of ratio-optimal forecasts relies on replacing the density $p(\hat{\lambda}_i(\theta),H_i|\theta)$ and the common parameter $\theta$ by consistent estimates.

\subsection{Parametric Estimation of Tweedie Correction}
\label{subsec:implementation.tweedie.parametric}

If the random-effects distribution $\pi(\lambda|h_i)$ is Gaussian, then it is possible to derive the marginal density of the sufficient statistic $p(\hat{\lambda}_i(\theta)|h_i,\theta)$ analytically. Let
\be
    \lambda_i | (H_i,\theta) \sim N \big( \Phi H_i, \underline{\Omega} \big).
    \label{eq.parametric.pi}
\ee
Moreover, define $\xi = \big( \mbox{vec}(\Phi), \, \mbox{vech}(\underline{\Omega}) \big)'$. To highlight the dependence of the correlated random-effects distribution on the hyperparameter $\xi$ we will write $\pi(\lambda_i|h_i,\xi)$. The marginal density (omitting the $i$ subscripts and the $\theta$-argument of $\hat{\lambda}$) is given by
\begin{eqnarray}
 p \big(\hat{\lambda}(\theta) \big| h,\theta,\xi \big)
  &=& \int p \big(\hat{\lambda}(\theta) | \lambda, h, \theta \big) \pi(\lambda|h,\xi) d \lambda \\
  &=& (2 \pi)^{-k_w/2} \big|\underline{\Omega}^{-1} \big|^{1/2} \big|w'\Sigma^{-1} w \big|^{1/2} \big|\bar{\Omega}\big|^{1/2} \nonumber \\
  & & \times \exp \left\{ -\frac{1}{2} \big( \hat{\lambda}'w'\Sigma^{-1} w \hat{\lambda} +
  h'\Phi' \underline{\Omega}^{-1} \Phi h - \bar{\lambda}' \bar{\Omega}^{-1} \bar \lambda \big)\right\}. \nonumber 
\end{eqnarray}
Here, we used the likelihood of $\hat{\lambda}$ in~(\ref{eq.lambdahat.likelihood}), the density associated with the Gaussian prior in~(\ref{eq.parametric.pi}), and then the properties of a multivariate Gaussian density to integrate out $\lambda$. The terms $\bar{\lambda}$ and $\bar{\Omega}$ are the posterior mean and variance of $\lambda$, respectively:
\[
  \bar{\Omega}^{-1} = \underline{\Omega}^{-1} + w'\Sigma^{-1} w, \quad
  \bar{\lambda}     = \bar{\Omega} \big( \underline{\Omega}^{-1} \Phi h  + w'\Sigma^{-1} w \hat{\lambda} \big).
\]

Conditional on $\theta$ the vector of hyperparameters $\xi$ can be estimated by maximizing the marginal likelihood
\be
        \hat{\xi}(\theta) = \mbox{argmax}_\xi \; \prod_{i=1}^N p(\hat{\lambda}_i(\theta)|h_i,\theta,\xi) 
        \label{eq.xihat.theta}
\ee
using the cross-sectional distribution of the sufficient statistic. Tweedie's formula can then be evaluated based on $p \big(\hat{\lambda}_i(\theta)|h_i,\theta,\hat{\xi}(\theta) \big)$. In principle it is possible to replace the Gaussian prior distribution with a more general parametric distribution. However, in general it will not be possible to derive an analytical formula for the marginal likelihood. 

\subsection{Nonparametric Estimation of Tweedie Correction}
\label{subsec:implementation.tweedie.nonparametric}

A nonparametric implementation of the Tweedie correction can be obtained by replacing
$p(\hat{\lambda}_i(\theta),h_i|\theta)$ and its derivative with respect to $\hat{\lambda}_i(\theta)$ with a Kernel density estimate, e.g.,
\begin{eqnarray}
 \lefteqn{ \hat{p}(\hat{\lambda}_i(\theta),h_i|\theta)} \label{eq.density.nonparametric}\\
  &=& \frac{1}{N} \sum_{j=1}^N \bigg[ (2\pi)^{-k_w/2}|B_N|^{-k_w} |V_{\hat{\lambda}}|^{-1/2} \exp \left\{ -\frac{1}{2B_N^2} \big( \hat{\lambda}_i(\theta) - \hat{\lambda}_j(\theta) \big)' V_{\hat{\lambda}}^{-1} \big( \hat{\lambda}_i(\theta) - \hat{\lambda}_j(\theta) \big) \right\} \nonumber\\
   && \times (2\pi)^{-k_h/2}|B_N|^{-k_h} |V_h|^{-1/2} \exp \left\{ -\frac{1}{2B_N^2} \big( h_i - h_j\big)' V_h^{-1} \big( h_i - h_j \big)  \right\} \bigg] \nonumber,
\end{eqnarray}
where $B_N$ is the bandwidth and $V_{\hat{\lambda}}$ and $V_h$ are tuning matrices.
Note that even if the prior distribution $\pi(\lambda)$ is a pointmass, the sufficient statistic $\hat{\lambda}$ in (\ref{def.sufficient.stat}) has a continuous distribution and one can use a kernel density estimator to construct the Tweedie correction. 

If the dimension of the conditioning variables $H_i$ is large, the nonparametric estimation suffers from the curse of dimensionality. In this case, one may reduce the dimension of the conditioning set with some smaller dimensional indices, e.g., by assuming that $\lambda_i$ and $H_i$ dependent only through $\bar{H}_{i} = \frac{1}{T} \sum_{t=1}^T H_{it}$, that is, $\pi(\lambda|h) = \pi(\lambda|\bar{h})$.
In Section~\ref{sec:ratio.optimality} we provide a detailed analysis of the Gaussian kernel estimator in the context of the basic dynamic panel data model in~(\ref{m.restricted.linear.panel.regression}) with time-homoskedastic innovations.

\subsection{QMLE Estimation of $\theta$}
\label{subsec:implementation.theta.qmle}

Notice that under Assumption \ref{as:error.normal}, $\hat{\lambda}_i(\theta)$ in (\ref{def.sufficient.stat}) is a sufficient statistic of $\lambda_i$ conditional on $\theta,h_i$, and $ \pi_{\lambda}(\lambda_i|h_i,\xi)$ is the parametric version of the correlated random effect density. 
Integrating out $\lambda$ under a parametric correlated random effect (or prior) distribution $\pi_{\lambda}(\lambda|x_{0},w_2,z,\xi)$, we have (omitting the $i$ subscripts)
\begin{eqnarray}
        \lefteqn{p(y,x_{2}|h,\theta,\xi)} \label{eq.qmleobj.thetahat.xihat} \\
        &=& \int p(y,x_{2}|h,\theta,\lambda) \pi_{\lambda}(\lambda|h,\hat{\xi}(\theta)) d\lambda  \nonumber \\
        &\propto& |\Sigma(\theta)|^{-1/2} \exp \left\{ - \frac{1}{2} \big(\tilde{y}(\theta)-w\hat{\lambda}(\theta)\big)'\Sigma^{-1}(\theta)\big(\tilde{y}(\theta)-w\hat{\lambda}(\theta)\big) \right\} \nonumber \\
        && \times \int \exp \left\{ -\frac{1}{2}\big(\hat{\lambda}(\theta)-\lambda\big)'w'\Sigma^{-1}(\theta)w\big(\hat{\lambda}(\theta)-\lambda\big)\right\}
        \pi_{\lambda}\big(\lambda(\theta)|h,\hat{\xi}(\theta)\big) d\lambda \nonumber \\
        &\propto& |\Sigma(\theta)|^{-1/2} \exp \left\{ - \frac{1}{2} \big(\tilde{y}(\theta)-w\hat{\lambda}(\theta)\big)'\Sigma^{-1}(\theta)\big(\tilde{y}(\theta)-w\hat{\lambda}(\theta)\big) \right\} \nonumber \\
        && \times \big| w'\Sigma^{-1} w \big|^{-1/2}
        p(\hat{\lambda}(\theta)|h,\theta,\xi). \nonumber
\end{eqnarray}
Here, we used the definition of $\tilde{y}(\theta)$ in (\ref{eq.likelihood.deftildey}) and the product of Gaussian likelihood and prior in (\ref{eq.likelihoodprior.Gaussian}). Note that the term $p(\hat{\lambda}(\theta)|h,\theta,\xi)$ in the last line of (\ref{eq.qmleobj.thetahat.xihat}) is identical to the objective function for $\xi$ used in (\ref{eq.xihat.theta}). Thus, we can now jointly determine $\theta$ and $\xi$ by maximizing the integrated likelihood as a function:
\be
\big( \hat{\theta}_{QMLE},\hat{\xi}_{QMLE} \big) = \mbox{argmax}_{\theta,\xi} \; \prod_{i=1}^N 
p(y_i,x_{2i}|h_i,\theta,\xi).
\label{eq.qmle.thetahat.xihat}
\ee
We refer to this estimator as {\em quasi} (Q) maximum likelihood estimator (MLE), because the correlated random effects distribution could be misspecified.

\subsection{GMM Estimation of $\theta$}
\label{subsec:implementation.theta.gmm}

Without a convenient assumption about the random effects distribution, one can estimate the parameter $\theta$ using a sample analogue of the moment conditions that were used in the identification analysis in Section \ref{sec:model}. 
For $t=1,\ldots,T-k_w$, define 
\be
        Y_{it}^{*} = Y_{it} - 
        \left( \sum_{s=t+1}^T Y_{is} W_{is-1}' \right) \left( \sum_{s=t+1}^T W_{is-1} W_{is-1}^{\prime}\right)^{-1} W_{it-1}. \label{eq.forwarddetrendy}
\ee
Moreover, define $X_{it-1}^{*}$ and $Z_{it-1}^{*}$ by replacing $Y_{i\cdot}$ in (\ref{eq.forwarddetrendy}) with $X_{i\cdot}$ and $Z_{i\cdot}$, respectively, and let
%\begin{eqnarray*}
%	Y_{it}^{*} &=& Y_{it} - 
%	\left( \sum_{s=t+1}^T Y_{is} W_{is-1}' \right) \left( \sum_{s=t+1}^T W_{is-1} W_{is-1}^{\prime}\right)^{-1} W_{it-1} \\
%	X_{it-1}^{*} &=& X_{it-1} - 
%	\left( \sum_{s=t+1}^T X_{is-1}W_{is-1}'\right) \left( \sum_{s=t+1}^T W_{is-1} W_{is-1}^{\prime}\right)^{-1} W_{it-1} \\
%	Z_{it-1}^{*} &=& Z_{it-1} - 
%	\left( \sum_{s=t+1}^T Z_{is-1} W_{is-1}' \right) \left( \sum_{s=t+1}^T W_{is-1} W_{is-1}^{\prime}\right)^{-1}  W_{it-1}
%\end{eqnarray*}
\[
g_{it}(\rho,\alpha) = (Y_{it}^{*} - \rho^{\prime}X_{it-1}^{*} - \alpha^{\prime}Z_{it-1}^{*})
\left[
\begin{array}{c}
        X_i^{0:t-1}\\ 
        Z_i^{0:T}
\end{array} \right], \quad
g_i(\rho,\alpha) = \big[ g_{i1}(\rho,\alpha)', \ldots ,g_{iT-k_{w}}(\rho,\alpha)' \big]'.
\]
%\begin{eqnarray*}
%	g_{it}(\rho,\alpha) &=& (Y_{it}^{*} - \rho^{\prime}X_{it-1}^{*} - \alpha^{\prime}Z_{it-1}^{*})
%	\left[
%	\begin{array}{c}
%		X_i^{0:t-1}\\ 
%		Z_i^{0:T}
%	\end{array} \right] \\
%	g_i(\rho,\alpha) &=& \big[ g_{i1}(\rho,\alpha)', \ldots ,g_{iT-k_{w}}(\rho,\alpha)' \big]'
%\end{eqnarray*}
The continuous-updating GMM estimator of $\rho$ and $\alpha$ solves
\begin{eqnarray}
(\hat{\rho}_{GMM},\hat{\alpha}_{GMM})  \label{def:rhoalphahatGMM} 
= \argmin_{\rho,\alpha} 
\left( \sum_{i=1}^N g_i(\rho,\alpha) \right)'
\left( \sum_{i=1}^N g_i(\rho,\alpha)g_i(\rho,\alpha)' \right)^{-1}
\left( \sum_{i=1}^N g_i(\rho,\alpha) \right). 
\end{eqnarray}
%\begin{eqnarray}
%\lefteqn{ (\hat{\rho}_{GMM},\hat{\alpha}_{GMM}) } \label{def:rhoalphahatGMM} \\
%&=& \argmin_{\rho,\alpha} 
%\left( \sum_{i=1}^N g_i(\rho,\alpha) \right)'
%\left( \sum_{i=1}^N g_i(\rho,\alpha)g_i(\rho,\alpha)' \right)^{-1}
%\left( \sum_{i=1}^N g_i(\rho,\alpha) \right). \nonumber
%\end{eqnarray}
This estimator was proposed by \cite{ArellanoBover1995} and we will refer to it as GMM(AB) estimator in the Monte Carlo simulations (Section~\ref{sec:monte-carlo-simulations}) and the empirical application (Section~\ref{sec:empiricalapplication}).\footnote{There exists a large literature on the estimation of dynamic panel data models. Alternative estimators include \cite{ArellanoBond1991} and \cite{BlundellBond1998}.}

To estimate the heteroskedasticity parameter $\gamma = [\gamma_1,...,\gamma_T]'$ in $\sigma^2_t(H_i,\gamma_t)$, define:
\begin{eqnarray*}
	\tilde{Y}_i(\hat{\rho},\hat{\alpha}) &=& Y_i - X_{i,-T} \hat{\rho} - Z_{i,-T} \hat{\alpha}, \quad
	\Sigma_i^{1/2}(\gamma) = \mbox{diag}\big( \sigma_1(h_i, \gamma_1), \ldots, \sigma_T(h_i, \gamma_T) \big), \\
	S_i(\gamma) &=& \Sigma_i^{-1/2}(\gamma) W_i, \quad M_i(\gamma) = I - S_i(S_i'S_i)^{-1} S_i',
\end{eqnarray*}
where $\hat{\rho}$ and $\hat{\alpha}$ could be the estimators in (\ref{def:rhoalphahatGMM}).
We use the sample analogue to a set of moment condition implied by a generalization of  (\ref{eq.identification.momcondgamma}):
\begin{eqnarray}
\hat{\gamma}_{GMM}  \label{def:gammahatGMM} 
&=& \mbox{argmin}_\gamma \; \frac{1}{N} \sum_{i=1}^N \bigg\| B \, \mbox{vec} \bigg( M_i(\gamma) \Sigma_i^{-1/2}(\gamma) \tilde{Y}_i(\hat{\rho},\hat{\alpha}) \\
&& \times \tilde{Y}_i'(\hat{\rho},\hat{\alpha}) \Sigma_i^{-1/2}(\gamma) M_i(\gamma) - M_i(\gamma) \bigg) \bigg\|^2, \nonumber
\end{eqnarray}
where $B$ is a selection matrix that can be used to eliminate off-diagonal elements of the covariance matrix. In population, these off-diagonal elements should be zero, because the $U_{it}$'s are assumed to be uncorrelated across time. 

\subsection{Extension to Multi-Step Forecasting}

 While this paper focuses on single-step forecasting, we briefly discuss in the context of the basic dynamic panel data model how the framework can
	be extended to multi-step forecasts. We can express
\[
   Y_{iT+h} =  \left(\sum_{s=0}^{h-1} \rho^s\right) \lambda_i + \rho^h Y_{iT}  + \sum_{s=0}^{h-1} \rho^2 U_{iT+h-s}. 
\] 
Under the assumption that the oracle knows $\rho$ and $\pi(\lambda_i,Y_{i0})$ we can express the oracle forecast
as
\[
  \widehat{Y}^{opt}_{iT+h} = \left(\sum_{s=0}^{h-1} \rho^s\right) \mathbb{E}_{\theta,{\cal Y}_i}^{\lambda_i}[\lambda_i] 
  + \rho^h Y_{iT}.
\]
As in the case of the one-step-ahead forecasts, the posterior mean $\mathbb{E}_{\theta,{\cal Y}_i}^{\lambda_i}[\lambda_i]$ can be replaced by an approximation based on Tweedie's formula and the $\rho$'s can be replaced by consistent estimates. A model with additional covariates would require external multi-step forecasts of the covariates, or the specification in (\ref{eq.yit}) would have to be modified such that all exogenous regressors appear with an $h$-period lag.

\section{Ratio Optimality in the Basic Dynamic Panel Model}\label{sec:ratio.optimality}

Throughout this section we will consider the basic dynamic panel data model with homoskedastic Gaussian innovations:
\be
   Y_{it} = \lambda_i + \rho Y_{it-1} + U_{it}, \quad U_{it} \sim iidN(0,\sigma^2), \quad
   (\lambda_i,Y_{i0}) \sim \pi(\lambda,y_{i0}).
   \label{m.restricted.linear.gaussian.panel.regression}
\ee
We will prove that ratio optimality for a general prior density $\pi(\lambda_i|h_i)$ can be achieved with a Kernel estimator of the joint density of the sufficient statistic and initial condition: $p(\hat{\lambda}_i(\theta),H_i|\theta)$. The proof of the main result  is a significant generalization of the proof in \cite{BrownGreenshtein2009} for a vector of means to the dynamic panel data model with estimated common coefficients.

For the model in (\ref{m.restricted.linear.gaussian.panel.regression}), 
the sufficient statistic is given by 
\be
\hat{\lambda}_i(\rho) = \frac{1}{T} \sum_{t=1}^{T} (Y_{it} -\rho Y_{it-1})
\label{eq.simplemodel.lambdahat}
\ee
and the posterior mean of $\lambda_i$ simplifies to
\be
  \mathbb{E}_{\theta,{\cal Y}_i}^{\lambda_i} [\lambda_i] =  \mu\big( \hat{\lambda}_i(\rho),\sigma^2/T,p(\hat{\lambda}_i,Y_{i0})\big)
    = \hat{\lambda}_i(\rho) + \frac{\sigma^2}{T} \frac{\partial}{\partial \hat{\lambda}_i(\theta)} \ln p(\hat{\lambda}_i(\rho),Y_{i0}).
    \label{eq.simplemodel.lambdapostmean}
\ee
The formula recognizes that the heterogeneous coefficient is a scalar intercept and that the errors are homoskedastic. We simplified the notation by writing 
$p(\hat{\lambda}_i(\rho),Y_{i0})$ instead of $p(\hat{\lambda}_i(\rho),Y_{i0}|\theta)$. This simplification is justified because we will estimate the density of $(\hat{\lambda}_i(\rho),Y_{i0})$ directly from the data; see (\ref{eq.kernel.plambday0}) below. We will use the notation $\mu(\cdot)$ to refer to the conditional mean as function 
of the sufficient statistic $\hat{\lambda}$, the scale factor $\sigma^2/T$, and the density $p(\hat{\lambda}_i,Y_{i0})$.

To facilitate the theoretical analysis, we make two adjustments to the 
posterior mean predictor of $Y_{iT+1}$. First, we replace the kernel density estimator 
of  $(\hat{\lambda}_i(\rho),Y_{i0})$ given in (\ref{eq.density.nonparametric}) by a leave-one-out estimator of the form:
\be
\hat{p}^{(-i)} ( \hat{\lambda}_i(\rho), Y_{i0})
= \frac{1}{N-1} \sum_{j\not=i} 
\frac{1}{B_N} \phi\left( \frac{\hat{\lambda}_j(\rho) - \hat{\lambda}_i(\rho) }{B_N}\right)
\frac{1}{B_N} \phi\left( \frac{Y_{j0} - Y_{i0}}{B_N}\right),
\label{eq.kernel.plambday0}
\ee   
where $\phi(\cdot)$ is the pdf of a $N(0,1)$. Using the fact that the observations are cross-sectionally independent and conditionally normally distributed one can directly compute the expected value of the leave-one-out  estimator:
\begin{eqnarray}
        \mathbb{E}_{\theta,{\cal Y}_i}^{{\cal Y}^{(-i)}}[\hat{p}^{(-i)}(\hat{\lambda}_i,y_{i0}) ]  
        &=& \int 
        \frac{1}{\sqrt{\sigma^2/T+B_N^2}} \phi\left( \frac{\hat{\lambda}_i - \lambda_i }{\sqrt{\sigma^2/T+B_N^2}}\right) \label{eq.kernel.plambday0.expectations} \\
        && \times
        \left[ \int \frac{1}{B_N} \phi\left( \frac{y_{i0} - \tilde{y}_{i0}}{B_N}\right)  p(\tilde{y}_{i0}|\lambda_i)  d \tilde{y}_{i0} \right]  p(\lambda_i) d\lambda_i. \nonumber
\end{eqnarray}
Taking expectations of the kernel estimator leads to a variance adjustment for conditional distribution of
$\hat{\lambda}_i|\lambda_i$ ($\sigma^2/T+B_N^2$ instead of $\sigma^2/T$) and the density of $y_{i0}|\lambda_i$ is replaced by a convolution. 

Second, we replace the scale factor $\hat{\sigma}^2/T$ in the posterior mean function $\mu(\cdot)$ by $\hat{\sigma}^2/T+B_N^2$, which is the term that appears in (\ref{eq.kernel.plambday0.expectations}).
Moreover, we truncate the absolute value of the posterior mean function from above.  
For $C > 0$ and for any $x \in \mathbb{R}$, define $\left[x\right]^C := \sgn(x) \min \{ |x|,C\}$.
Then
\be
\widehat{Y}_{iT+1}
= \left[
\mu \big( \hat{\lambda}_i(\hat{\rho}),\hat{\sigma}^2/T +B_N^2,\hat {p}^{-i}(\cdot) \big)
\right]^{C_N}  + \hat{\rho} Y_{iT},
\label{def:optimal.forecast}
\ee
where $C_N \longrightarrow \infty$ slowly.
Formally, we make the following technical assumptions.

\begin{assumption} [Marginal distribution of $\lambda_i$] \label{as:ratio.optimality.marginal.dist.lambda} \setstretch{1}
	The marginal density of $\lambda_i$, $\pi(\lambda)$ has support $\Lambda^{\pi} \subset [-C_N,C_N]$, where for any $\epsilon>0$, $C_N = o(N^{\epsilon})$.

\end{assumption}

\begin{assumption} [Bandwidth] \label{as:ratio.optimality.bandwidth} \setstretch{1}
	 Let $C_N' = (1+k) (\sqrt{\ln N} + C_N)$, where $k$ is a constant such that $ k> \max\{0,\sqrt{2\sigma^2/T}-1\}$.
     The bandwidth for the kernel density estimator, $B_N$, satisfies the following conditions: (i) for any $\epsilon > 0$, $1/B_N^2 = o(N^\epsilon)$; (ii) $B_N(C_N'+2C_N) = o(1)$.  
\end{assumption}

\begin{assumption}[Conditional distribution of $Y_{i0}|\lambda_i$] \setstretch{1} \label{as:tail.intial.condition} Let $\mathcal{Y}_{\lambda}^{\pi}$ be the  support of the conditional density $\pi(y_{i0}|\lambda_i)$. The conditional density of $Y_{i0}$ conditioning on $\lambda_i = \lambda$, $\pi(y|\lambda)$, satisfies the following three conditions: (i)   $0< \pi(y|\lambda) < M$ for $y \in \mathcal{Y}_{\lambda}^{\pi}$ and $\lambda \in \Lambda^{\pi}$. (ii) There exists a finite constant $\bar{C}$ such that for any large value $C > \bar{C},$
         \[
         \max\left\{ \int_{C}^{\infty} \pi(y|\lambda)dx, 
         \int_{-\infty}^{-C} \pi(y|\lambda)dy \right\} \leq \exp(- m(C,\lambda)),
         \]
         where the function $m(C,\lambda)>0$ satisfies the following: $m(C,\lambda)$ is an increasing function of $C$ for each $\lambda$ and there exists finite constants $K>0$ and $\epsilon \geq 0$ such that 
         $$ \liminf_{N \longrightarrow \infty} \, \inf_{| \lambda| \leq C_N} \, 
         \left( m \left(K ( \sqrt{\ln N} + C_N),\lambda \right) - (2+\epsilon)\ln N \right)  \geq 0.$$
(iii) The following holds uniformly in $y \in  \mathcal{Y}_{\lambda}^{\pi} \cap [-C_N',C_N]$ and $\lambda \in \Lambda^{\pi}$: 
    \[
    \int \frac{1}{B_N} \phi\left( \frac{\tilde{y} - y }{B_N} \right) \pi(\tilde{y}|\lambda)d \tilde{y}  = \big(1 +o(1)\big) \pi(y|\lambda).
    \]
\end{assumption}

\begin{assumption}[Estimators of $\rho$ and $\sigma^2$] \label{as:rhohat.ratio.optimality} \setstretch{1}
        There exist estimators $\hat{\rho}$ and $\hat{\sigma}^2$ such that for any $\epsilon > 0,$ (i) $\mathbb{E}_\theta^{{\cal Y}^N} \big[ |\sqrt{N} (\hat{\rho} -\rho)|^4 \big] \leq o(N^{\epsilon})$, (ii) $\mathbb{E}_\theta^{{\cal Y}^N} \big[ \hat{\sigma}^4 \big] \leq
        o(N^{\epsilon})$, and (iii) $\mathbb{E}_\theta^{{\cal Y}^N} \big[ |\sqrt{N} (\hat{\sigma}^2 -\sigma^2) |^2 \big] \leq 
        o(N^{\epsilon})$.
\end{assumption}

We factorize the correlated random effects distribution as $\pi(\lambda_i,y_{i0}) = \pi(\lambda_i)\pi(y_{i0}|\lambda_i)$ and impose regularity conditions on the marginal distribution of the heterogeneous coefficient and the conditional distribution of the initial condition. 
In Assumption~\ref{as:ratio.optimality.marginal.dist.lambda} we let the support of $\pi(\lambda_i)$ slowly expand with the sample size
by assuming that $C_N$ grows at a subpolynomial rate. Assumption~\ref{as:ratio.optimality.bandwidth} provides an upper and a lower bound for the rate at which the bandwidth of the kernel estimator shrinks to zero. Note that for technical reasons the assumed rate is much slower than in typical density estimation problems.\footnote{In a nutshell, we need to 
	control the behavior of $\hat{p}(\hat{\lambda}_i,Y_{i0})$ and its derivative uniformly, which, in certain steps of the proof, requires us to consider bounds of the form $M/B_N^2$, where $M$ is a generic constant. If the bandwidth shrinks too fast, the bounds diverge too quickly to ensure that it suffices to standardize the regret in Definition~\ref{def:ratio.optimality} by $N^{\epsilon_0}$ if the $\lambda_i$ coefficients are identical for each cross-sectional unit.}

Assumption \ref{as:tail.intial.condition} imposes regularity conditions on the conditional density of the initial observation.
In (i) we assume that $\pi(y_{i0}|\lambda_i)$ is bounded. In (ii) we control the tails of the distribution. In the first constraint on $m(C,\lambda)$ we essentially assume that the density of $y_{i0}$ has exponential tails. This also guarantees that the fourth moment of $Y_{i0}$ exists. In part (iii) we assume that $\pi(y|\lambda)$ is sufficiently smooth with respect to $y$ such that the convolution on the left-hand side uniformly converges to $\pi(y|\lambda)$ as the bandwidth $B_N$ tends to zero.
We verify in the Appendix that a $\pi(y|\lambda)$ that satisfies Assumption \ref{as:tail.intial.condition} is
$\pi(y|\lambda) = \phi( y - \lambda)$, where $\phi(x) = \exp(-\frac{1}{2}x^2)/\sqrt{2\pi}$. 
Finally, Assumption \ref{as:rhohat.ratio.optimality} postulates the existence of finite sample moments of the estimators of the common parameter. 
The main result is stated in the following theorem:

\begin{theorem}\label{thm:ratio.optimality}
            Suppose that Assumptions~\ref{as.likelihood}, \ref{as:error.normal}, and \ref{as:ratio.optimality.marginal.dist.lambda}  to \ref{as:rhohat.ratio.optimality}. Then, for the basic dynamic panel model the predictor $\widehat{Y}_{iT+1}$ defined in (\ref{def:optimal.forecast}) satisfies the ratio optimality in Definition~\ref{def:ratio.optimality}.
\end{theorem}

The result in Theorem~\ref{thm:ratio.optimality} is pointwise with respect to $\theta$. However, the convergence 
of the predictor $\widehat{Y}_{iT+1}$ to the oracle predictor is uniform with respect to the unobserved heterogeneity and the observed trajectory ${\cal Y}_i$ in the sense that the integrated risk (conditional on $\theta$) of the feasible predictor converges to the integrated risk of the oracle predictor. The proof of the theorem is a generalization of the proof in \cite{BrownGreenshtein2009}, allowing for the presence of estimated parameters in the sufficient statistic $\hat{\lambda}(\cdot)$. The remarkable aspect of the results is the acceleration of the convergence ($N^\epsilon_0$ instead of $N$ in the denominator of the standardized regret in Definition~\ref{def:ratio.optimality}) in cases in which the intercepts are identical across units and $\pi(\lambda)$ is a pointmass.

\section{Monte Carlo Simulations}
\label{sec:monte-carlo-simulations}

We will now conduct several Monte Carlo experiments to illustrate the performance of the empirical Bayes predictor. 

\subsection{Experiment 1: Gaussian Random Effects Model}
\label{subsec:monte-carlo-simulations.exp1}

The first Monte Carlo experiment is based on the basic dynamic panel data model in (\ref{m.restricted.linear.panel.regression}). The design of the experiment is summarized in Table~\ref{t_MC1design}. We assume that the $\lambda_i$'s are normally distributed and uncorrelated with the initial condition $Y_{i0}$.  The innovations $U_{it}$ and the heterogeneous intercepts $\lambda_i$ have unit variances. We consider two values for the autocorrelation parameter: $\rho \in \{0.5, 0.95\}$. The panel consists of $N=1,000$ cross-sectional units and the number of time periods is $T=3$. Generally, the smaller $T$ relative to number of right-hand-side variables with heterogeneous coefficients, the larger the gain from using a prior distribution to compute posterior mean estimates of the $\lambda_i$'s. We will compare the performance of the following predictors:

\begin{table}[t!]
	\caption{Monte Carlo Design 1}
	\label{t_MC1design}
	\begin{center}
	\scalebox{0.9}{
	\begin{tabular}{l} \hline \hline
	Law of Motion: $Y_{it} = \lambda_i + \rho Y_{it-1} + U_{it}$ where $U_{it} \sim iid N(0,\gamma^2)$. $\rho \in \{0.5, 0.95\}$, $\gamma=1$ \\
	Initial Observations: $Y_{i0} \sim N(0,1)$ \\
	Gaussian Random Effects: $\lambda_i|Y_{i0} \sim N(\phi_0+\phi_1Y_{i0},\underline{\Omega})$, $\phi_0=0$, $\phi_1=0$, $\underline{\Omega}=1$ \\
	Sample Size: $N=1,000$, $T=3$ \\
	Number of Monte Carlo Repetitions: $N_{sim}=1,000$ \\ \hline
    \end{tabular}
    }
    \end{center}
\end{table}

\noindent {\bf Oracle Forecast.} The oracle knows the parameters $\theta = (\rho,\gamma)$ as well as the random effects distribution $\pi(\lambda_i|Y_{i0},\xi)$, where $\xi = (\phi_0,\phi_1,\underline{\Omega})$. However, the oracle does not know the specific $\lambda_i$ values. Its forecast is given by (\ref{eq.oracle.forecast}).

\noindent {\bf Posterior Predictive Mean Approximation Based on QMLE}. The random effects distribution is correctly modeled as belonging to the family $\lambda_i|(Y_{i0},\xi) \sim N(\phi_0+\phi_1Y_{i0},\underline{\Omega})$. The estimators $\hat{\theta}_{QMLE}$ and $\hat{\xi}_{QMLE}$ are defined in (\ref{eq.qmle.thetahat.xihat}). Tweedie's formula (see (\ref{eq.simplemodel.lambdapostmean}) for the simplified version) is evaluated based on  
$p\big(\hat{\lambda}_i(\hat\theta_{QMLE})|y_{i0},\hat\theta_{QMLE},\hat{\xi}_{QMLE} \big)$.

\noindent {\bf Posterior Predictive Mean Approximation Based on GMM Estimator}. We use the Arellano-Bover estimator described in Section~\ref{subsec:implementation.theta.gmm}. The estimator for $\rho$ is given by (\ref{def:rhoalphahatGMM}) and the estimator for $\gamma$ by (\ref{def:gammahatGMM}). The formulas simplify considerably. We have $W_{it}=1$, $X_{it-1}=Y_{it-1}$, $Z_{it-1}=\emptyset$ and $\alpha = \emptyset$. Moreover, $\Sigma_i^{1/2} = \gamma I$, $M_i(\gamma) = I - \iota\iota'/T$, where $\iota$ is a $T\times 1$ vector of ones. Let $\bar{\tilde{Y}}_i(\hat{\rho})$ be the temporal average of $\tilde{Y}_i(\hat{\rho})$. Then
\[
\hat{\gamma}_{GMM}^2 = \frac{1}{NT} \frac{T}{T-1} 
\sum_{i=1} \mbox{tr}\big[ (\tilde{Y}_i(\hat{\rho}) - \iota \bar{\tilde{Y}}_i(\hat{\rho}) )(\tilde{Y}_i(\hat{\rho}) - \iota \bar{\tilde{Y}}_i(\hat{\rho}) )' \big].
\]
The estimator $\hat{\xi}(\hat{\theta}_{GMM})$ is obtained from (\ref{eq.xihat.theta}). Finally, Tweedie's formula is evaluated based on $p\big(\hat{\lambda}_i(\hat\theta_{GMM})|y_{i0},\hat\theta_{GMM},\hat{\xi}(\hat{\theta}_{GMM}) \big)$.

\noindent {\bf GMM Plug-In Predictor.} We use the Arellano-Bover estimator to obtain $\hat{\rho}_{GMM}$. Instead of using the posterior mean for $\lambda_i$, the plug-in predictor is based on the MLE $\hat{\lambda}_i(\hat{\rho}_{GMM})$. The resulting predictor is $\widehat{Y}_{iT+1} = \hat{\lambda}_i(\hat{\rho}_{GMM}) + \hat{\rho}_{GMM} Y_{iT}$.

\noindent {\bf Loss-Function-Based Predictor.} We construct an estimator of $(\rho,\lambda^N)$ based on the objective function:
\be
    \hat{\rho}_L = \mbox{argmin}_\rho \; \frac{1}{NT} \sum_{i=1}^N \sum_{t=1}^T \big(Y_{it} - \rho Y_{it-1} - \hat{\lambda}_i(\rho) \big)^2, \quad \hat{\lambda}_i(\rho) = \frac{1}{T} \sum_{t=1}^T Y_{it} - \rho Y_{it-1}.
    \label{eq.estimator.lossfunction}
\ee
This estimator minimizes the loss function under which the forecasts are evaluated in sample. It is well-known that due to the incidental parameter problem, the estimator $\hat{\rho}_L$ is inconsistent under fixed-$N$ asymptotics. The resulting predictor is
$\widehat{Y}_{iT+1} = \hat{\lambda}_i(\hat{\rho}_L) + \hat{\rho}_L Y_{iT}$.

\noindent {\bf Pooled-OLS Predictor.} Ignoring the heterogeneity in the $\lambda_i$'s and imposing that $\lambda_i=\lambda$ for all $i$, we can define
\be
   ( \hat{\rho}_P, \hat{\lambda}_P ) = \mbox{argmin}_{\rho,\lambda} \;
            \frac{1}{NT} \sum_{i=1}^N \sum_{t=1}^T \big(Y_{it} - \rho Y_{it-1} - \lambda \big)^2.
\ee   
The resulting predictor is $\widehat{Y}_{iT+1} = \hat{\lambda}_P + \hat{\rho}_P Y_{iT}$.

\noindent {\bf First-Difference Predictor.} In the panel data literature it is common to difference-out idiosyncratic intercepts, which suggests to predict $\Delta Y_{iT+1}$ based on $\Delta Y_{iT}$. We evaluate the first-difference predictor at the Arellano-Bover GMM estimator of $\rho$ to obtain $\widehat{Y}_{iT+1}^{FD}(\hat{\rho}_{GMM})$.

In Table~\ref{t_MC1results} we report the regret associated with each predictor relative to the posterior variance of $\lambda_i$, averaged over all trajectories ${\cal Y}^N$, as specified in Definition~\ref{def:ratio.optimality} (setting $N^\epsilon=1$). For the oracle predictor the regret is by definition zero and we tabulate the risk $R_N^{opt}$ instead (in parentheses). We also report the median forecast error $\widehat{e}_{iT+1|T} = Y_{iT+1}-\widehat{Y}_{iT+1}$ to highlight biases in the forecasts. 

\begin{sidewaystable}[t!]
\caption{Monte Carlo Experiment 1: Random Effects, Parametric Tweedie Correction, Selection Bias}
\label{t_MC1results}
\begin{center}
\scalebox{0.90}{
\begin{tabular}{lcccccccc}\hline\hline
& \multicolumn{2}{c}{All Units} & \multicolumn{2}{c}{Bottom Group}  & \multicolumn{2}{c}{Middle Group} & \multicolumn{2}{c}{Top Group} \\
&  & Median &  & Median  &  & Median &  & Median \\
Estimator / Predictor  & Regret & Forec.E. & Regret & Forec.E. & Regret & Forec.E & Regret & Forec.E. \\ \hline
\multicolumn{9}{c}{Low Persistence: $\rho=0.50$} \\ \hline
Oracle Predictor  & (1252.7) &   0.002 & (65.95) &  -0.037 & (62.48) &   0.003 & (62.10) &  -0.003  \\ \hline
Post. Mean ($\hat{\theta}_{QMLE}$, Parametric)  &   0.005 &   0.005 &   0.002 &  -0.030 &   0.002 &   0.006 &   0.018 &  -0.004   \\ 
Post. Mean ($\hat{\theta}_{GMM}$, Parametric)  &   0.030 &   0.004 &   0.015 &  -0.035 &   0.022 &   0.008 &   0.100 &   0.004   \\ 
Plug-In Predictor ($\hat{\theta}_{GMM}$, $\hat{\lambda}_i(\hat{\theta}_{GMM})$)  &   0.358 &   0.005 &   1.150 &   0.536 &   0.045 &   0.009 &   1.421 &  -0.558   \\ 
Loss-Function-Based Estimator  &   0.369 &   0.199 &   0.275 &   0.190 &   0.348 &   0.197 &   0.352 &   0.188   \\ 
Pooled OLS  &   0.656 &  -0.285 &   1.892 &  -0.663 &   0.491 &  -0.288 &   0.223 &   0.044   \\ 
First-Difference Predictor ($\hat{\theta}_{GMM}$) & 2.963 & 0.001 & 5.317 & 0.935 & 1.936 & 0.009 & 5.656 & -0.986 \\ \hline
\multicolumn{9}{c}{High Persistence: $\rho=0.95$} \\ \hline
Oracle Predictor  & (1252.7) &   0.002 & (67.36) &  -0.081 & (63.16) &   0.007 & (61.86) &  -0.002  \\ \hline
Post. Mean ($\hat{\theta}_{QMLE}$, Parametric)  &   0.009 &   0.011 &   0.003 &  -0.075 &   0.005 &   0.016 &   0.036 &   0.015   \\ 
Post. Mean ($\hat{\theta}_{GMM}$, Parametric)  &   0.046 &   0.003 &   0.019 &  -0.071 &   0.023 &   0.010 &   0.178 &  -0.005   \\ 
Plug-In Predictor ($\hat{\theta}_{GMM}$, $\hat{\lambda}_i(\hat{\theta}_{GMM})$)  &   0.380 &   0.004 &   1.036 &   0.498 &   0.039 &   0.017 &   1.546 &  -0.569   \\ 
Loss-Function-Based Estimator  &   0.623 &   0.357 &   0.014 &   0.033 &   0.522 &   0.357 &   1.358 &   0.597   \\ 
Pooled OLS  &   1.015 &  -0.454 &   1.066 &  -0.517 &   0.967 &  -0.459 &   0.872 &  -0.422   \\ 
First-Difference Predictor ($\hat{\theta}_{GMM}$) & 3.986 & 0.000 & 6.582 & 0.887 & 2.733 & 0.013 & 6.912 & -0.939 \\\hline
\end{tabular}
}
\end{center}
{\footnotesize {\em Notes:} The design of the experiment is summarized in Table~\ref{t_MC1design}. For the oracle predictor we report the compound risk (in parentheses) instead of the regret. The regret is standardized by the average posterior variance of $\lambda_i$, see Definition~\ref{def:ratio.optimality}.}\setlength{\baselineskip}{4mm}
\end{sidewaystable}

\afterpage{\clearpage}

The columns titled ``All Units'' correspond to $D_i({\cal Y}^N)=1$. As expected from the theoretical analysis, the posterior mean predictors have the lowest regret among the feasible predictors. The density of $\hat{\lambda}_i$ is estimated parametrically, using a family of distributions that nests the true random effects distribution. Because it is based on a correctly specified likelihood function, the predictor based on $\hat{\theta}_{QMLE}$ performs slightly better than the predictor based on $\hat{\theta}_{GMM}$. Consider $\rho=0.5$: for the QMLE-based predictor the regret is 0.5\% of the average posterior variance, whereas it is 3\% for the GMM-based predictor.
The plug-in predictor that replaces the unknown $\lambda_i$'s by the sufficient statistic $\hat{\lambda}_i$ (which is also the maximum likelihood estimator)  instead of the posterior mean is associated with a much larger relative regret, which is about 37\%. 

The remaining three predictors are also strictly dominated by the posterior mean predictors. Ignoring the serial correlation in $\Delta Y_{it}$, the first-difference predictor performs the worst for both choices of $\rho$. The second-to-worst predictor is the pooled-OLS predictor which ignores the cross-sectional heterogeneity in the $\lambda_i$'s. A reduction of the variance $\underline{\Omega}$ of the heterogeneous intercepts would improve the relative performance of the pooled-OLS predictor. Finally, the loss-function-based predictor dominates the pooled-OLS and the first difference predictor. As mentioned above, while conceptually appealing, the loss-function-based predictor relies on an inconsistent estimate of $\rho$, which in comparison to the GMM plug-in predictor is unappealing if the cross-sectional dimension $N$ is very large.

Across all units, the predictions under the loss-function-based estimator and the pooled-OLS estimator appear to be biased. To study this bias further we now consider level-based selection rules $D_i({\cal Y}^i)$. 
%We first compute the unconditional distribution of  based on the data generating process described (DGP)in Table~\ref{t_MC1design} and then calculate quantiles for this distribution. 
Using the 5\%, 47.5\%, 52.5\%, and 95\% quantiles of the population distribution of $Y_{iT}$, we define cut-offs for a bottom 5\% group, a middle 5\% group, and a top 5\% group. Because the cut-offs are computed from the population distribution of $Y_{iT}$, for unit $i$ the selection rules only depends on ${\cal Y}_{iT}$ and not on $Y_{jT} $ with $j\not=i$. 

\begin{figure}
	\caption{QMLE Estimation: Distribution of $\widehat{\mathbb{E}}_{\hat{\theta},{\cal Y}_i}^{\lambda_i}[\lambda_i]$ versus $\hat{\lambda}_i(\hat{\theta})$}
	\label{f_MC1_Elambdavslambdahat}
	\begin{center}
		\begin{tabular}{cccc}
		All Units & Bottom Group & Middle Group & Top Group \\
		\includegraphics[width=1.5in]{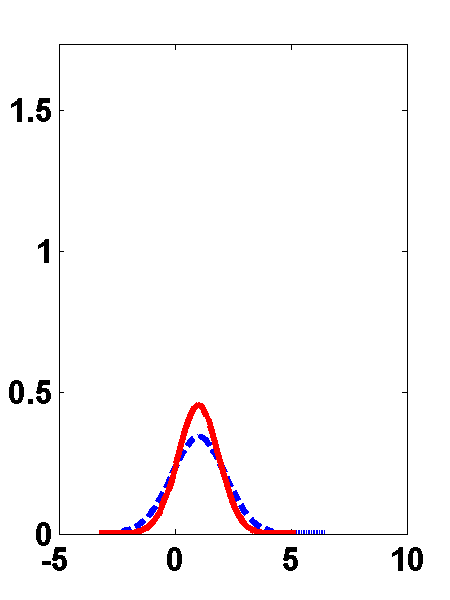} &
		\includegraphics[width=1.5in]{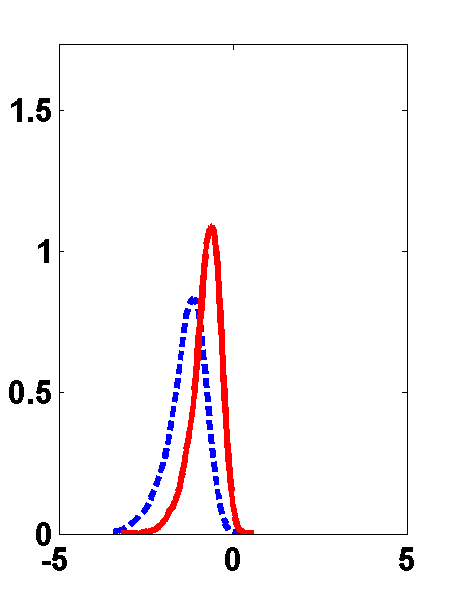} &
		\includegraphics[width=1.5in]{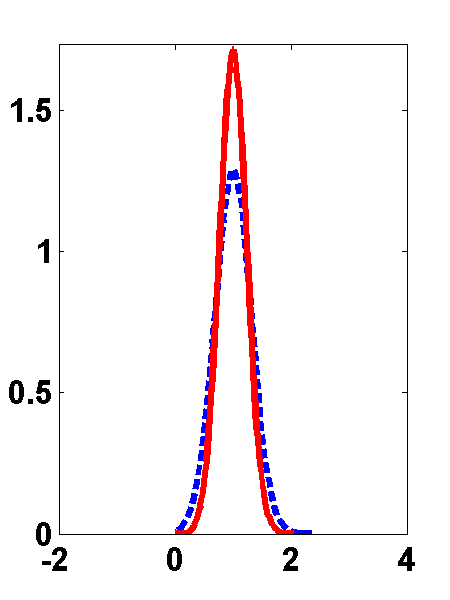} &
		\includegraphics[width=1.5in]{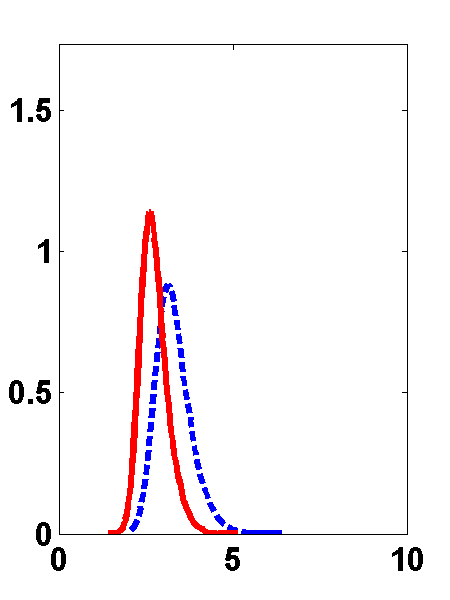} 
		\end{tabular}				
	\end{center}
	{\footnotesize {\em Notes:} Solid (red) lines depict cross-sectional densities of posterior mean estimates $\widehat{\mathbb{E}}_{\hat{\theta},{\cal Y}_i}^{\lambda_i}[\lambda_i]$. Dashed (blue) lines depict cross-sectional densities of sufficient statistic $\hat{\lambda}_i(\hat{\theta})$. The results are based on the QMLE estimator. The Monte Carlo design is described in Table~\ref{t_MC1design}.} \setlength{\baselineskip}{4mm}
\end{figure}

For the top and bottom groups only the posterior mean predictors lead to unbiased forecast errors. The sufficient statistic 
$\hat{\lambda}_i$ tends to overestimate (underestimate) $\lambda_i$ for the top (bottom) group, because it interprets a sequence of above-average (below-average) $U_{iT}$'s as evidence for a high (low) $\lambda_i$. This is reflected in the bias: the plug-in predictors' forecast errors for the top group are on average positive, whereas the forecast errors for the bottom group tend to be negative. The posterior mean tends to correct these biases because it shrinks toward the mean of the prior distribution of the $\lambda_i$'s. This reduces the regrets for the top and bottom groups, and is also reflected in the risk calculated across all units. 
The bias correction is illustrated in Figure~\ref{f_MC1_Elambdavslambdahat}, which compares the cross-sectional distribution of the sufficient statistics $\hat{\lambda}_i(\hat\theta)$ to the distribution of the posterior mean estimates $\widehat{\mathbb{E}}_{\hat{\theta},{\cal Y}_i}^{\lambda_i}[\lambda_i]$ obtained with Tweedie's formula. Due to the shrinkage effect of the prior, the distribution of the posterior means, in particular for the top and bottom groups, is more compressed.  

\subsection{Experiment 2: Non-Gaussian Correlated Random Effects Model}
\label{subsec:monte-carlo-simulations.exp2}

We now change the Monte Carlo design in two dimensions. First, we replace the Gaussian random effects specification with a non-Gaussian specification in which the heterogeneous coefficient $\lambda_i$ is correlated with the initial condition $Y_{i0}$. Second, we consider a Tweedie correction based on a kernel density estimate of $p(\hat{\lambda}_i|Y_{i0})$ as discussed in Section~\ref{subsec:implementation.tweedie.nonparametric}.

\begin{table}[t!]
	\caption{Monte Carlo Design 2}
	\label{t_MC2design}
	\begin{center}
	\scalebox{0.9}{
		\begin{tabular}{l} \hline \hline
			Law of Motion: $Y_{it} = \lambda_i + \rho Y_{it-1} + U_{it}$ where $U_{it} \sim iid N(0,\gamma^2)$; $\rho = 0.5$, $\gamma=1$ \\
			Initial Observation: $Y_{i0} \sim N \left( \frac{\underline{\mu}_\lambda}{1-\rho}, V_Y 
			+ \frac{\underline{V}_\lambda}{(1-\rho)^2} \right)$, $V_Y = \gamma^2/(1-\rho^2)$; $\underline{\mu}_\lambda = 1$, $\underline{V}_\lambda=1$ \\ 
			Non-Gaussian Correlated Random Effects:\\
    \hspace*{2cm}$
    \lambda_i | Y_{i0} \sim 
    \left\{
    \begin{array}{ll} 
    N\big(\phi_+(Y_{i0}),\underline{\Omega} \big)
    & \mbox{with probability } p_\lambda \\
    N\big(\phi_-(Y_{i0}),\underline{\Omega} \big)
    & \mbox{with probability } 1-p_\lambda
    \end{array}
    \right.,
    $\\
    \hspace*{2cm} $\phi_+(Y_{i0})=\phi_0+\delta+(\phi_1+\delta) Y_{i0}$, \\
    \hspace*{2cm} $\phi_-(Y_{i0})=\phi_0-\delta+(\phi_1-\delta) Y_{i0}$, \\
    \hspace*{2cm} $\underline{\Omega} = \left[ \frac{1}{(1-\rho)^2} V_Y^{-1} + \underline{V}_\lambda^{-1} \right]^{-1}$, 
    $\phi_0 = \underline{\Omega} \, \underline{V}_\lambda^{-1}\underline{\mu}_\lambda$, 
    $\phi_1 = \frac{1}{1-\rho} \underline{\Omega} V_Y^{-1}$,\\
	\hspace*{2cm} $p_\lambda=1/2$, $\underline{\Omega}=1$, $\delta \in \{1/5,\,1,\,5\}$ ($\delta = 1/\sqrt{\kappa}$)\\
			Sample Size: $N=1,000$, $T=3$ \\
			Number of Monte Carlo Repetitions: $N_{sim}=1,000$ \\ \hline
		\end{tabular}
	}
	\end{center}
\end{table}

\begin{figure}[t!]
	\caption{QMLE Estimation: Density $p(\hat{\lambda}_i|y_{i0},\theta)$ for $\delta=1/10$ versus $\delta=1$}
	\label{f_MC2_pofkappa}
	\begin{center}
		\begin{tabular}{ccc}
			$y_{i0}= -2.5$ & $y_{i0}= 2.0$ & $y_{i0}= 6.5$ \\
			\includegraphics[width=1.75in]{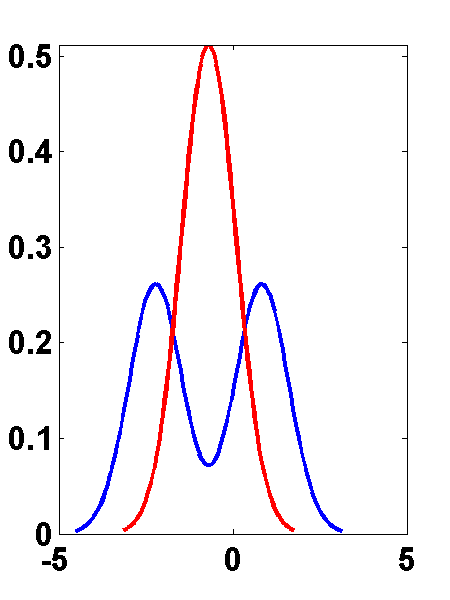} &
			\includegraphics[width=1.75in]{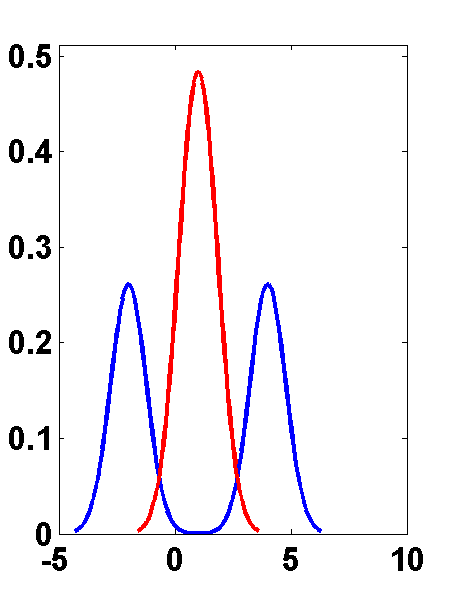} &
			\includegraphics[width=1.75in]{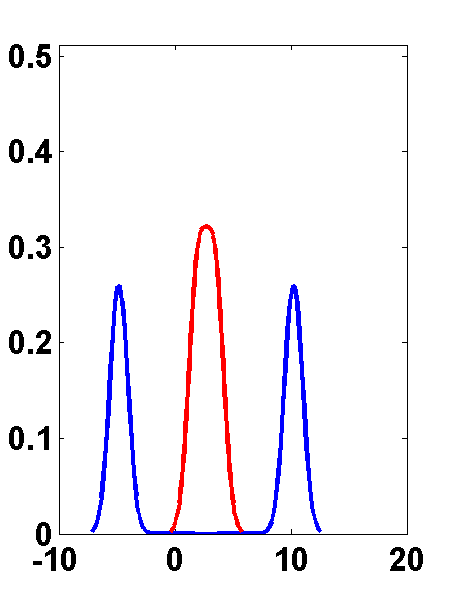} 
		\end{tabular}				
	\end{center}
	{\footnotesize {\em Notes:} Solid (blue) line is $\delta=1$ and solid (red) line is $\delta=1/10$. The Monte Carlo design is described in Table~\ref{t_MC2design}. }\setlength{\baselineskip}{4mm}
\end{figure}

The Monte Carlo design is summarized in Table~\ref{t_MC2design}. Starting point is a joint normal distribution for $(\lambda_i,Y_{i0})$, factorized into a marginal distribution $\pi_*(\lambda_i)$ and a conditional distribution $\pi_*(Y_{i0}|\lambda_i)$. We assumed $\lambda_i \sim N(\underline{\mu}_\lambda,\underline{V}_\lambda)$ and that $Y_{i0}|\lambda_i$ corresponds to the stationary distribution of $Y_{it}$ associated with its autoregressive law of motion. The implied marginal distribution for $Y_{i0}$ is used as $\pi(Y_{i0})$ in the Monte Carlo design. To obtain $\pi(\lambda_i|Y_{i0})$ we took $\pi_*(\lambda_i|Y_{i0})$ from the Gaussian model and replaced it with a mixture of normals described in Table~\ref{t_MC2design}. For $\delta=0$ the mixture reduces to $\pi_*(\lambda_i|Y_{i0})$, whereas for large values of $\delta$ it becomes bimodal. This bimodality also translates into the distribution of $\hat{\lambda}|Y_{i0}$, which is depicted in Figure~\ref{f_MC2_pofkappa} for $\delta=1/10$ (almost Gaussian) and $\delta=1$ (bimodal).

%In this experiment we will compare the performance of predictors based on the Tweedie correction from a misspecified parametric Gaussian correlated random effects density and from nonparametric kernel density estimates. The parametric estimator is identical to the one used in Experiment~1. 
In this experiment we consider a parametric Tweedie correction (same as in Experiment 1, but now misspecified in view of the DGP) and two nonparametric Tweedie corrections. 
First, we compute the correction based on the simple Gaussian kernel in (\ref{eq.density.nonparametric}). The bandwidth is chosen in accordance with the theory in Section~\ref{sec:ratio.optimality}. We set $B_N = c/(\ln N)^{0.55}$, which would be consistent with a truncation of the form $C_N = c \sqrt{\ln N}$, and let $c \in \{1/2, 1, 2\}$.\footnote{The tuning matrices $V_{\hat{\lambda}}$ and $V_h$ are set equal to the sample variances of $\hat{\lambda}_i$ and $y_{i0}$, respectively.} Second, we use the adaptive estimator proposed by \cite{BotevGrotowskiKroese2010}, henceforth BGK estimator, which is based on the solution of a diffusion partial differential equation. This estimator is associated with a plug-in bandwidth selection rule that requires no further tuning.\footnote{Our estimates are based on Algorithms~1 and~2 in BGK. We use the authors' MATLAB code to implement the density estimator.} Unless otherwise noted, the subsequent results are based on the BGK estimator.

\begin{figure}[t!]
	\caption{QMLE Estimation: ``True'' Density $p(\hat{\lambda}_i|y_{i0},\theta)$ versus Gaussian and Nonparametric Estimates}
	\label{f_MC2_phatlambdahat}
	\begin{center}
		\begin{tabular}{cccc}
			\multicolumn{4}{c}{Parametric Gaussian Estimates $p_*(\hat{\lambda}_i|y_{i0},\hat{\theta}_{QMLE},\hat{\xi}_{QMLE})$} \\
			\multicolumn{2}{c}{Misspecification $\delta=1/10$} & \multicolumn{2}{c}{Misspecification $\delta=1$} \\
			$y_{i0}= -2.5$ & $y_{i0}= 2.0$ & $y_{i0}= -2.5$ & $y_{i0}= 2.0$ \\
			\includegraphics[width=1.3in]{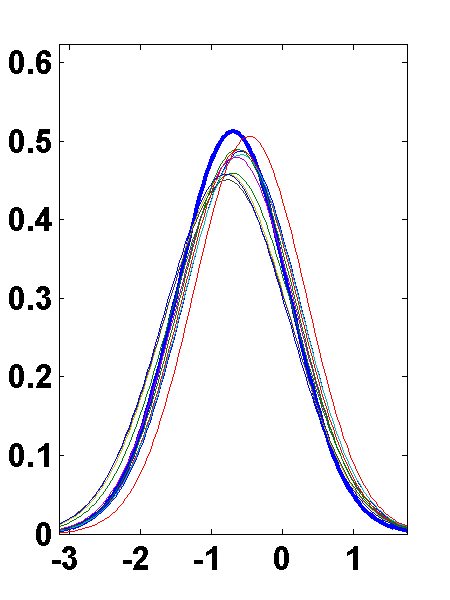} &
			\includegraphics[width=1.3in]{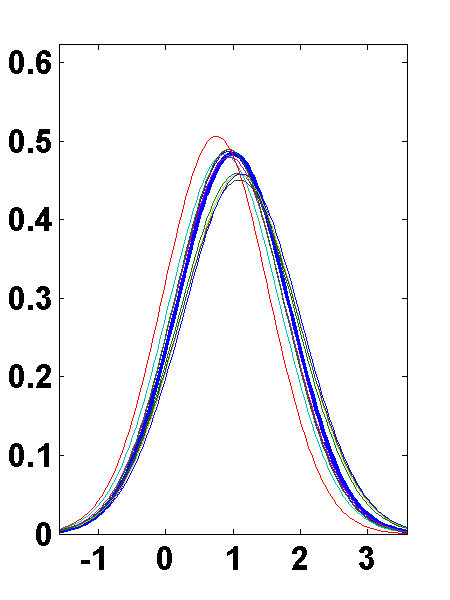} &
			\includegraphics[width=1.3in]{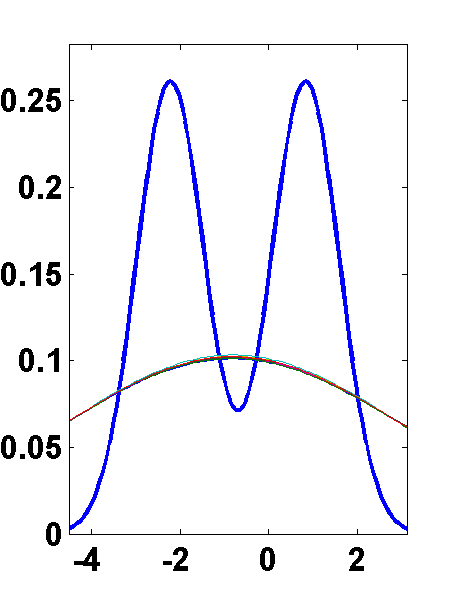}
			&
			\includegraphics[width=1.3in]{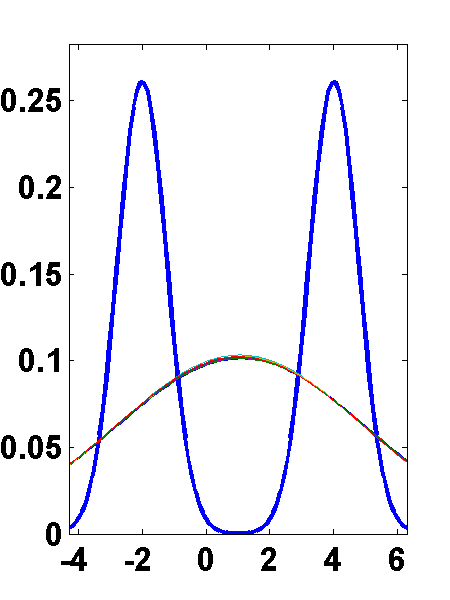} \\[1ex]
			\multicolumn{4}{c}{Nonparametric Kernel Estimates $\hat{p}(\hat{\lambda}_i|y_{i0},\hat{\theta}_{QMLE})$} \\
			\multicolumn{2}{c}{Misspecification $\delta=1/10$} & \multicolumn{2}{c}{Misspecification $\delta=1$} \\
			$y_{i0}= -2.5$ & $y_{i0}= 2.0$ & $y_{i0}= -2.5$ & $y_{i0}= 2.0$ \\			\includegraphics[width=1.3in]{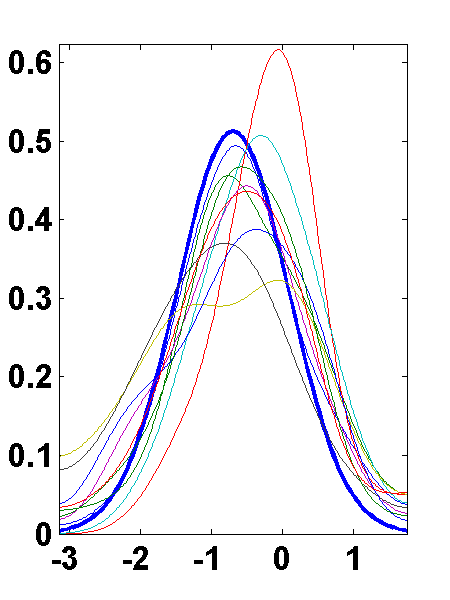} &
			\includegraphics[width=1.3in]{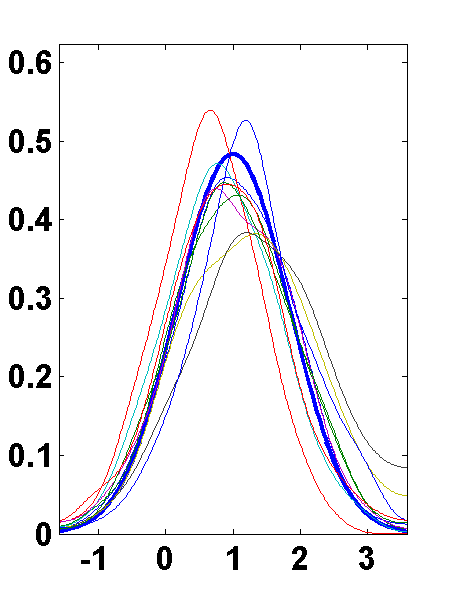} &
			\includegraphics[width=1.3in]{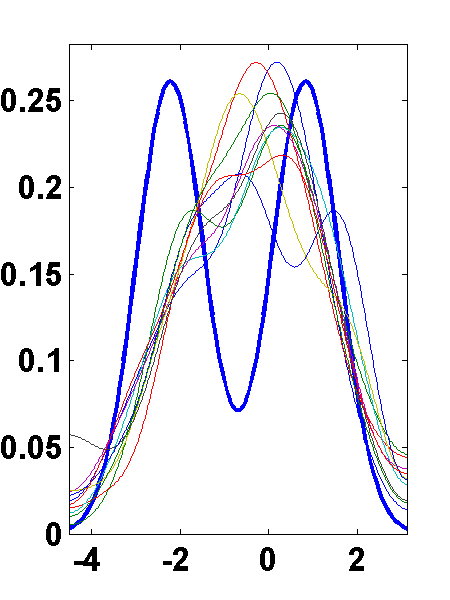} &
			\includegraphics[width=1.3in]{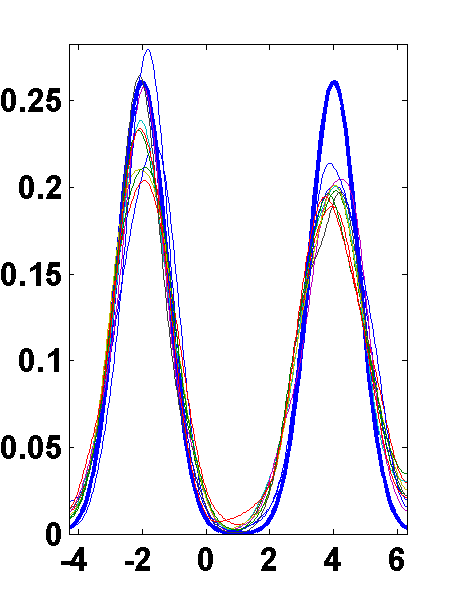} 
		\end{tabular}				
	\end{center}
	{\footnotesize {\em Notes:} Solid (blue) lines depict ``true'' $p(\hat{\lambda}_i|y_{i0},\theta)$. Colored ``hairs'' depict 10 estimates from the Monte Carlo repetitions. The nonparametric estimates are based on the BGK kernel estimator. The Monte Carlo design is described in Table~\ref{t_MC2design}. }\setlength{\baselineskip}{4mm}
\end{figure}

Figure~\ref{f_MC2_phatlambdahat} shows the ``true'' density $p(\hat{\lambda}_i|y_{i0},\theta)$ as well as Gaussian and nonparametric approximations. Under the Gaussian correlated random effects distribution we can directly calculate the conditional distribution of $\hat{\lambda}_i$ given $y_{i0}$. The nonparametric approximation is obtained by dividing an estimate of the joint density of $(\hat{\lambda}_i,y_{i0})$ by an estimate of the marginal density of $y_{i0}$ (this normalization is not required for the Tweedie correction).
Each hairline in Figure~\ref{f_MC2_phatlambdahat} corresponds to a density estimate from a different Monte Carlo run. For $\delta=1/10$ the Gaussian approximation is accurate and the variability of the estimates is much smaller than that of the kernel estimates. For $\delta=1$ the Gaussian density is unable to approximate the bimodal $p(\hat{\lambda}_i,y_{i0}|\theta)$, whereas the non-parametric approximation, at least for $y_{i0}=2.0$ captures the key features of the density of $\hat{\lambda}_i$.

\begin{figure}[t!]
	\caption{QMLE Estimation: Gaussian versus Nonparametric Estimates Tweedie Correction}
	\label{f_MC2_tweediecorrection}
	\begin{center}
		\begin{tabular}{cccc}
			\multicolumn{4}{c}{Parametric Gaussian Estimates $p_*(\hat{\lambda}_i|y_{i0},\hat{\theta}_{QMLE},\hat{\xi}_{QMLE})$} \\
			\multicolumn{2}{c}{Misspecification $\delta=1/10$} & \multicolumn{2}{c}{Misspecification $\delta=1$} \\
			$y_{i0}= -2.5$ & $y_{i0}= 2.0$ & $y_{i0}= -2.5$ & $y_{i0}= 2.0$ \\
			\includegraphics[width=1.3in]{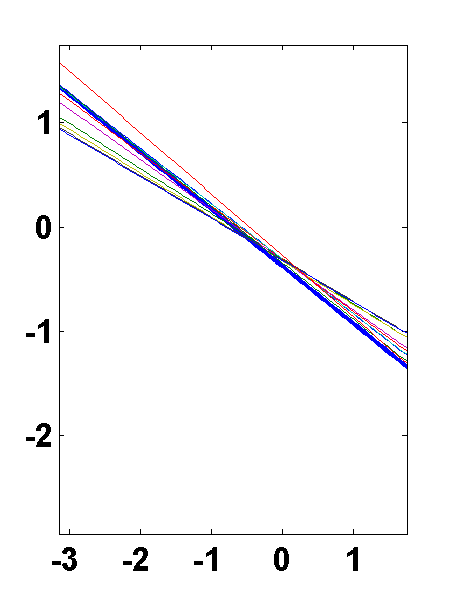} &
			\includegraphics[width=1.3in]{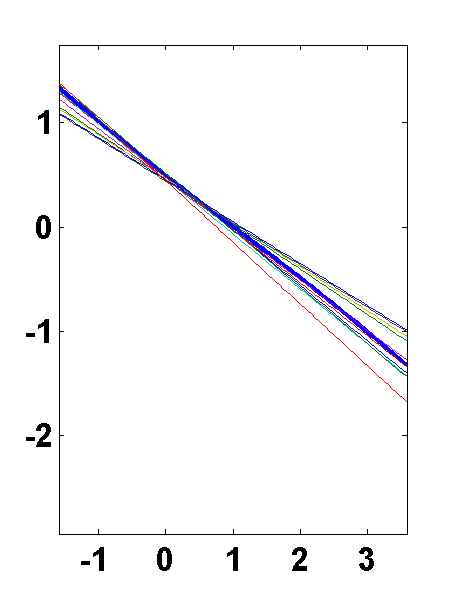} &
			\includegraphics[width=1.3in]{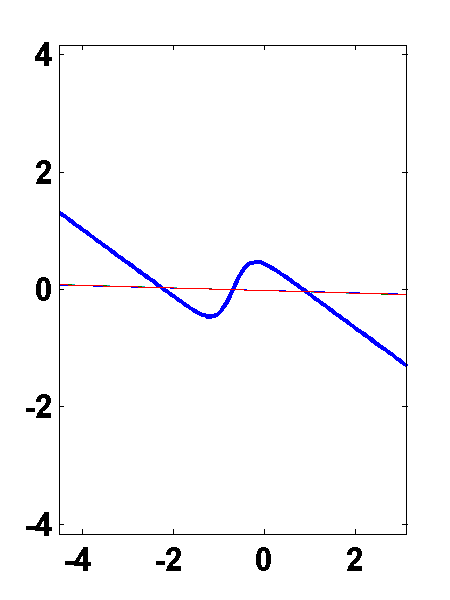}
			&
			\includegraphics[width=1.3in]{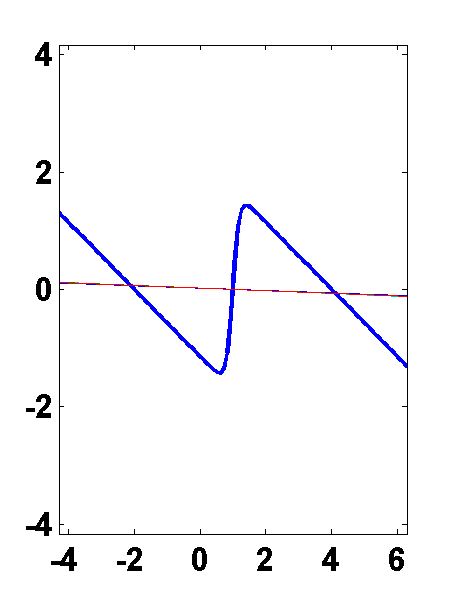} \\[1ex]
			\multicolumn{4}{c}{Nonparametric Kernel Estimates $\hat{p}(\hat{\lambda}_i|y_{i0},\hat{\theta}_{QMLE})$} \\
			\multicolumn{2}{c}{Misspecification $\delta=1/10$} & \multicolumn{2}{c}{Misspecification $\delta=1$} \\
			$y_{i0}= -2.5$ & $y_{i0}= 2.0$ & $y_{i0}= -2.5$ & $y_{i0}= 2.0$ \\			\includegraphics[width=1.3in]{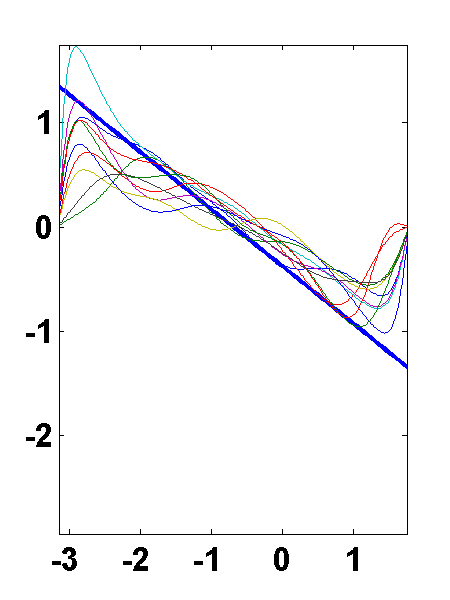} &
			\includegraphics[width=1.3in]{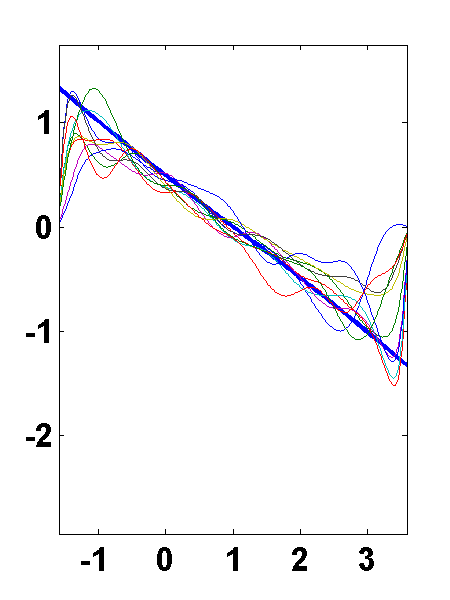} &
			\includegraphics[width=1.3in]{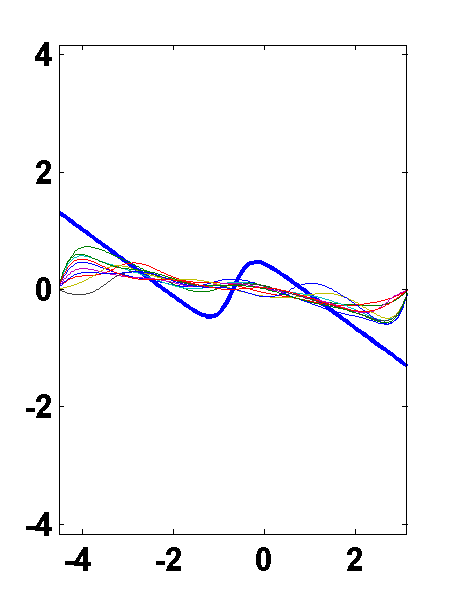} &
			\includegraphics[width=1.3in]{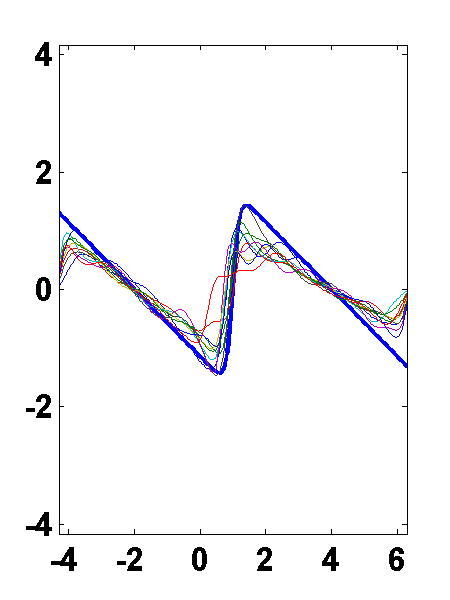} 
		\end{tabular}				
	\end{center}
	{\footnotesize {\em Notes:} Solid (blue) lines depict Tweedie correction based on $p(\hat{\lambda}_i|y_{i0},\theta)$. Colored ``hairs'' depict 10 estimates from the Monte Carlo repetitions. The nonparametric estimates are based on the BGK kernel estimator. The Monte Carlo design is described in Table~\ref{t_MC2design}. }\setlength{\baselineskip}{4mm}
\end{figure}

For the prediction, the relevant object is the correction $(\sigma^2/T)\partial \ln p(\hat{\lambda}_i,y_{i0}|\theta) / \partial \hat{\lambda}_i$, which is depicted in Figure~\ref{f_MC2_tweediecorrection}. Under a Gaussian correlated random effects distribution, the Tweedie correction is linear in $\hat{\lambda}_i$ because the posterior mean is a linear combination of the prior mean and the maximum of the likelihood function.  
Thus, the corrections based on the Gaussian density estimate are linear regardless of $\delta$. For $\delta=1/10$ the correction under the ``true'' random effects distribution is nearly linear, and thus well approximated by the Gaussian correction. The nonparametric correction is fairly accurate for values of $\hat{\lambda}$ in the center of the conditional distribution $\hat{\lambda}_i|(y_{i0},\theta)$, but it becomes less accurate in the tails. For $\delta=1$, on the other hand, the kernel-based correction provides a much better approximation of the optimal correction than the Gaussian correction.

Table~\ref{t_MC2results} compares the performance of twelve predictors; half of them based on QMLE and the other half based on GMM. It is well-known that the GMM estimator of $\theta$ is consistent under the DGP described in Table~\ref{t_MC2design}. We show in the Appendix that the QMLE estimator is also consistent for $\theta$ under this DGP, despite the fact that the correlated random effects distribution is misspecified. For each of the two $\theta$ estimators we construct posterior mean predictors using four different nonparametric Tweedie corrections as well as the Gaussian Tweedie correction. Moreover, we compute the plug-in predictor based on $\hat{\lambda}_i(\hat{\theta})$.

\begin{table}[t!]
	\caption{Monte Carlo Experiment 2: Correlated Random Effects, Non-parametric versus Parametric Tweedie Correction}
	\label{t_MC2results}
	\begin{center}
		\scalebox{0.80}{
			\begin{tabular}{lcccccccc}\hline\hline
& \multicolumn{2}{c}{All Units} & \multicolumn{2}{c}{Bottom Group}  & \multicolumn{2}{c}{Top Group} \\
&  & Median &  & Median  &  & Median \\
Estimator / Predictor  & Regret & Forec.E. & Regret & Forec.E. & Regret & Forec.E \\ \hline
\multicolumn{7}{c}{$\delta= 1/10 $} \\ \hline 
Oracle Predictor  & (1177.6) &   0.003 & (54.92) &  -0.046 & (63.97) &  -0.010  \\ \hline
Post. Mean ($\hat{\theta}_{QMLE}$, BGK Kernel)  &   0.179 &  -0.001 &   0.737 &   0.159 &   0.543 &  -0.119   \\ 
Post. Mean ($\hat{\theta}_{QMLE}$, Gaussian Kernel $c=0.5$)  &   0.635 &   0.001 &   1.711 &   0.438 &   1.157 &  -0.360   \\ 
Post. Mean ($\hat{\theta}_{QMLE}$, Gaussian Kernel $c=1.0$)  &   0.454 &   0.000 &   1.126 &   0.345 &   0.779 &  -0.279   \\ 
Post. Mean ($\hat{\theta}_{QMLE}$, Gaussian Kernel $c=2.0$)  &   0.416 &0.000 &   0.826 &   0.267 &   0.568 &  -0.183   \\ 
Post. Mean ($\hat{\theta}_{QMLE}$, Parametric)  &   0.048 &   0.001 &   0.053 &   0.060 &   0.130 &   0.127   \\ 
Plug-in Predictor ($\hat{\theta}_{QMLE}$, $\hat{\lambda}_i(\hat{\theta}_{QMLE})$) &   0.915 &   0.001 &   2.323 &   0.527 &   1.549 &  -0.437   \\ 
Post. Mean ($\hat{\theta}_{GMM}$, BGK Kernel)  &   0.217 &   0.002 &   0.766 &   0.135 &   0.566 &  -0.095   \\ 
Post. Mean ($\hat{\theta}_{GMM}$, Gaussian Kernel $c=0.5$)  &   0.693 &   0.002 &   1.761 &   0.423 &   1.182 &  -0.336   \\ 
Post. Mean ($\hat{\theta}_{GMM}$, Gaussian Kernel $c=1.0$)  &   0.509 &   0.001 &     1.180 & 0.333 &   0.813 &  -0.255   \\ 
Post. Mean ($\hat{\theta}_{GMM}$, Gaussian Kernel $c=2.0$)  &   0.459 &   0.002 &   0.866 &   0.252 &   0.601&-0.160   \\ 
Post. Mean ($\hat{\theta}_{GMM}$, Parametric)  &   0.091 &   0.002 &   0.079 &   0.043 &   0.192 &   0.146   \\ 
Plug-in Predictor ($\hat{\theta}_{GMM}$, $\hat{\lambda}_i(\hat{\theta}_{GMM})$)  &   0.968 &   0.003 &   2.356 &   0.511 &   1.558 &  -0.413   \\ 
\hline\hline
\multicolumn{7}{c}{$\delta=    1 $} \\ \hline 
Oracle Predictor  & (1161.7) &  -0.003 & (54.43) &  -0.056 & (65.78) &  -0.024  \\ \hline
Post. Mean ($\hat{\theta}_{QMLE}$, BGK Kernel)  &   0.298 &   0.006 &   0.756 &   0.181 &   0.735 &  -0.073   \\ 
Post. Mean ($\hat{\theta}_{QMLE}$, Gaussian Kernel $c=0.5$)  &   0.526 &   0.001 &   0.857 &   0.240 &   0.855 &  -0.089   \\ 
Post. Mean ($\hat{\theta}_{QMLE}$, Gaussian Kernel $c=1.0$)  &   0.661 &   0.002 &   0.894 &   0.226 &   0.936 &  -0.050   \\ 
Post. Mean ($\hat{\theta}_{QMLE}$, Gaussian Kernel $c=2.0$)  &   0.833 &   0.005 &   1.080 &   0.225 &   1.100 &   0.000   \\ 
Post. Mean ($\hat{\theta}_{QMLE}$, Parametric)  &   1.025 &   0.001 &   1.292 &   0.233 &   1.256 &  -0.012   \\ 
Plug-in Predictor ($\hat{\theta}_{QMLE}$, $\hat{\lambda}_i(\hat{\theta}_{QMLE})$)  &   1.068 &   0.001 &   1.852 &   0.388 &   1.468 &  -0.158   \\ 
Post. Mean ($\hat{\theta}_{GMM}$, BGK Kernel)  &   0.343 &   0.006 &   0.906 &   0.171 &   0.874 &  -0.068   \\ 
Post. Mean ($\hat{\theta}_{GMM}$, Gaussian Kernel $c=0.5$)  &   0.571 &   0.001 &   1.015 &   0.234 &   0.994 &  -0.086   \\ 
Post. Mean ($\hat{\theta}_{GMM}$, Gaussian Kernel $c=1.0$)  &   0.706 &   0.002 &   1.050 &   0.217 &   1.076 &  -0.046   \\ 
Post. Mean ($\hat{\theta}_{GMM}$, Gaussian Kernel $c=2.0$)  &   0.930 &   0.005 &   1.235 &   0.218 &   1.242 &   0.006   \\ 
Post. Mean ($\hat{\theta}_{GMM}$, Parametric)  &   1.071 &   0.001 &   1.443 &   0.228 &   1.392 &  -0.005   \\ 
Plug-in Predictor ($\hat{\theta}_{GMM}$, $\hat{\lambda}_i(\hat{\theta}_{GMM})$)  &   1.115 &   0.001 &   2.011 &   0.383 &   1.609 &  -0.154   \\ 
\hline
\end{tabular}
}
\end{center}
{\footnotesize {\em Notes:} The design of the experiment is summarized in Table~\ref{t_MC2design}. For the oracle predictor we report the compound risk (in parentheses) instead of the regret. The regret is standardized by the average posterior variance of $\lambda_i$, see Definition~\ref{def:ratio.optimality}. The BGK estimator relies on a adaptive bandwidth choice. For the Gaussian kernel estimator in (\ref{eq.density.nonparametric}) we set $B_N = c / (\ln N)^{0.49}$.} \setlength{\baselineskip}{4mm}
\end{table}

Among the nonparametric predictors, the one based on the BGK density estimator clearly dominates the ones derived from the simple kernel density estimator. If the random effects distribution is almost normal, i.e., $\delta=1/10$, setting $c=2$ is preferable to the other choices of $c$. For the bimodal random effects distribution, i.e., $\delta=1$, the best performance of the simple kernel estimator is attained for $c=1/2$. The predictors that rely on posterior mean approximations generally outperform the naive predictors based on $\hat{\lambda}_i(\hat{\theta})$. 
The benefits from shrinkage are most pronounced for the bottom and top groups.
If the misspecification is small $(\delta=1/10)$, the parametric correction leads to more precise forecasts than the nonparametric correction because it is based on a more efficient density estimator. As the degree of misspecification increases, the nonparametric correction starts to perform better and for $\delta=1$ it clearly dominates the parametric competitor. This is consistent with the accuracy of the underlying density estimators shown in Figures~\ref{f_MC2_phatlambdahat} and \ref{f_MC2_tweediecorrection}. 

%\afterpage{\clearpage}

\subsection{Experiment 3: Misspecified Likelihood Function}
\label{subsec:monte-carlo-simulations.exp3}

\begin{table}[t!]
	\caption{Monte Carlo Design 3}
	\label{t_MC3design}
	\begin{center}
		\scalebox{0.9}{
		\begin{tabular}{l} \hline \hline
			Law of Motion: $Y_{it} = \lambda_i + \rho Y_{it-1} + U_{it}$, $\rho =0.5$, $\mathbb{E}[U_{it}]=0$, $\mathbb{V}[U_{it}]=1$ \\ 
			Scale Mixture: $U_{it} \sim iid \left\{ \begin{array}{ll} N(0,\gamma_+^2) & \mbox{with probability } p_u \\
			N(0,\gamma_-^2) & \mbox{with probability } 1-p_u
			\end{array} \right.$,\\
			\hspace*{2.7cm} $\gamma_+^2 = 4$, $\gamma_-^2=1/4$, $p_u = (1-\gamma_-^2)/(\gamma_+^2-\gamma_-^2)=1/5$  \\
			Location Mixture: $U_{it} \sim iid \left\{ \begin{array}{ll} N(\mu_+,\gamma^2) & \mbox{with probability } p_u \\
			N(-\mu_-,\gamma^2) & \mbox{with probability } 1-p_u
			\end{array} \right.$,\\
			\hspace*{3.4cm} $\mu_-=1/4$, $\mu_+=2$, $p_u=\mu_u^-/(\mu_u^-+\mu_u^+)=1/9$,\\
			\hspace*{3.4cm} $\gamma^2= 1 - p_u(\mu_u^+)^2 -(1-p_u)(\mu_u^-)^2=1/2$\\
			Initial Observations: $Y_{i0} \sim N(0,1)$ \\
			Gaussian Random Effects: $\lambda_i|Y_{i0} \sim N(\phi_0+\phi_1Y_{i0},\underline{\Omega})$, $\phi_0=0$, $\phi_1=0$, $\underline{\Omega}=1$ \\
			Sample Size: $N=1,000$, $T=3$ \\
			Number of Monte Carlo Repetitions: $N_{sim}=1,000$ \\
	\hspace*{4cm}\includegraphics[width=2.5in]{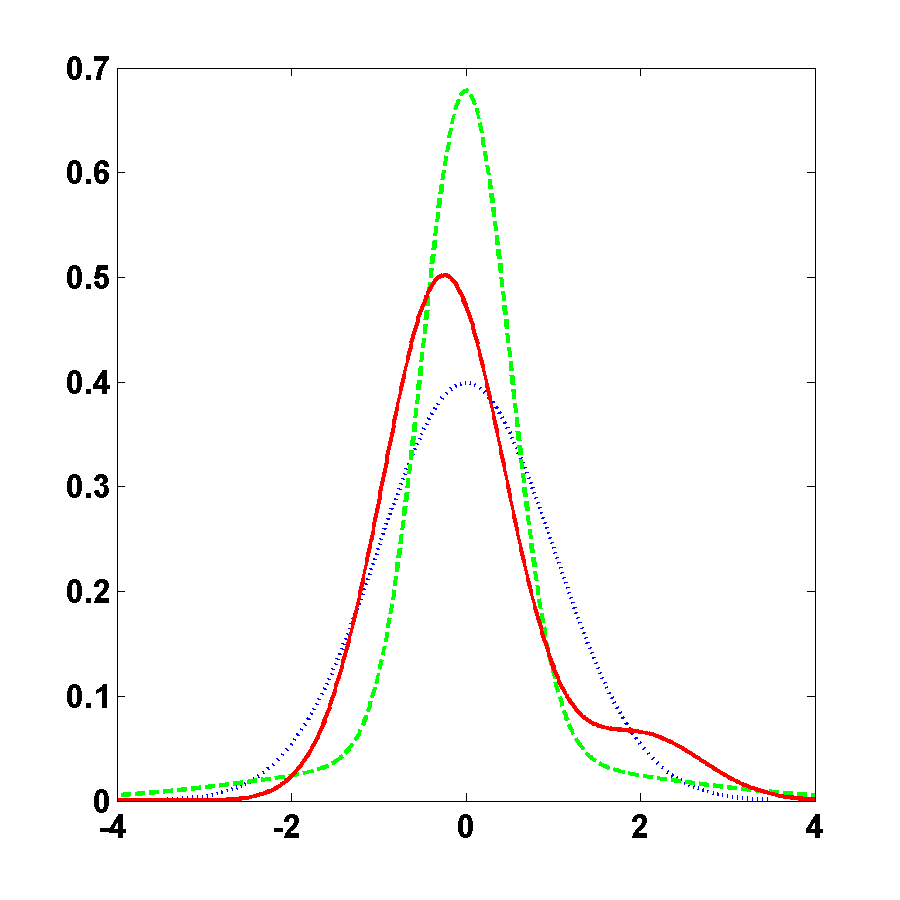} \\The plot overlays a $N(0,1)$ density (blue, dotted), the scale mixture \\(green, dashed), and the location mixture (red, solid).	\\			
			 \hline
		\end{tabular}
	} 
	\end{center}
\end{table}

%\begin{figure}[t!]
%	\caption{Densities of $U_{it}$ in Experiment 3}
%	\label{f_MC3_Udensity}
%	\begin{center}
%		\includegraphics[width=2.5in]{figures/mc3/mixture.png}	
%	\end{center}
%	{\footnotesize {\em Notes:} The plot overlays a $N(0,1)$ density (blue,dotted), the scale mixture (green,dashed), and the location mixture (red,solid). All three densities imply $\mathbb{E}[U_{it}]=0$ and $\mathbb{V}[U_{it}]=1$. The Monte Carlo design is summarized in Table~\ref{t_MC3design}.}\setlength{\baselineskip}{4mm}
%\end{figure}

In the third experiment, summarized in Table~\ref{t_MC3design}, we consider a misspecification of the Gaussian likelihood function by replacing the Normal distribution in the DGP with two mixtures. We consider a scale mixture that generates excess kurtosis and a location mixture that generates skewness. The innovation distributions are normalized such that $\mathbb{E}[U_{it}]=0$ and $\mathbb{V}[U_{it}]=1$. For the heterogeneous intercepts $\lambda_i$ we adopt the Gaussian random effects specification of Experiment~1. 
%We plot the two mixture distributions together with a standard normal distribution in Figure~\ref{f_MC3_Udensity}. 
In this experiment we compute the relative regret for five predictors:\footnote{The computation of the oracle predictor and the normalization of the regret by the posterior variance of $\lambda$ require a Gibbs sampler which is described in the Appendix.} the posterior mean predictor based on the non-parametric Tweedie correction and the plug-in predictor based on $\hat{\theta}_{QMLE}$ and $\hat{\theta}_{MLE}$, respectively. Note that both the QMLE and the GMM estimator of $\theta$ remain consistent under the likelihood misspecification. However, the (non-parametric) Tweedie correction no longer delivers a valid approximation of the posterior mean. 

\begin{table}[t!]
	\caption{Monte Carlo Experiment 3: Misspecified Likelihood Function}
	\label{t_MC3results}
	\begin{center}
		\scalebox{0.90}{
			\begin{tabular}{lcccccc}\hline\hline
				& \multicolumn{2}{c}{All Units} & \multicolumn{2}{c}{Bottom Group}  & \multicolumn{2}{c}{Top Group} \\
				&  & Median &  & Median  &  & Median \\
				Estimator / Predictor  & Regret & Forec.E. & Regret & Forec.E & Regret & Forec.E. \\ \hline
				\multicolumn{7}{c}{Scale Mixture -- Excess Kurtosis} \\ \hline
				Oracle Predictor  & (1153.7) &   0.000 & (67.98) &   0.002 & (55.99) &  -0.033  \\ \hline
				Post. Mean ($\hat{\theta}_{QMLE}$, BGK Kernel)  &   0.977 &  -0.002 &   2.031 &   0.170 & 2.226 & -0.227  \\ 
				Post. Mean ($\hat{\theta}_{GMM}$, BGK Kernel)  &   1.033 &  -0.000 &   2.055 &   0.162 & 2.388 & -0.211   \\ 
				Plug-In Predictor ($\hat{\theta}_{GMM}$, $\hat{\lambda}_i(\hat{\theta}_{GMM})$)  &   1.605 &   0.002 &   3.666 &   0.555 & 4.396 & -0.642  \\ 
				Loss-Function-Based Estimator  &   1.615 &   0.197 &   1.423 &   0.206 & 1.198 & 0.146   \\ 
				Pooled OLS  &   2.244 &  -0.286 &   4.295 &  -0.644 & 2.516 & -0.020  \\ \hline
				\multicolumn{7}{c}{Location Mixture -- Skewness} \\ \hline
				Oracle Predictor  & (1200.2) &  -0.146 & (63.29) &  -0.167 & (62.31) & -0.162 \\  \hline
				Post. Mean ($\hat{\theta}_{QMLE}$, BGK Kernel)  &   0.359 &  -0.106 &   0.338 &  -0.077 & 0.962 & -0.410   \\ 
				Post. Mean ($\hat{\theta}_{GMM}$, BGK Kernel)  &   0.398 &  -0.105 &   0.362 &  -0.080 & 1.086 & -0.399  \\ 
				Plug-In Predictor ($\hat{\theta}_{GMM}$, $\hat{\lambda}_i(\hat{\theta}_{GMM})$)  &   0.810 &  -0.091 &   1.359 &   0.330 & 2.784 & -0.818  \\ 
				Loss-Function-Based Estimator  &   0.807 &   0.099 &   0.461 &   0.030 & 0.497 & -0.006   \\ 
				Pooled OLS  &   1.240 &  -0.391 &   3.902 &  -0.889 & 0.828 & -0.235  \\  \hline
			\end{tabular}
		}
	\end{center}
	{\footnotesize {\em Notes:} The design of the experiment is summarized in Table~\ref{t_MC3design}. For the oracle predictor we report the compound risk (in parentheses) instead of the regret. The regret is standardized by the average posterior variance of $\lambda_i$, see Definition~\ref{def:ratio.optimality}.}\setlength{\baselineskip}{4mm}
\end{table}

%\afterpage{\clearpage}

The results are summarized in Table~\ref{t_MC3results}. The risk of the oracle predictors can be compared to that reported in Table~\ref{t_MC1design}. The excess kurtosis of the scale mixture and the skewness of the location mixture slightly reduce the posterior variance of $\lambda$ compared to the standard normal benchmark in Experiment~1. Due to the misspecification of the likelihood function, the relative regret of the various predictors increases considerably, but the relative ranking is essentially unchanged. The posterior mean predictors based on the nonparametric Tweedie correction dominate all the other predictor, attaining a relative regrets of about 1 and 0.4, respectively. Compared to the plug-in and loss-function based predictors, the Tweedie correction still reduces the regret 40\% to 50\%. The predictor based on the pooled OLS estimation performs the worst among the five predictors in this experiment.  

\section{Empirical Application}
\label{sec:empiricalapplication}

We will now use the previously-developed predictors to forecast pre-provision net revenues (PPNR) of bank holding companies (BHC). The stress tests that have become mandatory under the 2010 Dodd-Frank Act require banks to establish how PPNR varies in stressed macroeconomic and financial scenarios. A first step toward building and estimating models that provide trustworthy projections of PPNR and other bank-balance-sheet variables under hypothetical stress scenarios, is to develop models that generate reliable forecasts under the observed macroeconomic and financial conditions. Because of changes in the regulatory environment in the aftermath of the financial crisis as well as frequent mergers in the banking industry our large $N$ small $T$ panel-data-forecasting framework seems particularly attractive for stress-test applications.  

We generate a collection of panel data sets in which pre-provision net revenue as a fraction of consolidated assets (the ratio is scaled by 400 to obtain annualized percentages) is the key dependent variable. The data sets are based on the FR Y-9C consolidated financial statements for bank holding companies for the years 2002 to 2014, which are available through the website of the Federal Reserve Bank of Chicago. Because the balance sheet data exhibit strong seasonal features, we time-aggregate the quarterly observations into annual observations and take the time period $t$ to be one year. 

We construct rolling samples that consist of $T+2$ observations, where $T$ is the size of the estimation sample and varies between $T=3$ and $T=11$ years. The additional two observations in each rolling sample are used, respectively, to initialize the lag in the first period of the estimation sample and to compute the error of the one-step-ahead forecast. For instance, with data from 2002 to 2014 we can construct $M=9$ samples of size $T=3$ with forecast origins running from $\tau = 2005$ to $\tau = 2013$. Each rolling sample is indexed by the pair $(\tau,T)$. The cross-sectional dimension $N$ varies from sample to sample and ranges from approximately $=460$ to 725. Further details about the data as well as a description of our procedure to create balanced panels and eliminate outliers are provided in the Appendix.

In Section~\ref{subsection:empiricalapplication.simplemodel} we use the basic dynamic panel data model to generate PPNR forecasts. In Section~\ref{subsection:empiricalapplication.covariates} we extend the model to include covariates and compare forecasts under the actual realization of the covariates and stressed scenarios in which we set the covariantes to counterfactual levels. 

\subsection{Results from the Basic Dynamic Panel Model}
\label{subsection:empiricalapplication.simplemodel}

We begin by evaluating forecasts from the basic dynamic panel model in (\ref{m.restricted.linear.gaussian.panel.regression}). The parametric Tweedie correction is based on $\lambda_i | (H_i,\theta) \sim N(\phi_0 + \phi_1 Y_{i0}, \underline{\omega}^2)$.
The forecast evaluation criterion is the mean-squared error (MSE) computed across institutions and across time:
\be
MSE = \frac{1}{M} \sum_{\tau=\tau_1}^{\tau_1+M-1} \left(  \frac{\frac{1}{N_\tau} \sum_{i=1}^{N_\tau}
	D_i({\cal Y}_{i\tau})\big( Y_{i\tau+1} - \widehat{Y}_{i\tau+1} \big)^2}{\frac{1}{N_\tau} \sum_{i=1}^{N_\tau} D_i({\cal Y}_{i\tau})} \right),
\ee
where $M$ is the number of rolling samples. 
Table~\ref{t_mse_simple_byT} summarizes the MSEs for different estimators and different sizes $T$ of the estimation samples. Recall that the unit of $\widehat{Y}_{i\tau}$ is annual revenue as fraction of total assets converted into annualized percentages.

\begin{table}[t!]
	\caption{MSE for Basic Dynamic Panel Model}
	\label{t_mse_simple_byT}
	\begin{center}
		\begin{tabular}{lccccc}\hline\hline
			& \multicolumn{5}{c}{Rolling Samples} \\
			& $T=3$ & $T=5$ & $T=7$ & $T=9$ & $T=11$ \\ \hline
			Post. Mean ($\hat{\theta}_{QMLE}$, Parametric)  & 0.74 & 0.69 & 0.58 & 0.48 & 0.45 \\
			Post. Mean ($\hat{\theta}_{QMLE}$, BGK Kernel)  & 0.84 & 0.74 & 0.59 & 0.50 & 0.46 \\
			Plug-In Predictor ($\hat{\theta}_{QMLE}$, $\hat{\lambda}_i(\hat{\theta}_{QMLE})$)
			& 0.90 & 0.79 & 0.60 & 0.51 & 0.48 \\ 
			Post. Mean ($\hat{\theta}_{GMM}$, Parametric)   & 1.08 & 0.83 & 0.60 & 0.49 & 0.43 \\
			Post. Mean ($\hat{\theta}_{GMM}$, BGK Kernel)   & 1.16 & 0.93 & 0.61 & 0.50 & 0.44 \\
			Plug-In Predictor ($\hat{\theta}_{GMM}$, $\hat{\lambda}_i(\hat{\theta}_{GMM})$)  
			& 1.17 & 0.89 & 0.61 & 0.51 & 0.46 \\
			Loss-Function-Based Estimator                   & 0.91 & 0.84 & 0.63 & 0.53 & 0.42 \\
			Pooled OLS                                      & 0.71 & 0.68 & 0.57 & 0.48 & 0.45 \\ \hline
		\end{tabular}
	\end{center}
	{\footnotesize {\em Notes:} The MSEs are computed across the different forecast origins $\tau$ associated with each sample size $T$.}\setlength{\baselineskip}{4mm}
\end{table}

\begin{figure}
	\caption{Tweedie Corrections for $T=5$ and $\tau=2012$}
	\label{f_empirical_tweedie_T5}
	\begin{center}
		\begin{tabular}{ccc}
			%			\multicolumn{3}{c}{Rolling Sample $\tau=2007$}\\
			%			$Y_{i0} = 0$ & $Y_{i0} = -2$ & $Y_{i0} = -3$ \\
			%			\includegraphics[width=2in]{figures/empirical/T5tau2007/tweedie-lambdahat-y00.png} &
			%			\includegraphics[width=2in]{figures/empirical/T5tau2007/tweedie-lambdahat-y0m2.png} &	   	
			%			\includegraphics[width=2in]{figures/empirical/T5tau2007/tweedie-lambdahat-y0m3.png} \\
			%			\multicolumn{3}{c}{Rolling Sample $\tau=2012$}\\
			$Y_{i0} = 0$ & $Y_{i0} = -2$ & $Y_{i0} = -3$ \\
			\includegraphics[width=1.75in]{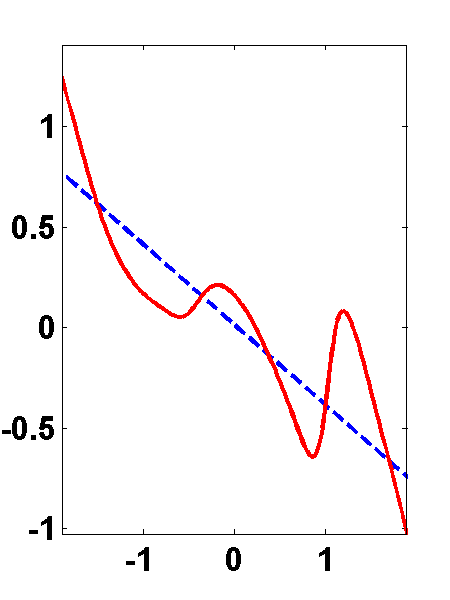} &
			\includegraphics[width=1.75in]{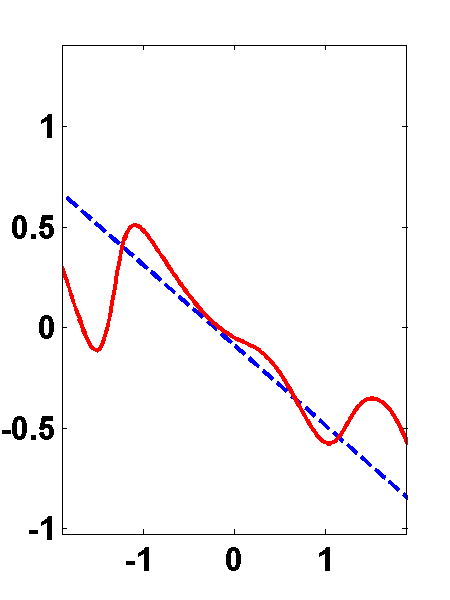} &	   	
			\includegraphics[width=1.75in]{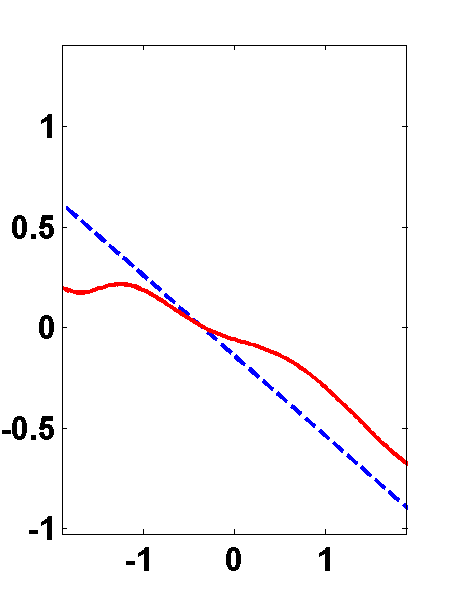} 	   	
		\end{tabular}
	\end{center}
	{\footnotesize {\em Notes:} Each panel shows the parametric (dashed blue) and the non-parametric (solid red) Tweedie correction for $\hat{\theta}_{QMLE}$.}\setlength{\baselineskip}{4mm}	
\end{figure}

For the short samples, i.e., $T=3$ and $T=5$, the QMLE-based predictors are more accurate than the GMM-based predictors. This 
discrepancy vanishes as the sample size is increased to $T=11$. The posterior mean predictors computed with the Tweedie correction are more accurate than the plug-in predictors. As expected, the MSE differential is largest in the small $T$ samples, because the unit-specific likelihood function contains fairly little information and the prior strongly influences the posterior. The parametric Tweedie correction delivers more accurate predictions than the non-parametric Tweedie correction, in particular for small $T$. In Figure~\ref{f_empirical_tweedie_T5} we compare the Tweedie corrections for $T=5$ and $\tau=2012$. While the corrections are quite similar for values of the sufficient statistic $\hat{\lambda}_i(\rho) = \frac{1}{T} \sum_{t=1}^T (Y_{it}-\rho Y_{it-1})$ between -1\% and 1\%, the non-parametric correction behaves somewhat erratic outside of this interval which hurts the predictive performance.

Returning to the MSE results in Table~\ref{t_mse_simple_byT}, the posterior mean predictor yields roughly the same MSE as pooled OLS. This suggests that {\em a posteriori} the data sets contain only weak evidence for heterogeneous intercepts. In this regard, the parametric specification is more efficient in shrinking the intercept estimates toward a common value. Finally, for all sample sizes except $T=11$, the posterior-mean predictor based on $\hat{\theta}_{QMLE}$ and the parametric Tweedie correction is more accurate than the loss-function-based predictor.   

In Table~\ref{t_mse_simple_T5} we focus on the sample size $T=5$. In addition to averaging forecast errors across all $T=5$ samples, we also report results for specific forecast origins, namely choices of $\tau$ that correspond to the years 2007, the onset of the Great Recession, and 2012, which is during the recovery period. Moreover, we compute MSEs based on cross-sectional selection rules that depend on the level of PPNR at the forecast origin $\tau$. We focus on institutions with PPNR less than 0\%, -1\%, -2\%, and -3\%, respectively. 
Because the QMLE predictors dominate the GMM predictors and the parametric Tweedie correction was preferable to the nonparametric correction, we now restrict our attention to the posterior-mean predictor based on $\hat{\theta}_{QMLE}$ and the parametric Tweedie correction, the $\hat{\theta}_{QMLE}$ plug-in predictor, and predictors constructed from loss-function-based estimates and pooled OLS, respectively.  

For the 2007 sample, the plug-in and the loss-function-based predictor are dominated by the other two predictors. The performance of the posterior-mean and the pooled-OLS predictor are essentially identical. For the 2012 sample, the posterior-mean predictor performs better than the plug-in predictor if we average across all institutions or if we condition on 
BCHs with PPNR of less than -3\%. In the other cases the ranking is reversed. Across all rolling samples, the posterior mean predictor dominates. Across all institutions its performance is only slightly better than pooled OLS, but if we condition on BCHs with PPNR of less than -1\%, -2\%, or -3\% then the accuracy relative to pooled OLS is more pronounced.

\begin{table}[t!]
	\caption{MSE for Basic Dynamic Panel Model for $T=5$}
	\label{t_mse_simple_T5}
	\begin{center}
		\scalebox{0.9}{
		\begin{tabular}{lccccc}\hline\hline
			& \multicolumn{5}{c}{Selection $D_i({\cal Y}_{i\tau})$} \\
			& All & $y_{i\tau} \le 0$ & $y_{i\tau} \le -1$ & $y_{i\tau} \le -2$ & $y_{i\tau} \le -3$ \\ \hline
		\multicolumn{5}{c}{Rolling Sample $\tau = 2007$} \\ \hline
			Post. Mean ($\hat{\theta}_{QMLE}$, Parametric)  & 0.90 & 0.90 & 1.04 & 1.29 & 1.72 \\
			Plug-In Predictor ($\hat{\theta}_{QMLE}$, $\hat{\lambda}_i(\hat{\theta}_{QMLE})$)
			& 1.26 & 1.21 & 1.39 & 1.65 & 2.08 \\ 
			Loss-Function-Based Estimator                   & 1.17 & 1.17 & 1.54 & 2.31 & 1.99 \\
			Pooled OLS                                      & 0.91 & 0.91 & 1.04 & 1.28 & 1.71 \\ \hline
	    \multicolumn{5}{c}{Rolling Sample $\tau = 2012$} \\ \hline
			Post. Mean ($\hat{\theta}_{QMLE}$, Parametric)  & 0.51 & 0.56 & 0.83 & 0.91 & 1.01 \\
			Plug-In Predictor ($\hat{\theta}_{QMLE}$, $\hat{\lambda}_i(\hat{\theta}_{QMLE})$)
			& 0.55 & 0.51 & 0.75 & 0.85 & 1.05 \\ 
			Loss-Function-Based Estimator                   & 0.63 & 0.69 & 0.98 & 1.02 & 1.00 \\
			Pooled OLS                                      & 0.48 & 0.57 & 0.85 & 0.97 & 1.12 \\ \hline
		\multicolumn{5}{c}{All Rolling Samples $\tau = 2007,\ldots,2013$} \\ \hline
			Post. Mean ($\hat{\theta}_{QMLE}$, Parametric)  & 0.69 & 0.88 & 1.12 & 1.43 & 1.69 \\
			Plug-In Predictor ($\hat{\theta}_{QMLE}$, $\hat{\lambda}_i(\hat{\theta}_{QMLE})$)
			& 0.79 & 1.00 & 1.32 & 1.72 & 2.16 \\ 
			Loss-Function-Based Estimator                   & 0.84 & 1.00 & 1.24 & 1.54 & 1.63 \\
			Pooled OLS                                      & 0.71 & 0.90 & 1.16 & 1.50 & 1.80 \\ \hline			    	
		\end{tabular}
	}
	\end{center}
	{\footnotesize {\em Notes:} For the last panel (all rolling samples) the MSEs are computed across the different forecast origins $\tau$.}\setlength{\baselineskip}{4mm}
\end{table}

Table~\ref{t_parameters_simple_T5} in the Appendix provides point estimates of the parameters of the basic dynamic panel model and the parametric correlated random effects distribution for $T=5$ and $\tau = 2007, \ldots, 2013$.
Until 2010 the estimated variance of the correlated random effects distribution is essentially zero, which implies that $\lambda_i \approx \phi_0+\phi_1 Y_{i0}$. Because of a non-zero $\hat{\phi}_1$ the resulting predictor is not exactly pooled OLS but it is very similar as we have seen from the results in Table~\ref{t_mse_simple_T5}. Starting in 2011, we obtain non-trivial estimates of $\hat{\underline{\omega}}^2$ which imply non-trival {\em a priori} dispersion of the intercepts (that is not due to the dispersion in initial conditions). Overall, the estimates $\hat{\underline{\omega}}^2$ imply a large degree of shrinkage.
The positive estimate $\hat{\phi}_1$ generates positive correlation between $\lambda_i$ and $Y_{i0}$.
The intercept of the correlated random effects distribution drops during the Great Recession\footnote{Recall that the $\tau=2010$ estimation sample comprises the observations for 2006-2010.}, which is consistent with the fact that bank revenues eroded during the financial crisis.  
The estimated common autoregressive coefficients range from 0.7 to 0.9.

\begin{table}
	\caption{Parameter Estimates for $T=5$: $\hat{\theta}_{QMLE}$, Parametric Tweedie Correction}
	\label{t_parameters_simple_T5}
	\begin{center}
		\scalebox{0.9}{
		\begin{tabular}{lcccccc} \hline \hline
		$\tau$ & $\hat{\rho}$ & $\hat{\sigma}^2$ & $\hat{\phi}_0$ & $\hat{\phi}_1$ & $\hat{\underline{\omega}}^2$ & N \\ \hline
		2007 & 0.90 & 0.61 & 0.03 & 0.01 & 6E-8 & 537 \\ 
		2008 & 0.83	& 0.55 & 0.11 & 0.05 & 2E-8 & 598 \\
		2009 & 0.76	& 0.76 & 0.01 &	0.10 & 4E-8 & 613 \\
		2010 & 0.80 & 0.67 &-0.05 & 0.09 & 2E-7 & 606 \\
		2011 & 0.79	& 0.58 &-0.02 &	0.07 & 0.07 & 582 \\
		2012 & 0.71	& 0.53 & 0.04 &	0.13 & 0.16 & 587 \\
		2013 & 0.79	& 0.58 &-0.05 &	0.12 & 0.09 & 608 \\ \hline
	    \end{tabular}
	}
	\end{center}
	{\footnotesize {\em Notes:} Point estimates for the model $Y_{it+1} = \lambda_i + \rho Y_{it} + U_{it+1}$, $U_{it+1} \sim N(0,\sigma^2)$, $		
	\lambda_i|Y_{i0} \sim N(\phi_0 + \phi_1 Y_{i0}, \underline{\omega}^2)$.}\setlength{\baselineskip}{4mm}
\end{table}

\subsection{Results from Models with Covariates}
\label{subsection:empiricalapplication.covariates}

To analyze the performance of the banking sector under stress scenarios it is necessary to add predictors to the dynamic panel data model that reflect macroeconomic and financial conditions. We consider three aggregate variables: the unemployment rate, the federal funds rate, and the spread between the federal funds rate and the 10-year treasury bill. Because these predictors are not bank-specific, the effect of the predictors on PPNR has to be identified from time-series variation, which is challenging given the short time-dimension of our panels. We consider two specifications: the first model only includes the unemployment rate as additional predictor and we focus on the $T=5$ data sets. The second model includes all three aggregate predictors and we estimated it based on the $T=11$ sample. 

We generate forecasts using the actual values of the aggregate predictors (which we can evaluate based on the actual PPNR realizations for the forecast perior) and compare these forecasts to predictions under a stressed scenario, in which we use hypothetical values for the predictors.
When analyzing stress scenarios, one is typically interested
in the effect of stressed economic conditions on the current performance of the banking sector. For this reason, we are changing the timing convention slightly and include the time $t$ macroeconomic and financial variables into the vector $W_{it-1}$. We are implicitly assuming that there is no feedback from disaggregate BCH revenues to aggregate conditions. While this assumption is inconsistent with the notion that the performance of the banking sector affects macroeconomic outcomes, elements of the Comprehensive Capital Analysis and Review (CCAR) conducted by the Federal Reserve Board of Governors have this partial equilibrium flavor. 

\noindent {\bf Results From a Model with Unemployment.} We use the unemployment rate (UNRATE) from the FRED database maintained by the Federal Reserve Bank of St. Louis and convert it to annual frequency by temporal averaging. We begin by computing MSEs, which are reported in Table~\ref{t_mse_unemployment_T5}. This table has the same format as Table~\ref{t_mse_simple_T5}: we consider MSEs for 2007, 2012, and averaged across all rolling samples. Moreover, we compute MSEs conditional on the level of PPNR at the forecast origin. A few observations stand out. First, the MSE for the posterior mean predictor is slightly reduced by including unemployment for the 2007 and 2012 samples, but across all of the rolling samples it slightly increases. Second, the gain of using the Tweedie correction, that is, the MSE differential between the plug-in predictor and the posterior mean predictor, becomes larger as we include unemployment. This is very intuitive: the more coefficients need to be estimated based on a given time-series dimension, the more important the shrinkage induced from the prior distribution. Third, the performance of the posterior-mean predictor and the pooled-OLS predictors remain very similar, meaning that the Tweedie correction shrinks toward pooled OLS.\footnote{This is supported by the estimates of $\hat{\underline{\omega}}_1^2$ and $\hat{\underline{\omega}}_2^2$ reported in the Online Appendix.}

%Data source descriptions:
%
%All data are downloaded from FRED and averaged to get annual rates.
%
%1. unemployment:
%Source: US. Bureau of Labor Statistics
%Release: Employment Situation
%
%The unemployment rate represents the number of unemployed as a percentage of the labor force. Labor force data are restricted to people 16 years of age and older, who currently reside in 1 of the 50 states or the District of Columbia, who do not reside in institutions (e.g., penal and mental facilities, homes for the aged), and who are not on active duty in the Armed Forces.
%
%This rate is also defined as the U-3 measure of labor underutilization.
%
%The series comes from the 'Current Population Survey (Household Survey)'
%
%The source code is: LNS14000000
%
%Suggested Citation:
%US. Bureau of Labor Statistics, Civilian Unemployment Rate [UNRATE], retrieved from FRED, Federal Reserve Bank of St. Louis; https://fred.stlouisfed.org/series/UNRATE, August 31, 2016.
   
\begin{table}[t!]
	\caption{MSE for Model with Unemployment for $T=5$}
	\label{t_mse_unemployment_T5}
	\begin{center}
		\scalebox{0.9}{
		\begin{tabular}{lccccc}\hline\hline
			& \multicolumn{5}{c}{Selection $D_i({\cal Y}_{i\tau})$} \\
			& All & $y_{i\tau} \le 0$ & $y_{i\tau} \le -1$ & $y_{i\tau} \le -2$ & $y_{i\tau} \le -3$ \\ \hline
			\multicolumn{5}{c}{Rolling Sample $\tau = 2007$} \\ \hline
			Post. Mean ($\hat{\theta}_{QMLE}$, Parametric)  & 0.88 & 0.95 & 1.11 & 1.40 & 1.72 \\
			Plug-In Predictor ($\hat{\theta}_{QMLE}$, $\hat{\lambda}_i(\hat{\theta}_{QMLE})$)
			& 1.38 & 1.62 & 2.23 & 2.61 & 3.29 \\ 
			Loss-Function-Based Estimator                   & 1.44 & 1.23 & 1.55 & 2.14 & 1.92 \\
			Pooled OLS                                      & 0.88 & 0.93 & 1.06 & 1.31 & 1.70 \\ \hline
			\multicolumn{5}{c}{Rolling Sample $\tau = 2012$} \\ \hline
			Post. Mean ($\hat{\theta}_{QMLE}$, Parametric)  & 0.49 & 0.55 & 0.80 & 0.92 & 1.09 \\
			Plug-In Predictor ($\hat{\theta}_{QMLE}$, $\hat{\lambda}_i(\hat{\theta}_{QMLE})$)
			& 0.64 & 0.67 & 0.98 & 1.27 & 1.73 \\ 
			Loss-Function-Based Estimator                   & 0.84 & 1.12 & 1.56 & 1.66 & 1.60 \\
			Pooled OLS                                      & 0.49 & 0.58 & 0.85 & 0.97 & 1.12 \\ \hline
			\multicolumn{5}{c}{All Rolling Samples $\tau = 2007,\ldots,2013$} \\ \hline
			Post. Mean ($\hat{\theta}_{QMLE}$, Parametric)  & 0.72 & 0.92 & 1.16 & 1.45 & 1.70 \\
			Plug-In Predictor ($\hat{\theta}_{QMLE}$, $\hat{\lambda}_i(\hat{\theta}_{QMLE})$)
			& 2.52 & 3.90 & 4.39 & 6.07 & 5.88 \\ 
			Loss-Function-Based Estimator                   & 2.14 & 3.22 & 3.71 & 4.91 & 4.56 \\
			Pooled OLS                                      & 0.72 & 0.96 & 1.23 & 1.56 & 1.86 \\ \hline			    	
		\end{tabular}
	}
	\end{center}
	{\footnotesize {\em Notes:} For the last panel (all rolling samples) the MSEs are computed across the different forecast origins $\tau$.}\setlength{\baselineskip}{4mm}
\end{table}
   
\begin{figure}[t!]
	\caption{Predictions under Actual and Stressed Scenario for $T=5$}
	\label{f_stress_unemployment_T5}
	\begin{center}
	   \begin{tabular}{cc}
	   	Post. Mean ($\hat{\theta}_{QMLE}$, Parametric) & Plug-In Predictor ($\hat{\theta}_{QMLE}$, $\hat{\lambda}_i(\hat{\theta}_{QMLE})$) \\
	   	\multicolumn{2}{c}{Rolling Sample $\tau=2007$} \\	   	
	   	\includegraphics[width=2.2in]{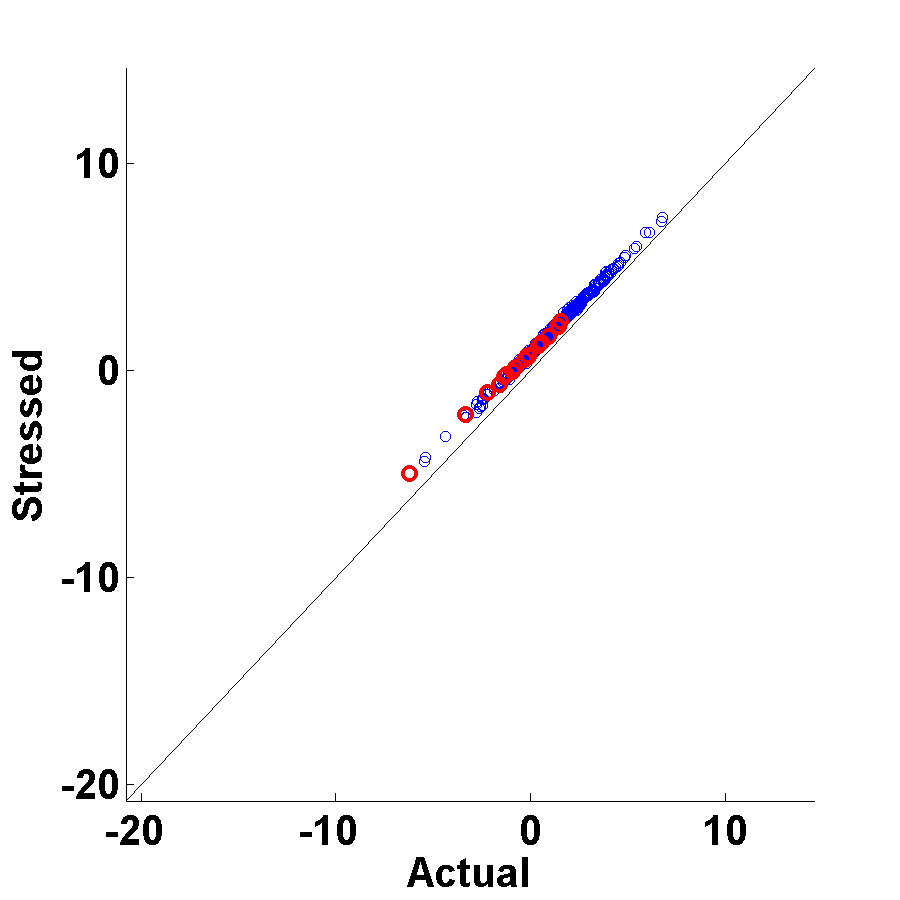} &
	   	\includegraphics[width=2.2in]{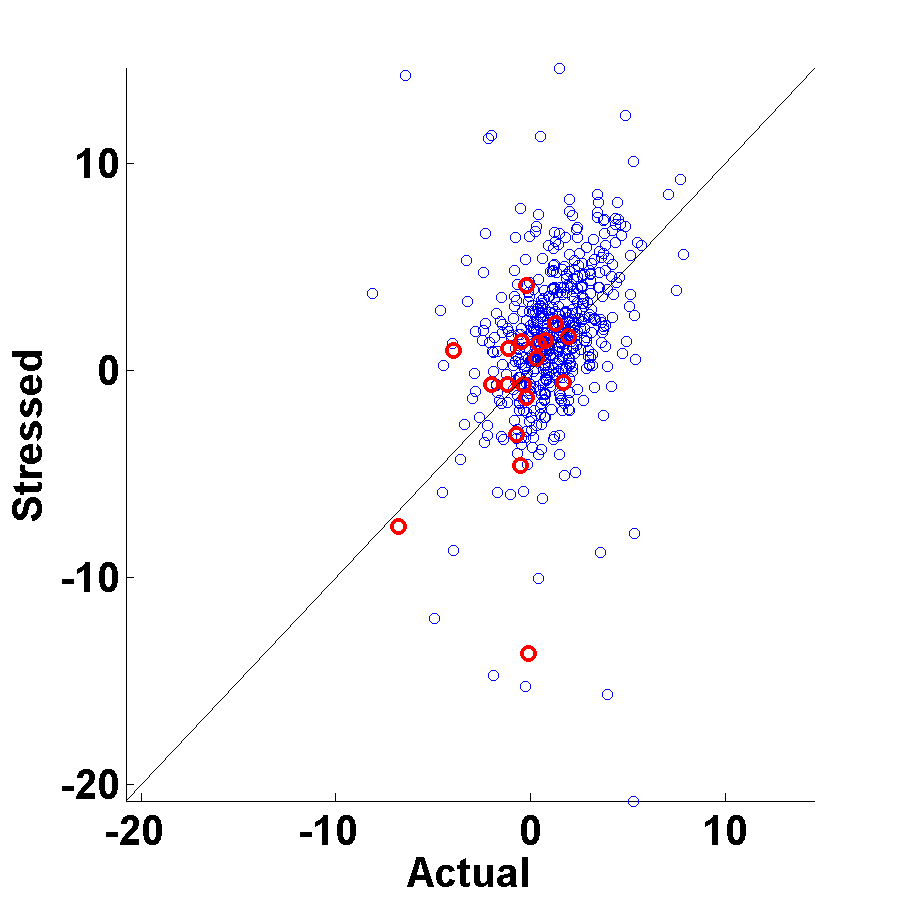} \\
	   	\multicolumn{2}{c}{Rolling Sample $\tau=2012$} \\
	   	\includegraphics[width=2.2in]{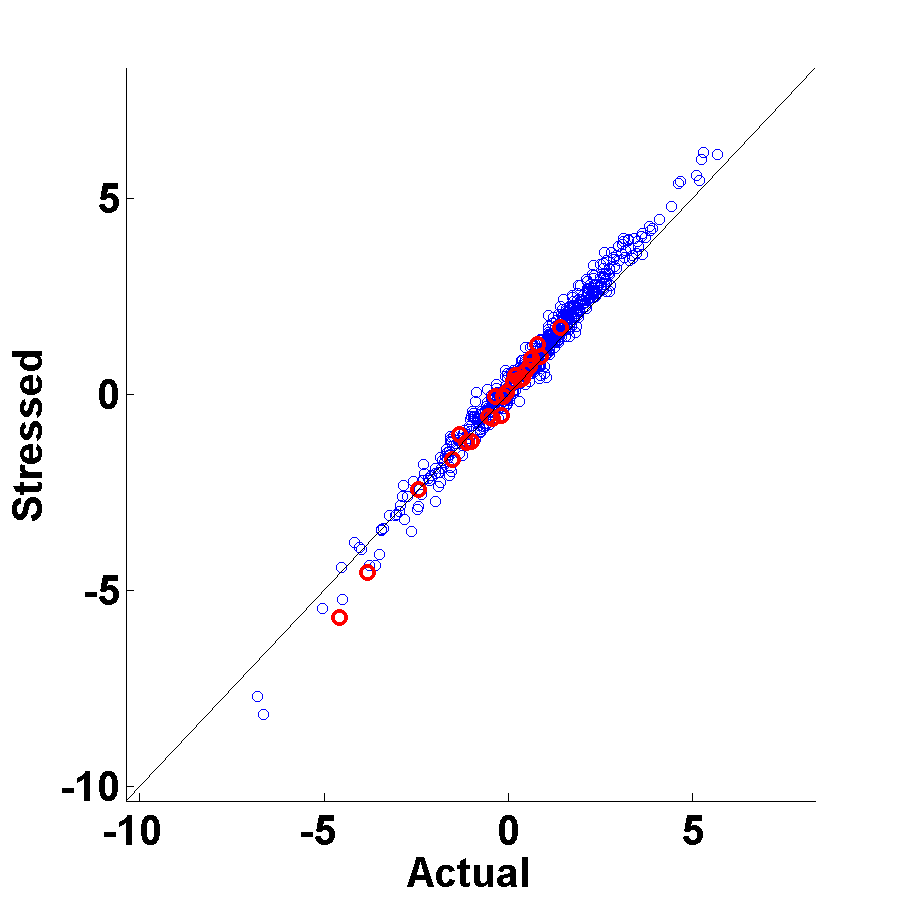} &
	   	\includegraphics[width=2.2in]{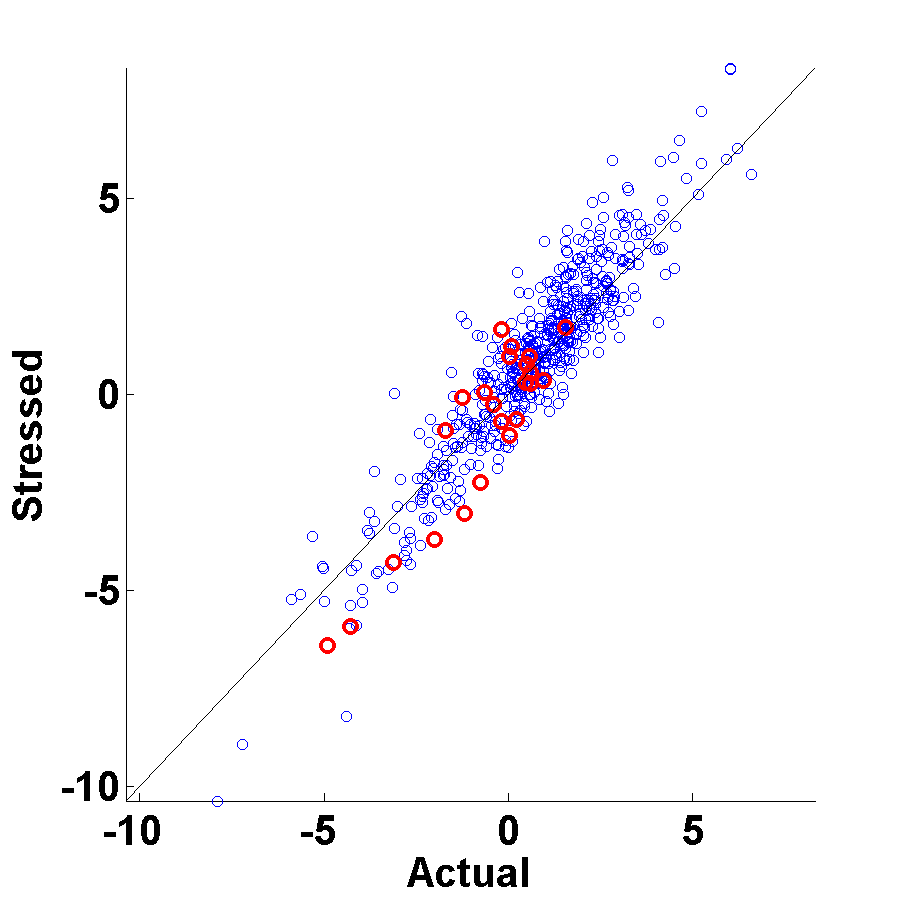} 	
	   \end{tabular}   	
	\end{center}
	{\footnotesize {\em Notes:} Each dot corresponds to a BHC in our dataset. We plot point predictions of PPNR under the actual macroeconomic conditions (the unemployment rate is at its observed level in period $\tau+1$) and a stressed scenario (unemployment rate is 5\% higher than its actual level).}\setlength{\baselineskip}{4mm}
		
\end{figure}

We now impose stress by increasing the unemployment rate by 5\%. This corresponds to the unemployment movement in the {\em severely adverse} macroeconomic scenario in the Federal Reserve's CCAR 2016. In Figure~\ref{f_stress_unemployment_T5} we are comparing one-year-ahead predictions for forecast origins $\tau=2007$ and $\tau=2012$ under the actual period $\tau+1$ unemployment rate and the stressed unemployment rate. Each circle in the graphs corresponds to a particular BHC. We indicate institutions with assets greater than 50 billion dollars\footnote{These are the BHCs that are subject to the CCAR requirements.} by red circles, while the other BHCs appear as blue circles. The large institutions have in general smaller revenues than the smaller BHCs. According to the plug-in predictor (the two right panels), the response to the unemployment shock is very heterogeneous. For about half of the intitutions a rise in unemployment leads to a drop in revenues, whereas for the other half higher unemployment is associated with larger revenues. However, we know from Table~\ref{t_mse_simple_T5} that forecasts from the plug-in predictor are fairly inaccurate. The stress-test implications of the posterior mean predictor are markedly different. Due to the strong shrinkage the effect is more homogeneous across institutions and appears to be slightly positive. 

\noindent {\bf A Model with Unemployment, Federal Funds Rate, and Spread.} We now expand the list of covariates and in addition to the unemployment rate include the federal funds rate and the spread between the federal funds rate and the 10-year treasury bill. Both series are obtained from the FRED database (FEDFUNDS and DGS10). We convert the series into annual frequency by temporal averaging. Because we now have three regressors that do not vary across units (meaning all BHCs are operating within the same macroeconomic conditions, but may have hetereogeneous responses to these conditions), we focus on the data set with the largest time series dimension, namely $T=11$. 
MSEs are presented in Table~\ref{t_mse_unemploymentffrspread_T11}. The forecast origin is $\tau=2013$. As before, the posterior mean predictor with the Tweedie correction strongly dominates the plug-in predictor. Moreover, the posterior mean predictor is also slightly more accurate than the predictor based on pooled OLS.\footnote{While the estimates of the conditional variances of the $\lambda_{ij}$ coefficients are close to zero, the estimated conditional means of $\lambda_{ij}$ vary with $Y_{i0}$. This explains the difference between the posterior mean and the pooled-OLS predictor.} Unlike in the previous cases, the predictor constructed from the loss-function-based estimate of the model coefficients now performs slightly better than the posterior mean predictor.

\begin{table}[t!]
	\caption{MSE for Model with Unemployment, Fed Funds Rate, and Spread for $T=11$}
	\label{t_mse_unemploymentffrspread_T11}
	\begin{center}
		\scalebox{0.9}{
		\begin{tabular}{lccccc}\hline\hline
			& \multicolumn{5}{c}{Selection $D_i({\cal Y}_{i\tau})$} \\
			& All & $y_{i\tau} \le 0$ & $y_{i\tau} \le -1$ & $y_{i\tau} \le -2$ & $y_{i\tau} \le -3$ \\ \hline
			Post. Mean ($\hat{\theta}_{QMLE}$, Parametric)  & 0.49 & 0.64 & 0.94 & 1.00 & 1.08 \\
			Plug-In Predictor ($\hat{\theta}_{QMLE}$, $\hat{\lambda}_i(\hat{\theta}_{QMLE})$)
			& 0.78 & 1.35 & 2.14 & 2.04 & 1.61 \\ 
			Loss-Function-Based Estimator                   & 0.47 & 0.61 & 0.88 & 0.88 & 0.78 \\
			Pooled OLS                                      & 0.50 & 0.68 & 1.00 & 1.04 & 1.10 \\ \hline
		\end{tabular}
	}
	\end{center}
	{\footnotesize {\em Notes:} The MSEs are computed for the forecast origin $\tau=2013$.}\setlength{\baselineskip}{4mm}
\end{table}

\begin{figure}[t!]
	\caption{Predictions under Actual and Stressed Scenario for $T=11$ and $\tau=2013$}
	\label{f_stress_unemploymentffrspread_T11}
	\begin{center}
		\begin{tabular}{cc}
			Post. Mean ($\hat{\theta}_{QMLE}$, Parametric) & Plug-In Predictor ($\hat{\theta}_{QMLE}$, $\hat{\lambda}_i(\hat{\theta}_{QMLE})$) \\
			\includegraphics[width=2.2in]{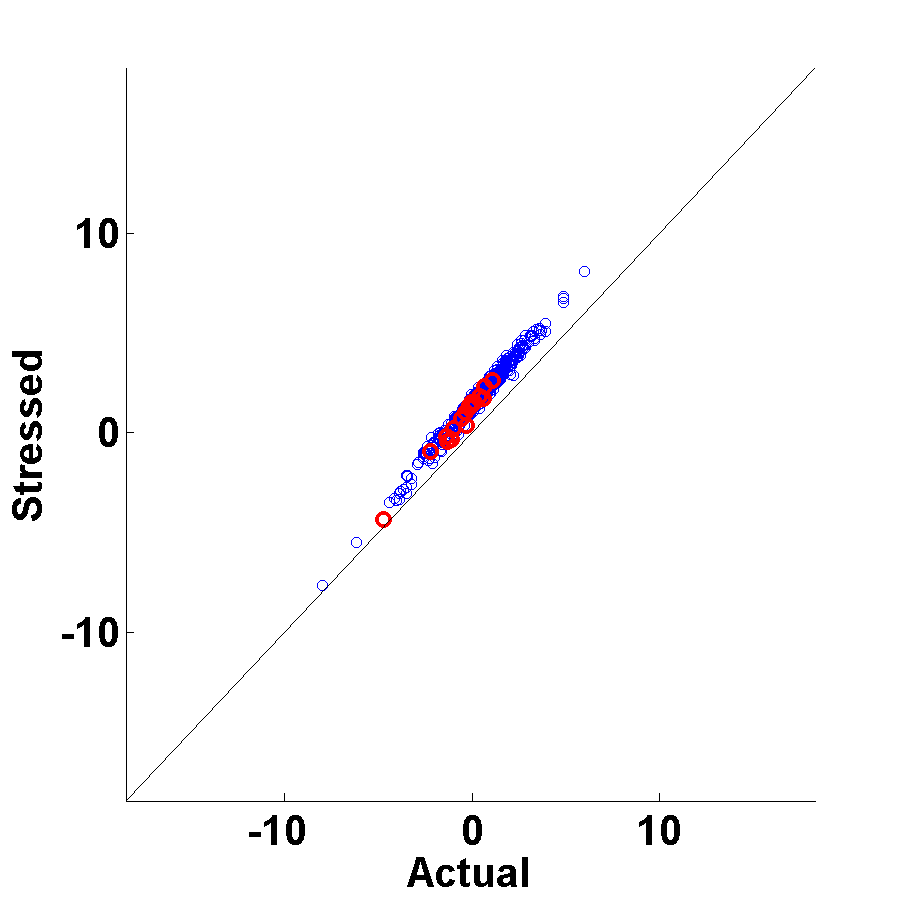} &
			\includegraphics[width=2.2in]{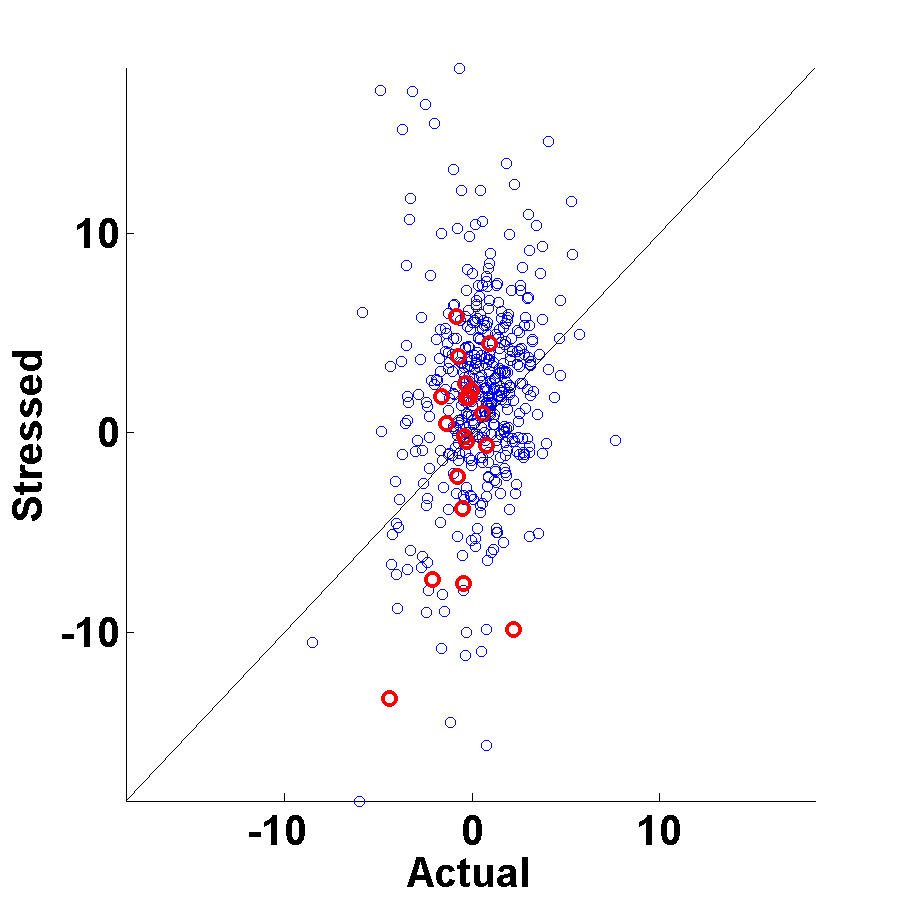} 	
		\end{tabular}   	
	\end{center}
	{\footnotesize {\em Notes:} Each dot corresponds to a BHC in our dataset. We plot point predictions of PPNR under the actual macroeconomic conditions (the unemployment rate, federal funds rate, and spread are at their observed 2014 levels) and a stressed scenario (the unemployment rate, federal funds rate, and spread are 5\% higher than their actual level in 2014).}\setlength{\baselineskip}{4mm}	
\end{figure}

Figure~\ref{f_stress_unemploymentffrspread_T11} compares PPNR predictions under the actual macroeconomic conditions and a stressed macroeconomic scenario. The stressed scenario comprises an increase in the unemployment rate by 5\% (as before) and an increase in nominal interest rates and spreads by 5\%. This scenario could be interpreted as an aggressive monetary tightening that induced a sharp drop in macroeconomic activity. The plug-in predictor generates very heterogeneous responses to the macroeconomic stress scenario. Some banks benefit from the monetary tightening and others experience a substantial fall in revenues. The posterior mean predictor implies a much more homogeneous response of the banking sector under which there is a very small (relative to the cross-sectional dispersion) increase in predicted revenues.  

\noindent {\bf Discussion.} We view this analysis as a first-step toward applying state-of-the-art panel data forecasting techniques to stress tests. First, it is important to ensure that the empirical model is able to accurately predict bank revenues and balance sheet characteristics under observed macroeconomic conditions. Our analysis suggests that there are substantial performance differences among various plausible estimators and predictors. Second, a key challenge is to cope with model complexity in view of the limited information in the sample. There is a strong temptation to over-parameterize models that are used for stress tests. We decided to time-aggregate the revenue data to smooth out irregular and non-Gaussian features of the accounting data at the quarterly frequency. This limits the ability to precisely measure the potentially heterogeneous effects of macroeconomic conditions on bank performance. Prior information is used to discipline the inference. In our empirical Bayes procedure, this prior information is essentially extracted from the cross-sectional variation in the data set. While we {\em a priori} allowed for heterogeneous responses, it turned out {\em a posteriori}, trading-off model complexity and fit, that the estimated coefficients exhibited very little heterogeneity. Third, our empirical results indicate that relative to the cross-sectional dispersion of PPNR, the effect of severely adverse scenarios on revenue point predictions are very small. We leave it future research to explore richer empirical models that focus on specific revenue and accounting components and consider a broader set of covariates. Finally, it would be desirable to allow for a feedback from the performance of the banking sector into the aggregate conditions.

\section{Conclusion}
\label{sec:conclusion}

The literature on panel data forecasting in settings in which the cross-sectional dimension is large and the time-series dimension is small is very sparse. Our paper contributes to this literature by developing an empirical Bayes predictor that uses the cross-sectional information in the panel to construct a prior distribution that can be used to form a posterior mean predictor for each cross-sectional unit. The shorter the time-series dimension, the more important this prior becomes for forecasting and the larger the gains from using the posterior mean predictor instead of a plug-in predictor. We consider a particular implementation of this idea for linear models with Gaussian innovations that is based on Tweedie's posterior mean formula. It can be implemented by estimating the cross-sectional distribution of sufficient statistics for the heterogeneous coefficients in the forecast model. We consider both parametric and nonparametric techniques to estimate this distribution. We provide a theorem that establishes a ratio-optimality property for the nonparametric estimator of the Tweedie correction. The nonparametric estimation works well in environments in which the cross-sectional distribution of heterogeneous coefficients is irregular. If it is well approximated by a Gaussian distribution, then a parametric implementation of the Tweedie correction is preferable. We illustrate in an application that our forecasting techniques may be useful to execute bank stress tests.
Our paper focuses on one-step-ahead point forecasts. We leave extensions to multi-step forecasting and density forecasting for future work.

\setstretch{1}

\ifx\undefined\BySame
\newcommand{\BySame}{\leavevmode\rule[.5ex]{3em}{.5pt}\ }
\fi
\ifx\undefined\textsc
\newcommand{\textsc}[1]{{\sc #1}}
\newcommand{\emph}[1]{{\em #1\/}}
\let\tmpsmall\small
\renewcommand{\small}{\tmpsmall\sc}
\fi

%\bibliography{ref}

\clearpage
\setstretch{1.3}
\appendix
\setcounter{saveeqn}{\value{section}}\renewcommand{\theequation}{\mbox{%
                \Alph{saveeqn}.\arabic{equation}}} \setcounter{saveeqn}{1} %
\setcounter{equation}{0}
\renewcommand*\thepage{A-\arabic{page}}
\setcounter{page}{1}
\renewcommand*\thetable{A-\arabic{table}}
\setcounter{table}{0}
\renewcommand*\thefigure{A-\arabic{figure}}
\setcounter{figure}{0}

\bc

{\Large {\bf Supplemental Appendix to ``Forecasting with Dynamic Panel Data Models'' }}

{\bf Laura Liu, Hyungsik Roger Moon, and Frank Schorfheide}

\ec

\section{Theoretical Derivations and Proofs}

\subsection{Proofs for Section~\ref{sec:model}} 

\begin{lemma} \label{lemma.id.sigmas.full.rank}
        Suppose that $T \geq k_w + 1 \geq 2$. Suppose that $W$ is a $T \times k_w$ matrix with $\rm{rank}(W) = k_w$. Let $\Sigma$ be a $T \times T$ matrix of rank $T$. Let $S= \Sigma W$. Then, $\rm{rank}(M_{S \otimes S} B) = T$, where $M_{S \otimes S}$ and $B$ are defined in the proof of Theorem \ref{thm.identification}.       
\end{lemma}

\noindent \textbf{Proof of Lemma \ref{lemma.id.sigmas.full.rank}.}
Notice that the matrix $B$ is a $T^2 \times T$ selection matrix that has one at positions $(1,1), (T+2,2), (2T+3,3),...,(T^2,T)$ and zeros at the other positions. Notice that since $\Sigma$ is full rank, $\mbox{rank}(S) = \mbox{rank}(\Sigma W) = \mbox{rank}(W)=k_w$. 
If $\mbox{rank}(S) = k_w$, then $ \mbox{rank}(S \otimes S) = k_w^2$. Since the rank of the projection matrix is the same as its trace, we have $\mbox{rank}(M_{S \otimes S}) = tr(M_{S \otimes S}) = T^2 - k_w^2.$ 

By the spectral decomposition, we can decompose $M_{S \otimes S} = F \Lambda F^{\prime}$, where $F$ is a $T^2 \times T^2$ orthogonal matrix and $\Lambda$ is a $T^2 \times T^2$ diagonal matrix whose first $T^2 - k_w^2$ elements are one and the rest are zero.
Since $F$ is full rank, $\mbox{rank}(M_{S \otimes S}B) = \mbox{rank}(F \Lambda F^{\prime} B) = \mbox{rank}(\Lambda F^{\prime}B)$. Notice that $F^{\prime}B$ is a $T^2 \times T$ matrix that collects the columns of $F^{\prime}$ in the positions of $1, T+2, 2T+3,...,T^2$. Since the columns of $F^{\prime}$ are linearly independent, $\mbox{rank}(F^{\prime}B) = T$.
Notice that $ \Lambda F^{\prime}B$ is a submatrix of $F^{\prime}B$ that selects the first $T^2 - k_w^2$ rows. 
Since $T - 1 \geq k_w$ and $T \geq 2$ implies that $T^2 - k_w^2 \geq 2T - 1 > T$, the $( T^2 - k_w^2) \times T$ submatrix of $F^{\prime}B$, $ \Lambda F^{\prime}B$, has rank $T$. $\Box$  

The matrix $\mathbb{E}\big[ (W_{it}',X_{it}',Z_{it}')'(W_{it}',X_{it}',Z_{it}') \big]$ has full rank for 
$t=1,\ldots,T$. The matrices $\sum_{s=t+1}^T W_{is-1}W_{is-1}'$ are invertible with probability one for all $t=1,\ldots,T-k_w$ and $i=1,\ldots,N$.

\noindent {\bf Proof of Theorem~\ref{thm.identification}.} 
(i) The parameters $\alpha$ and $\rho$ are identifiable by Assumption~\ref{as.identification}.

(ii) Let $Y_i$, $W_i$, $X_{i}$, $Z_{i}$ and $U_i$ denote the matrices vectors that stack $Y_{it}$, $W_{it-1}'$, $X_{it-1}'$, $Z_{it-1}'$, and $U_{it}$, respectively, for $t=1,\ldots,T$. Define
\begin{eqnarray*}
        \Sigma_i^{1/2}(\gamma) &=& \mbox{diag}\big( \sigma_1(h_i, \gamma_1), \ldots, \sigma_T(h_i, \gamma_T) \big), \\
        S_i(\gamma) &=& \Sigma_i^{-1/2}(\gamma) W_i, \quad M_i(\gamma) = I - S_i(S_i'S_i)^{-1} S_i'.
\end{eqnarray*}
Using the same manipulation as in the main text, we obtain the condition
\be
M_i(\tilde{\gamma}) \big( \Sigma_i^{-1/2}(\tilde{\gamma}) \Sigma_i(\gamma)\Sigma_i^{-1/2}(\tilde{\gamma}) -I \big) M_i'(\tilde{\gamma}) = 0.
\ee
for each $h_i$. Taking expectations with respect to $H_i$ and using Assumption~\ref{as.identification}(ii), we deduce that 
\be
\mathbb{E}\big[ M_i(\tilde{\gamma}) \big( \Sigma_i^{-1/2}(\tilde{\gamma}) \Sigma_i(\gamma)\Sigma_i^{-1/2}(\tilde{\gamma}) -I \big) M_i'(\tilde{\gamma}) \big] = 0.
\ee
if and only if $\tilde{\gamma} = \gamma$.

(iii) The subsequent argument is similar to the proof of Theorem~2 in \citet{ArellanoBonhomme2012}. Conditional on $\rho$, $\alpha$, and $\gamma$ we can remove the effect of $X_i$ and $Z_i$ from $Y_i$ and define
\be
\tilde{Y}_i = \Sigma_i^{-1/2}(\gamma)(Y_i - X_i \rho - Z_i \alpha) = S_i(\gamma) \lambda_i + V_i.
\label{appeq.tildey}
\ee
To simplify the notation, we will omit the $i$ subscripts and the $\gamma$ argument in the remainder of the proof.

Because $S(\gamma)$, $\lambda$ and $V$ are independent conditional on $H$ (and $\gamma$), we have
\begin{equation}
\ln \Psi_{\tilde{Y}} (\tau|h) = \ln \Psi_\lambda(S'\tau|h) + \ln \Psi_{V}(\tau) \label{appeq.psitildey}
\end{equation}  
Taking the second derivative with respect to $\tau$ leads to
\begin{eqnarray}
\frac{\partial^2}{\partial \tau \partial \tau'} \ln \Psi_{\tilde{Y}} (\tau|h) 
&=& \frac{\partial^2}{\partial \tau \partial \tau'} \left(  \ln \Psi_\lambda(S'\tau|h) \right) 
+  \frac{\partial^2}{\partial \tau \partial \tau'} \ln \Psi_{V}(\tau) \\ 
&=& S \left( \frac{\partial^2}{\partial \xi \partial \xi'} \ln \Psi_\lambda(S'\tau|h) \right)S' 
+  \frac{\partial^2}{\partial \tau \partial \tau'} \ln \Psi_{V}(\tau). \nonumber
\end{eqnarray}
Using the assumption that the $V_{t}$s are independent over $t$, we can write 
\[
\ln \Psi_V(\tau) = \sum_{t=1}^T \ln \Psi_{V_t}(\tau_t),
\]
where $\Psi_{V_t}$ is the characteristic function of $V_t$.
Then,
\begin{eqnarray}
\text{vec}
\left( \frac{\partial^2}{\partial \tau \partial \tau^{\prime}} 
\ln \Psi_V(\tau) \right) 
&=& \text{vec} 
\left(
\diag\left( 
\frac{\partial^2}{\partial \tau_1^2 } 
\ln \Psi_{V_1}(\tau_1),...,
\frac{\partial^2}{\partial \tau_T^2 } 
\ln \Psi_{V_T}(\tau_T)
\right)
\right) \\
&=& B \left( \frac{\partial^2}{\partial \tau_1^2 } 
\ln \Psi_{V_1}(\tau_1),...,
\frac{\partial^2}{\partial \tau_T^2 } 
\ln \Psi_{V_T}(\tau_T) \right)^{\prime} \nonumber
\end{eqnarray}
for a suitably chosen matrix $B$. 
Let 
\[
M_{S \otimes S} = I - S(S'S)^{-1}S' \otimes S(S'S)^{-1}S'. 
\]
Then, 
\begin{equation}
M_{S \otimes S} \text{vec}( \ln \Psi_{\tilde{Y}}(\tau|h) ) = M_{S \otimes S}B \left( \frac{\partial^2}{\partial \tau_1^2 } 
\ln \Psi_{V_1}(\tau_1),...,
\frac{\partial^2}{\partial \tau_T^2 } 
\ln \Psi_{V_T}(\tau_T) \right)^{\prime}. \label{eq.id.ln.char.func.v}
\end{equation}
Because $\Sigma(\gamma)$ is of full rank $T$ (Assumption \ref{as.identification}(iii)) and $W$ is of full rank of $k_w$ (Assumption \ref{as.identification}(iv)), $S(\gamma)$ has full rank $k_w$. Notice that $T \geq k_w + 1$. Then, according to Lemma \ref{lemma.id.sigmas.full.rank}, $M_{S \otimes S}B$ is also full rank. In turn, from $(\ref{eq.id.ln.char.func.v})$, we can identify $\ln \Psi_{V_t}(\tau_t)$ uniquely for $t=1,...,T$. 
Also using the restrictions that $\frac{\partial}{\partial \tau_t} 
\ln \Psi_{V_t}(0) = 0 \; (\mathbb{E}(V_{it})=0)$ and $\ln \Psi_{V_t}(0) = 0$, we can deduce that the characteristic function of $V_{t}$ is uniquely identified.

Next, we show how to identify $\ln \Psi_{\lambda}(\tau|h).$ Because $\ln \Psi_{\tilde{Y}} (\tau|h)$ and $ \ln \Psi_{V}(\tau)$ are identified, from (\ref{appeq.psitildey}) we obtain
\be 
\ln \Psi_{\tilde{Y}} (\tau|h) - \ln \Psi_{V}(\tau) = \ln \Psi_\lambda(S'\tau|h).
\ee
Taking second derivatives, we obtain
\be
\frac{ \partial^2}{\partial \tau \partial \tau'} \left(
\ln \Psi_{\tilde{Y}} (\tau|h) - \sum_{t=1}^T \ln \Psi_{V}(\tau_t) \right)
= S \left( \frac{ \partial^2}{\partial \xi \partial \xi'} 
\ln \Psi_\lambda(S'\tau|h)  \right) S'. 
\ee
Because $S$ is of full rank, we can identify
\be
\frac{ \partial^2}{\partial \xi \partial \xi'} 
\ln \Psi_\lambda(S'\tau|h)
= (S'S)^{-1}S' 
\left[
\frac{ \partial^2}{\partial \tau \partial \tau'} \left(
\ln \Psi_{\tilde{Y}} (\tau|h) - \sum_{t=1}^T \ln \Psi_{V}(\tau_t) \right)
\right]
S(S'S)^{-1}.
\ee

The mean $\mathbb{E}(\lambda|h)$ can be identified as follows. Note that 
\be
\hat{\lambda} = (S'S)^{-1}S'\tilde{Y} = \lambda + (S'S)^{-1}S'V.
\ee
Taking expectations yields
\be
\mathbb{E}(\lambda|h) = \mathbb{E}[\hat{\lambda}|h],
\ee 
because $\mathbb{E}[(S'S)^{-1}S'V|h] = (S'S)^{-1}S'\mathbb{E}[V|h] = 0$. 
Once the mean has been determined, we can identify $\ln \Psi_\lambda(\xi|h)$ using $\frac{ \partial}{\partial \xi } 
\ln \Psi_\lambda(0|h) = \mathbb{E}(\lambda|h)$ and $\ln \Psi_\lambda(0|h) = 0$.
$\blacksquare$

\noindent {\bf Discussion of Assumption~\ref{as.identification}(i).} We discuss an example of how to identify $\alpha$ and $\rho$  based on moment conditions in the general model (\ref{eq.yit}). Under the model (\ref{eq.yit}) we can remove the effect of $\lambda_i$ with the following within projections:
\begin{eqnarray*}
	Y_{it}^{*} &=& Y_{it} - 
	\left( \sum_{s=t+1}^T Y_{is} W_{is-1}' \right) \left( \sum_{s=t+1}^T W_{is-1} W_{is-1}^{\prime}\right)^{-1} W_{it-1} \\
	X_{it-1}^{*} &=& X_{it-1} - 
	\left( \sum_{s=t+1}^T X_{is-1}W_{is-1}'\right) \left( \sum_{s=t+1}^T W_{is-1} W_{is-1}^{\prime}\right)^{-1} W_{it-1} \\
	Z_{it-1}^{*} &=& Z_{it-1} - 
	\left( \sum_{s=t+1}^T Z_{is-1} W_{is-1}' \right) \left( \sum_{s=t+1}^T W_{is-1} W_{is-1}^{\prime}\right)^{-1}  W_{it-1}
\end{eqnarray*}
for $t=1,\ldots,T-k_w$. Because $\mathbb{E}[U_{it}|Y_i^{1:t-1},H_i,\lambda_i] = 0$, we obtain the moment condition
\be
\mathbb{E}\left[\left( Y_{it}^{*} - \big[ \begin{array}{cc} \tilde{\rho}' & \tilde{\alpha}' \end{array} \big]
\left[ \begin{array}{c} X_{it-1}^{*} \\ Z_{it-1}^{*} \end{array} \right]  \right)
\big[ \begin{array}{cc} X_{it-s-1}' & Z_{it-s-1}' \end{array} \big] \right] = 0
\ee  
for $s \geq 0$. To simplify the exposition, suppose that we choose $[X_{it-1},Z_{it-1}]$ as instrumental variables. In this case, for the moment conditions
to be only satisfied only at $\tilde{\rho} = \rho$ and $\tilde{\alpha} = \alpha$ it is necessary 
that the matrix 
\be
\mathbb{E} \left[ \begin{array}{cc} X_{it-1}^*X_{it-1}' & X_{it-1}^*Z_{it-1}' \\
Z_{it-1}^*X_{it-1}' & Z_{it-1}^*Z_{it-1}' \end{array} \right] \label{eq.rank.identify.alpharho}
\ee
is full rank. Consider, for instance, the upper-left element. We can write
\begin{eqnarray*}
	\lefteqn{\mathbb{E}[X_{it-1}^* X_{it-1}']} \\ 
	&=& \mathbb{E}\left[ \left( X_{it-1} - 
	\left( \sum_{s=t+1}^T X_{is-1}W_{is-1}'\right) \left( \sum_{s=t+1}^T W_{is-1} W_{is-1}^{\prime}\right)^{-1} W_{it-1} \right) X_{it-1}'  \right] \\
	&=& \mathbb{E} \left[ \mathbb{E}\left[ \left( X_{it-1} - 
	\left( \sum_{s=t+1}^T X_{is-1}W_{is-1}'\right) \left( \sum_{s=t+1}^T W_{is-1} W_{is-1}^{\prime}\right)^{-1} W_{it-1} \right) X_{it-1}' \; \bigg| W_{i}^{t:T-1} \right] \right] \\
	&=& \mathbb{E}[X_{it-1} X_{it-1}']  
	- \frac{1}{T-h} \bigg( \sum_{s=t+1}^T \mathbb{E} \bigg[ \mathbb{E}[X_{is-1} X_{it-1} | W_i^{t:T-1}]  \\
	&& \times   W_{is-1}' \bigg( \frac{1}{T-h} \sum_{s=t+1}^T W_{is-1} W_{is-1}^{\prime}\bigg)^{-1} W_{it-1}  \bigg] \bigg) \\
	&=& \mathbb{E}[X_{it-1} X_{it-1}']  
	- \frac{1}{T-h}  \sum_{s=t+1}^T \kappa_s \mathbb{E}[X_{is-1} X_{it-1}'] = I + II, \; \mbox{say}.   
\end{eqnarray*}
The fourth equality is based on the assumption that the $W_{it}$'s are strictly exogenous. The completion of the identification argument requires a moment bound for
\[
   \kappa_s = \mathbb{E} \left[ W_{is-1}' \bigg( \frac{1}{T-h} \sum_{s=t+1}^T W_{is-1} W_{is-1}^{\prime}\bigg)^{-1} W_{it-1} \right],
\]
a full rank condition on $\mathbb{E}[X_{it-1} X_{it-1}']$, and a condition that ensures that term $II$ does not induce a rank deficiency in term $I$. Similar conditions need to be imposed on the terms that appear in 
the other submatrices of (\ref{eq.rank.identify.alpharho}).

\subsection{Proofs for Section~\ref{sec:ratio.optimality}}
\label{appsubsec:proofs.ratio.optimality}

\subsubsection{Sufficient Conditions for Assumption~\ref{as:tail.intial.condition}(iii)}

The high-level condition in Assumption~\ref{as:tail.intial.condition}(iii) is satisfied 
if the following two conditions hold:
          
\noindent (a) There exists a sequence $D_N \rightarrow \infty$ such that $B_N D_N = o(1)$ and
         \[ 
         \exp\left( -\frac{D_N^2}{2}\right) = o(1) \left( \inf_{y \in \mathcal{Y}_{\lambda}^{\pi} \cap [-C_N', C_N],  \lambda \in \Lambda^{\pi}}\pi(y|\lambda)\right).
         \]
          (b) There exists a shrinking neighborhood of $y$ and a function $\delta(y,\lambda)$ such that for any $|a| \leq \kappa_N \rightarrow 0$,    
         $$  | \pi(y| \lambda) - \pi(y+a|\lambda) | \leq \delta(y,\lambda) | a |, $$ where
         $$ \sup_{y \in \mathcal{Y}_{\lambda}^{\pi} \cap [-C_N', C_N],  \lambda \in \Lambda^{\pi} } \left| 
         B_N \frac{\delta(y,\lambda)}{\pi(y|\lambda)} \right| = o(1).$$ 	 

\noindent The claim can be verified as follows. 	
For $|y| \leq \mathcal{Y}_{\lambda}^{\pi} \cap [-C_N',C_N]$ and $\lambda \in \Lambda^{\pi}$, by the change-of-variable with $y^{*}=\frac{\tilde{y}-y}{B_N}$, we have 
\begin{eqnarray*}
	\int \frac{1}{B_N} \phi\left( \frac{\tilde{y} - y }{B_N} \right) \left( \frac{\pi(\tilde{y}|\lambda) }{\pi(y|\lambda)} - 1\right) d \tilde{y} 
	&=& \int \phi(y^{*}) \left( \frac{\pi(y+B_Ny^{*}|\lambda) - \pi(y|\lambda) }{\pi(y|\lambda)}\right) d y^{*}.
\end{eqnarray*}
Split the integration into two, one over $|y^{*}| \leq D_N$ and other one over $|y^{*}| > D_N$. By Assumption \ref{as:tail.intial.condition}(i) and (iii)-(a),  uniformly in $|y^{*}| \leq D_N$ and other one over $|y^{*}| > D_N$,
\begin{eqnarray*}
\left| \int_{|y^{*}| > D_N} \phi(y^{*}) \left( \frac{\pi(y+B_Ny^{*}|\lambda) - \pi(y|\lambda) }{\pi(y|\lambda)}\right) d y^{*} \right| 
&\leq& \frac{ M \int_{|y^{*}| > D_N} \phi(y^{*}) d y^{*} }{\inf_{y \in \mathcal{Y}_{\lambda}^{\pi} \cap [-C_N', C_N],  \lambda \in \Lambda^{\pi}}\pi(y|\lambda)} \\
&\leq& \frac{ M \exp\left( -\frac{D_N^2}{2} \right) }{\inf_{y \in \mathcal{Y}_{\lambda}^{\pi} \cap [-C_N', C_N],  \lambda \in \Lambda^{\pi}}\pi(y|\lambda)} \\
&=& o(1)
\end{eqnarray*}
Also, notice that since $|y^{*}| \leq D_N$, $|B_N y^{*}| \leq B_N D_N = o(1).$  Then,  by Assumption (iii)-(b), 
\begin{eqnarray*}
\left| \int_{|y^{*}| \leq D_N} \phi(y^{*}) \left( \frac{\pi(y+B_Ny^{*}|\lambda) - \pi(y|\lambda) }{\pi(y|\lambda)}\right) d y^{*} \right| 
&\leq&  \int \phi(y^{*}) y^{*} d y^{*}
 \left| \frac{\delta(y, \lambda)}{\pi(y|\lambda)} B_N\right| \\
&=& M o(1) = o(1)   
\end{eqnarray*}  
uniformly in $y \in  \mathcal{Y}_{\lambda}^{\pi} \cap [-C_N',C_N]$ and $\lambda \in \Lambda^{\pi}.$ 

\subsubsection{An Example of a $\pi(y|\lambda)$ That Satisfies Assumption \ref{as:tail.intial.condition}}

Consider
$\pi(y|\lambda) = \phi( y - \lambda)$, where $\phi(x) = \exp(-\frac{1}{2}x^2)/\sqrt{2\pi}$. 
First, since $0<\phi(x)<1$, Assumption \ref{as:tail.intial.condition}(i) is satisfied. To verify Assumption \ref{as:tail.intial.condition}(ii), notice that because $Y_{i0}|\lambda_i \sim N(\lambda_i, 1)$, we have for $C \geq 0$,
\[
\mathbb{P}\{ Y_{i0} \geq C | \lambda_i = \lambda \} \leq \exp\left(-\frac{(C- \lambda)^2}{2}\right).
\] 
In this case, $m(C,\lambda) = (C-\lambda)^2/2$. Choose $K \geq \max\{ 1, \, \sqrt{2(2+\epsilon)} \} $ with any  $\epsilon \geq 0$. Then,
$$ \liminf_{N \longrightarrow \infty} \, \inf_{| \lambda| \leq C_N} \, 
( m(K (\sqrt{ \ln N} + C_N),\lambda) - (2+\epsilon) \ln N )  \geq 0,$$ as required for Assumption \ref{as:tail.intial.condition}(ii), regardless of the specific rate of $C_N$.
To verify Assumption \ref{as:tail.intial.condition}(iii) we can use the closed-form expression for the convolution:
\[
\int \frac{1}{B_N} \phi\left( \frac{\tilde{y} - y }{B_N} \right) \pi(\tilde{y}|\lambda)d \tilde{y} = \frac{1}{\sqrt{1+B_N^2}} \phi \left( \frac{ y - \lambda}{\sqrt{1+B_N^2}} \right).
\]
Note that we can write
\[
\phi\left( \frac{ y - \lambda}{\sqrt{1+B_N^2}} \right)
= \phi\big( y - \lambda \big) \exp\left( \frac{(B_N(y-\lambda) )^2}{2(1+B_N^2)} \right).
\]
Thus,
\[
\sup_{y \in  \mathcal{Y}_{\lambda}^{\pi} \cap [-C_N',C_N], \, \lambda \in \Lambda^{\pi}}	\; 
\exp\left( \frac{(B_N(y-\lambda) )^2}{2(1+B_N^2)} \right) - 1 
\le \exp\left( (B_N(C_N'+C_N) )^2 \right) - 1 = o(1),
\]
according to Assumption~\ref{as:ratio.optimality.bandwidth}.

\subsubsection{Main Theorem}
\label{appsubsubsec:proofs.main.theorem}

\noindent {\bf Proof of Theorem \ref{thm:ratio.optimality}.} The goal is to prove that for a given $\epsilon_0>0$
\be
\limsup_{N \rightarrow \infty}
\;  \frac{ R_{N}(\widehat{Y}^N_{T+1}) - R_N^{\text{opt}} }{N \mathbb{E}_\theta^{{\cal Y}^i,\lambda_i}\big[ (\lambda_i - \mathbb{E}_{\theta,{\cal Y}^i}^{\lambda_i}[\lambda_i] )^2\big]  + N^{\epsilon_0}} \leq  0,      
\ee
where
\begin{eqnarray*}
        R_{N}(\widehat{Y}^N_{T+1})
        &=&  N \mathbb{E}_\theta^{{\cal Y}^N,\lambda_i} \left[
        \left( \lambda_i + \rho Y_{iT} - \widehat{Y}_{iT+1} \right)^2 \right] + N \sigma^2 \\ 
        R_N^{\text{opt}} &=&  N \mathbb{E}_\theta^{{\cal Y}_i,\lambda_i} \left[ \left(\lambda_i -\mathbb{E}_{\theta,{\cal Y}^i}^{\lambda_i}[\lambda_i] \right)^2 \right] + N \sigma^2.
\end{eqnarray*}
Here we used the fact that there is cross-sectional independence and symmetry in terms of $i$. 
The statement is equivalent to 
\be
\limsup_{N \rightarrow \infty}
\;  \frac{ N \mathbb{E}_\theta^{{\cal Y}^N,\lambda_i} \left[
        \left( \lambda_i + \rho Y_{iT} - \widehat{Y}_{iT+1} \right)^2 \right] }{N \mathbb{E}_\theta^{{\cal Y}^i,\lambda_i}\big[ (\lambda_i - \mathbb{E}_{\theta,{\cal Y}^i}^{\lambda_i}[\lambda_i] )^2\big]  + N^{\epsilon_0}} \leq  1.      
\ee

\noindent {\bf Forecast Error Decomposition.} We decompose the forecast error as follows:
Using the previously developed notation, we expand the prediction error due to parameter estimation as follows:
\begin{eqnarray*}
	\lefteqn{ \widehat{Y}_{iT+1} - \lambda_i - \rho Y_{iT}} \\
	&=&  \left[ \mu\big( \hat{\lambda}_i(\hat\rho),\hat{\sigma}^2/T+B_N^2,\hat{p}^{(-i)}(\hat{\lambda}_i(\hat\rho),Y_{i0}) \big) \right]^{C_N} 
	- \mu \big( \hat{\lambda}_i(\rho),\sigma^2/T+B_N^2, p_*(\hat{\lambda}_i(\rho),Y_{i0}) \big) \\
	&& + \mu \big( \hat{\lambda}_i(\rho),\sigma^2/T+B_N^2, p_*(\hat{\lambda}_i(\rho),Y_{i0}) \big) -
	\lambda_i \\
	&&+ (\hat{\rho} - \rho) Y_{iT} \\
	&=& A_{1i} + A_{2i} + A_{3i}, \; \mbox{say}.
\end{eqnarray*}

We define the density $p_*(\hat{\lambda}_i(\rho),Y_{i0})$ as the expected value of the kernel density estimator:
\be
  p_*(\hat{\lambda}_i,y_{i0}) = \mathbb{E}_{\theta,{\cal Y}_i}^{{\cal Y}^{(-i)}}[\hat{p}^{(-i)}(\hat{\lambda}_i,y_{i0}) ].
\ee
It can be calculated as follows. Taking expectations with respect to $(\hat{\lambda}_j,y_{j,0})$ for $j \not= i$ yields
\begin{eqnarray*}
	\lefteqn{ \mathbb{E}_{\theta,{\cal Y}_i}^{{\cal Y}^{(-i)}}[\hat{p}^{(-i)}(\hat{\lambda}_i,y_{i0}) ] } \\
	&=& \sum_{j \not=i} \int \int
	\frac{1}{B_N} \phi\left( \frac{\hat{\lambda}_i - \hat{\lambda}_j }{B_N}\right)
	\frac{1}{B_N} \phi\left( \frac{y_{i0} - y_{j0}}{B_N}\right) p(\hat{\lambda}_j,y_{j0}) d\hat{\lambda}_j d y_{j0} \\
	&=&     \int \int \frac{1}{B_N} \phi\left( \frac{\hat{\lambda}_i - \hat{\lambda}_j }{B_N}\right)
	\frac{1}{B_N} \phi\left( \frac{y_{i0} - y_{j0}}{B_N}\right) p(\hat{\lambda}_j,y_{j0}) d\hat{\lambda}_j d y_{j0}.   
\end{eqnarray*}
The second equality follows from the symmetry with respect to $j$ and the fact that we integrate out $(\hat{\lambda}_j,y_{j0})$.
We now substitute in 
\[
p(\hat{\lambda}_j,y_{j0}) = \int p(\hat{\lambda}_j|\lambda_j)\pi(\lambda_j,y_{j0}) d\lambda_j,
\]
and change the order of integration. This leads to:
\begin{eqnarray*}
	\lefteqn{ \mathbb{E}_{\theta,{\cal Y}_i}^{{\cal Y}^{(-i)}}[\hat{p}^{(-i)}(\hat{\lambda}_i,y_{i0}) ] } \\
	&=&  \int \int \left[ \int
	\frac{1}{B_N} \phi\left( \frac{\hat{\lambda}_i - \hat{\lambda}_j }{B_N}\right) p(\hat{\lambda}_j|\lambda_j) d\hat{\lambda}_j \right]
	\frac{1}{B_N} \phi\left( \frac{y_{i0} - y_{j0}}{B_N}\right) \pi(\lambda_j,y_{j0}) d\lambda_j d y_{j0}  \\
	&=&  \int \int 
	\frac{1}{\sqrt{\sigma^2/T+B_N^2}} \phi\left( \frac{\hat{\lambda}_i - \lambda_j }{\sqrt{\sigma^2/T+B_N^2}}\right) 
	\frac{1}{B_N} \phi\left( \frac{y_{i0} - y_{j0}}{B_N}\right) \pi(\lambda_j,y_{j0}) d\lambda_j d y_{j0} \\
	&=&  \int 
	\frac{1}{\sqrt{\sigma^2/T+B_N^2}} \phi\left( \frac{\hat{\lambda}_i - \lambda_j }{\sqrt{\sigma^2/T+B_N^2}}\right) 
	\left[ \int \frac{1}{B_N} \phi\left( \frac{y_{i0} - y_{j0}}{B_N}\right)  \pi(y_{j0}|\lambda_j)  d y_{j0} \right]  \pi(\lambda_j) d\lambda_j.
\end{eqnarray*} 
Now re-label $\lambda_j$ and $\lambda_i$ and $y_{j0}$ as $\tilde{y}_{i0}$ to obtain:
\begin{eqnarray*}
	\lefteqn{ p_*(\hat{\lambda}_i,y_{i0}) } \\
	&=& \int 
	\frac{1}{\sqrt{\sigma^2/T+B_N^2}} \phi\left( \frac{\hat{\lambda}_i - \lambda_i }{\sqrt{\sigma^2/T+B_N^2}}\right) 
	\left[ \int \frac{1}{B_N} \phi\left( \frac{y_{i0} - \tilde{y}_{i0}}{B_N}\right)  \pi(\tilde{y}_{i0}|\lambda_i)  d \tilde{y}_{i0} \right]  \pi (\lambda_i) d\lambda_i.
\end{eqnarray*}

\noindent {\bf Risk Decomposition.} Write 
\[
N \mathbb{E}_\theta^{{\cal Y}^N} \left[
\left( \lambda_i + \rho Y_{iT} - \widehat{Y}_{iT+1} \right)^2 \right]
 = N \mathbb{E}_\theta^{{\cal Y}^N} \big[(A_{1i} + A_{2i} + A_{3i})^2\big].
\]
We deduce from the $C_r$ inequality that the statement of the theorem follows if we can show that for the $\epsilon_0 > 0$ given in Definition~\ref{def:ratio.optimality}: 
\begin{eqnarray*}
\mbox{(i)}  &&  N \mathbb{E}_\theta^{{\cal Y}^N} \big[A_{1i}^2 \big] = o(N^{\epsilon_0})\\
\mbox{(ii)} &&  \limsup_{N \rightarrow \infty}
\;  \frac{ N \mathbb{E}_\theta^{{\cal Y}^N,\lambda_i} \big[A_{2i}^2 \big] }{N \mathbb{E}_\theta^{{\cal Y}^i,\lambda_i}\big[ (\lambda_i - \mathbb{E}_{\theta,{\cal Y}^i}^{\lambda_i}[\lambda_i] )^2\big]  + N^{\epsilon_0}} \leq  1 \\
\mbox{(iii)} && N \mathbb{E}_\theta^{{\cal Y}^N} \big[A_{3i}^2 \big] = o(N^{\epsilon_0}).
\end{eqnarray*}
The required bounds are provided in Lemmas \ref{lemma:ap.ratio.optimality.phatpstar} (term $A_{1i}$), \ref{lemma:ap.ratio.optimality.pstarlambda} (term $A_{2i}$),  \ref{lemma:ap.ratio.optimality.rhohat} (term $A_{3i}$). $\blacksquare$

\subsubsection{Three Important Lemmas}
\label{appsubsubsec:three.lemmas}

\noindent {\bf Truncations.} The remainder of the proof involves a number of truncations that we will apply when analyzing the risk terms. For now, $L_N = o(N^\epsilon)$ will be a sequence such that $L_N \longrightarrow \infty$ as $N \longrightarrow \infty$. We will specify the rate at which $L_N$ diverges below. 

\begin{enumerate}
	\item Define the truncated region
	$\mathcal{T}_1 = \{ | \hat{\sigma}^2 - \sigma^2 | \leq 1/L_N \}$. 
	By Chebyshev's inequality and Assumption \ref{as:rhohat.ratio.optimality}, we can bound 
	\[ 
	N \mathbb{P} (\mathcal{T}_1^c) = N \mathbb{P} \{ | \hat{\sigma}^2 - \sigma^2 | > 1/L_N  \} \leq L_N^2 \mathbb{E}[N(\hat{\sigma}^2 - \sigma^2)^2] =  o(N^\epsilon),
	\]
	provided that $L_N^2 = o(N^\epsilon)$ for any $\epsilon$.
	\item Define the truncated region $\mathcal{T}_2 = \{ |\hat{\rho} -\rho | \leq 1/L_N^2 \}$.
	By Chebyshev's inequality and Assumption \ref{as:rhohat.ratio.optimality}, we can bound 
	\[
	N \mathbb{P} (\mathcal{T}_2^c) = N \mathbb{P} \{ | \hat{\rho} - \rho | > 1/L_N^2  \} \leq L_N^4 \mathbb{E}\left[ N(\hat{\rho} - \rho)^2 \right] =  o(N^\epsilon),
	\]
	provided that $L_N^4=o(N^\epsilon)$ for any $\epsilon$.
	\item Let $ \bar{U}_{i,-1}(\rho) = \frac{1}{T} \sum_{t=2}^{T} U_{it-1}(\rho)$ and $U_{it}(\rho) = U_{it} + \rho U_{it-1} + \cdots + \rho^{t-1} U_{i1}$. Define 
	the truncated region $\mathcal{T}_3 = \left\{  \max_{1 \leq i \leq N} | \bar{U}_{i,-1}(\rho) | \leq M_3 L_N  \right\}$ for some constant $M_3$. Notice that 
	$\bar{U}_{i,-1}(\rho) \sim iid N(0,\sigma^2_{\bar{U}})$ with $0 < \sigma^2_{\bar{U}} < \infty$. Thus, we have 
	\begin{eqnarray}
	N\mathbb{P}(\mathcal{T}_3^c) &=& N \mathbb{P} \{ \max_{1 \leq i \leq N} | \bar{U}_{i,-1}(\rho) | \geq  L_N \} \nonumber \\
	&\leq & N \sum_{i=1}^{N} \mathbb{P} \{ | \bar{U}_{i,-1}(\rho) | \geq  L_N \} \nonumber \\
	&=& N^2 \mathbb{P} \{ | \bar{U}_{i,-1}(\rho) | \geq L_N \} \nonumber \\
	&\leq & 2 \exp \left( - \frac{ L_N^2}{2 \sigma^2_{\bar{U}}}+ 2\ln N \right). \label {eq.trunction.t3} 
	\end{eqnarray}   
	\item Define the truncated region $\mathcal{T}_4=\left\{ \max_{1\leq i \leq N} | Y_{i0}| \leq L_N \right\}$. 
	Then,
	\begin{eqnarray}
	N \mathbb{P}\mathcal{T}_4^c &=& N \mathbb{P} \{ \max_{1\leq i \leq N} | Y_{i0}| \geq L_N \}  \nonumber \\
	&\leq& N \sum_{i=1}^N  \mathbb{P} \{ | Y_{i0}| \geq L_N \} \nonumber \\
	&=& N^2 \int \left[ 
	\int_{ L_N}^{\infty}  \pi(y_0| \lambda ) dy_0
	+ \int_{-\infty}^{- L_N}  \pi(y_0| \lambda )  dy_0
	\right] \pi_{\lambda}(\lambda) d\lambda \nonumber \\
	&\leq& 2N^2 \int \exp\left[ -m\left( L_N,\lambda \right) \right] \pi(\lambda) d\lambda  \nonumber \\
	&\leq& 2 C_N \left( \sup_{|\lambda| \leq C_N} \exp\left[ -m\left( L_N,\lambda \right) + 2\ln N \right] \right), \label {eq.trunction.t4}  
	\end{eqnarray}
	where the last three lines hold by Assumptions \ref{as:ratio.optimality.marginal.dist.lambda} and \ref{as:tail.intial.condition}.
	
	\item Let $\bar{Y}_{i,-1} = C_1(\rho) Y_{i0} + C_2(\rho) \lambda_i + \bar{U}_{i,-1}(\rho)$, 
	where $C_1(\rho) = \frac{1}{T} \sum_{t=1}^T \rho^{t-1} $,  
	$C_2(\rho) = \frac{1}{T} \sum_{t=2}^{T} (1+\cdots+\rho^{t-2})$.
	According to Assumption~\ref{as:ratio.optimality.marginal.dist.lambda} the support of $\lambda_i$ is contained in $[-C_N, C_N]$. Moreover, because $T$ is finite, 
	$|C_1(\rho)| \leq 1 $ and $|C_2(\rho)| < T$.
	Then, in the region $\mathcal{T}_3 \cap \mathcal{T}_4$:
	\begin{eqnarray*}
	\max_{1 \leq i \leq N} | \bar{Y}_{i,-1} | 
	&\leq&  | C_1(\rho) | \max_{1 \leq i \leq N} | \lambda_i| 
	+  | C_2(\rho)| \max_{1 \leq i \leq N} | Y_{i0}| 
	+   \max_{1 \leq i \leq N}  |\bar{U}_{i,-1}(\rho) | \\
	&\leq& C_N + T L_N + \exp \left( - \frac{ L_N^2}{2 \sigma^2_{\bar{U}}}+ 2\ln N \right)  
	\end{eqnarray*}
	which leads to 
	\be
	\max_{1 \leq i,j \leq N} | \bar{Y}_{j,-1} - \bar{Y}_{i,-1}| 
	\leq 2 \max_{1 \leq i \leq N} | \bar{Y}_{i,-1} | 
	\leq 2 \left(  C_N + T L_N + \exp \left( - \frac{ L_N^2}{2 \sigma^2_{\bar{U}}}+ 2\ln N \right) \right). \label{eq.truncation.Ybar_j - Ybar_i}
	\ee
	\item For the region $\mathcal{T}_2 \cap \mathcal{T}_3 \cap \mathcal{T}_4$ we obtain the bound 
	\begin{eqnarray}
	\max_{1 \leq i,j \leq N} | (\hat{\rho} - \rho) (\bar{Y}_{j,-1} - \bar{Y}_{i,-1} ) | \leq \frac{2\left(  C_N + T L_N + \exp \left( - \frac{ L_N^2}{2 \sigma^2_{\bar{U}}}+ 2\ln N \right) \right)}{ L_N^2}  \label{eq.truncation.rhohat-rho.Ybarj-Ybari}.
	\end{eqnarray}
\end{enumerate} 

\noindent Recall that $C_N = o(N^\epsilon)$ is the truncation for the support of the prior of $\lambda$ (Assumption~\ref{as:ratio.optimality.marginal.dist.lambda}). We will choose
\be
L_N = o(N^\epsilon) \; \mbox{such that} \; L_N = \max\;  \left\{ \sigma_{\bar{U}} \sqrt{2(2+\epsilon) \ln N}, K (\sqrt{\ln N}+C_N), \frac{1}{B_N} , C_N \right\} ,
\label{eq.choosing.LN}
\ee
so that we can deduce 
\begin{eqnarray}
&& N\mathbb{P} \mathcal{T}_1^c = o(N^\epsilon), \quad N\mathbb{P} \mathcal{T}_2^c = o(N^\epsilon), \quad N \mathbb{P} \mathcal{T}_3^c = o(N^\epsilon), \quad N \mathbb{P} \mathcal{T}_4^c = o(N^\epsilon) \nonumber \\
&& (\ref{eq.truncation.Ybar_j - Ybar_i}) =o(N^\epsilon), \quad   (\ref{eq.truncation.rhohat-rho.Ybarj-Ybari}) = o(N^\epsilon).
\label{eq.probT1-4c}
\end{eqnarray}
 for any $\epsilon$.
\paragraph{Term $A_{1i}$}

\begin{lemma} \label{lemma:ap.ratio.optimality.phatpstar} Suppose the assumptions in Theorem \ref{thm:ratio.optimality} hold. Then, 
\begin{eqnarray*}
   \lefteqn{N \mathbb{E}_\theta^{{\cal Y}^N} \bigg[ \bigg( \left[ \mu\big( \hat{\lambda}_i(\hat\rho),\hat{\sigma}^2/T+B_N^2,\hat{p}^{(-i)}(\hat{\lambda}_i(\hat\rho),Y_{i0}) \big) \right]^{C_N} }\\
   &&
  \hspace*{2cm} - \mu \big( \hat{\lambda}_i(\rho),\sigma^2/T+B_N^2, p_*(\hat{\lambda}_i(\rho),Y_{i0}) \big)    \bigg)^2  \bigg] = o(N^{\epsilon_0}).
\end{eqnarray*}
\end{lemma}

\noindent {\bf Proof of Lemma~\ref{lemma:ap.ratio.optimality.phatpstar}}. We begin
with the following bound:
\begin{eqnarray}
  |A_{1i}| &=&\ \left| \left[ \mu\big( \hat{\lambda}_i(\hat\rho),\hat{\sigma}^2/T+B_N^2,\hat{p}^{(-i)}(\hat{\lambda}_i(\hat\rho),Y_{i0}) \big) \right]^{C_N} 
 	- \mu \big( \hat{\lambda}_i(\rho),\sigma^2/T+B_N^2, p_*(\hat{\lambda}_i(\rho),Y_{i0}) \big)  \right| \nonumber \\
 &\le&  \left| \left[ \mu\big( \hat{\lambda}_i(\hat\rho),\hat{\sigma}^2/T+B_N^2,\hat{p}^{(-i)}(\hat{\lambda}_i(\hat\rho),Y_{i0}) \big) \right]^{C_N} \right| 
 + \bigg| \mu \big( \hat{\lambda}_i(\rho),\sigma^2/T+B_N^2, p_*(\hat{\lambda}_i(\rho),Y_{i0}) \big) \bigg| \nonumber \\
 &\le& 2 C_N. \label{eq.bound.A1i}
\end{eqnarray}
The last equality follows from the fact that the second term can be interpreted as a posterior mean under the likelihood function
\begin{eqnarray*}
\lefteqn{p_*(\hat{\lambda}_i,y_{i0}|\lambda_i)} \\
&=&     \frac{1}{\sqrt{\sigma^2/T+B_N^2}} \phi\left( \frac{\hat{\lambda}_i - \lambda_i }{\sqrt{\sigma^2/T+B_N^2}}\right) 
\left[ \int \frac{1}{B_N} \phi\left( \frac{y_{i0} - \tilde{y}_{i0}}{B_N}\right)  p(\tilde{y}_{i0}|\lambda_i)  d \tilde{y}_{i0} \right]. \nonumber 
\end{eqnarray*}
and the prior distribution $\pi(\lambda)$. Because, according to Assumption~\ref{as:ratio.optimality.marginal.dist.lambda}, the prior has support on the interval $[-C_N,\;C_N]$, we can deduce that the posterior mean has to be bounded by $C_N$ as well. Then,
\begin{eqnarray}
N \mathbb{E}_\theta^{{\cal Y}^N} [A_{1i}^2] 
  &\leq& N \mathbb{E}_\theta^{{\cal Y}^N} [A_{1i}^2 \mathbb{I}(\mathcal{T}_1) \mathbb{I}(\mathcal{T}_2) \mathbb{I}(\mathcal{T}_3) \mathbb{I}(\mathcal{T}_4)] 
+  C_N^2 N \left( \mathbb{P} \mathcal{T}_1^{c} +  \mathbb{P} \mathcal{T}_2^{c} +  \mathbb{P} \mathcal{T}_3^{c} +  \mathbb{P} \mathcal{T}_4^{c} \right) \nonumber \\
&\leq& N \mathbb{E}_\theta^{{\cal Y}^N} [A_{1i}^2 \mathbb{I}(\mathcal{T}_1) \mathbb{I}(\mathcal{T}_2) \mathbb{I}(\mathcal{T}_3) \mathbb{I}(\mathcal{T}_4)]   +  o(N^{\epsilon_0}). \label{eq.step1.bound}
\end{eqnarray}
The bound for the second term follows from the fact that (\ref{eq.probT1-4c}) and (\ref{eq.bound.A1i}) hold for any $\epsilon > 0$, including $\epsilon_0$.  In the remainder of the proof we will construct a bound for the first term on the right-hand side of (\ref{eq.step1.bound}). We proceed in two steps.

\noindent {\bf Step 1.}
We introduce two additional trunctation regions, $\mathcal{T}_{5i}$ and $\mathcal{T}_{6i}$, which are defined as follows:
\begin{eqnarray*}
	{\cal T}_{5i} &=& \big\{ (\hat{\lambda}_i,Y_{i0}) \, \big| \, -C_N^{\prime} \le \hat{\lambda}_i \le C_N^{\prime}, \, -C_N^{\prime} \le Y_{i0} \le C_N^{\prime}\big\}\\
	{\cal T}_{6i} &=& \left\{ (\hat{\lambda}_i,Y_{i0}) \, \bigg| \, p(\hat{\lambda}_i,Y_{i0}) \geq \frac{N^{\epsilon^{\prime}}}{N} \right\},
\end{eqnarray*}
where 
$C_N^{\prime} > C_N $ will be defined in (\ref{eq.ap.densityestimator.truncation.cprime}) below and it is assumed that $0<\epsilon'<\epsilon_0$. In the first truncation region both $\hat{\lambda}_i$ and $Y_{i0}$ are bounded by $C_N$. In the second truncation region the density $p(\hat{\lambda}_i,Y_{i0})$ is not ``high.'' We will show that
\begin{eqnarray} 
	N \mathbb{E}_\theta^{{\cal Y}^N} [A_{1i}^2 \mathbb{I}(\mathcal{T}_{5i})\mathbb{I}( \mathcal{T}_{6i}^c)] &\leq& o(N^{{\epsilon_0}}) \label{eq.Step 2.desired.1} \\
	N \mathbb{E}_\theta^{{\cal Y}^N} [ A_{1i}^2 \mathbb{I}(\mathcal{T}_{5i}^c)] &\leq& o(N^{{\epsilon_0}}). \label{eq.Step 2.desired.2}
\end{eqnarray}

\noindent {\bf Step 1.1.} First, we consider the case where $(\hat{\lambda}_i,y_{i0})$ are bounded and the density $p(\hat{\lambda}_i,y_{i0})$ is ``low'' in (\ref{eq.Step 2.desired.1}). Using the bound for $|A_{1i}|$ in (\ref{eq.bound.A1i}) we obtain:
\begin{eqnarray*}
	N \mathbb{E}_\theta^{{\cal Y}^N} \left[A_{1i}^2 \mathbb{I}(\mathcal{T}_{5i})\mathbb{I}( \mathcal{T}_{6i}^c)] \right] 
	&\leq& 4N C_N^2 \mathbb{P}(\mathcal{T}_{5i} \cap \mathcal{T}_{6i}^c) \\
	&=& 4NC_N^2 \int_{\hat{\lambda}_i = - C_N'}^{C_N'} \int_{y_{i0} = - C_N'}^{C_N'} 
	\mathbb{I}\left\{ p(\hat{\lambda_i},y_{i0}) < \frac{N^{\epsilon'}}{N} \right\} p(\hat{\lambda}_i,y_{i0}) d(\hat{\lambda}_i,y_{i0}) \\
	&\leq&  4 N C_N^2 \int_{\hat{\lambda}_i = - C_N'}^{C_N'} \int_{y_{i0} = - C_N'}^{C_N'} \left(\frac{N^{\epsilon'}}{N} \right) d y_{i0} d\hat{\lambda}_i \\
	&\le& 4 C_N^2 ( C'_N)^2 N^{\epsilon'} \\
	&=& o(N^{{\epsilon_0}}).
\end{eqnarray*}
The last equality holds by the definition of $C_N'$ found in (\ref{eq.ap.densityestimator.truncation.cprime}) below.
This establishes (\ref{eq.Step 2.desired.1}).

\noindent {\bf Step 1.2.} Next, we consider the case where $(\hat{\lambda}_i,y_{i0})$ exceed the $C_N'$ bound and the density $p(\hat{\lambda}_i,y_{i0})$ is ``high:'' 
\begin{eqnarray*}
	\lefteqn{N \mathbb{E}_\theta^{{\cal Y}^N}\left[A_{1i}^2 \mathbb{I}(\mathcal{T}_{5i}^c ) \right]}\\
		& \le &  4 N C_N^2 \int_{{\cal T}_5^c} p(\hat{\lambda}_i,y_{i0}) d(\hat{\lambda}_i,y_{i0}) \\
		& = & 4 N C_N^2 \int_{{\cal T}_5^c} \left[ \int_{\lambda_i} \frac{1}{\sigma/\sqrt{T}} \phi \left( \frac{\hat{\lambda}_i-\lambda_i}{\sigma/\sqrt{T}} \right) \pi(y_{i0}|\lambda_i) \pi(\lambda_i) d \lambda_i \right] d(\hat{\lambda}_i,y_{i0}) \\
		& \leq & 4 N C_N^2 \int_{\lambda_i} \bigg[  \int_{|\hat{\lambda}_i|> C_N'} \frac{1}{\sigma/\sqrt{T}} \phi \left( \frac{\hat{\lambda}_i-\lambda_i}{\sigma/\sqrt{T}} \right) \pi(y_{i0}|\lambda_i) d(\hat{\lambda}_i,y_{i0}) \\
		&&+  \int_{|y_{i0}| > C_N'} \frac{1}{\sigma/\sqrt{T}} \phi \left( \frac{\hat{\lambda}_i-\lambda_i}{\sigma/\sqrt{T}} \right) \pi(y_{i0}|\lambda_i) d(\hat{\lambda}_i,y_{i0}) \bigg]  \pi(\lambda_i) d \lambda_i \\
		&=&  4 N C_N^2 \int_{|\lambda_i|< C_N}  \left[  \int_{|\hat{\lambda}_i|> C_N'} \frac{1}{\sigma/\sqrt{T}} \phi \left( \frac{\hat{\lambda}_i-\lambda_i}{\sigma/\sqrt{T}} \right) d\hat{\lambda}_i \right]  \pi(\lambda_i) d \lambda_i \\
		&& + 4 N C_N^2 \int_{|\lambda_i|< C_N}  \left[	
		\int_{|y_{i0}| > C_N'} \pi(y_{i0}|\lambda_i)  dy_{i0}  \right]  \pi(\lambda_i) d \lambda_i \\
		&=& B_1 + B_2, \quad \mbox{say.}
	\end{eqnarray*}
The second equality is obtained by integrating out $y_{i0}$ and $\hat{\lambda}_i$, recognizing that the integrant is a properly scaled probability density function that integrates to one. We are able to restrict the range of integration for $\lambda_i$ to the set $|\lambda_i|<C_N$ because, by assumption, that is the support of the prior density $\pi(\lambda)$

We will first analyze term $B_1$. Note that 
\begin{eqnarray*}
	\lefteqn{\int_{|\hat{\lambda}_i|> C_N'} \frac{1}{\sigma/\sqrt{T}} \phi \left( \frac{\hat{\lambda}_i-\lambda_i}{\sigma/\sqrt{T}} \right) d\hat{\lambda}_i} \\
	&=& \int_{-\infty}^{-\sqrt{T}(C_N'+ \lambda_i)/\sigma} \phi(\tilde{\lambda}_i) d\tilde{\lambda}_i 
	+ \int_{\sqrt{T}(C_N'-\lambda_i)/\sigma}^\infty \phi(\tilde{\lambda}_i) d\tilde{\lambda}_i \\
	&\le& \int_{-\infty}^{-\sqrt{T}(C_N'- |\lambda_i|)/\sigma} \phi(\tilde{\lambda}_i) d\tilde{\lambda}_i 
	+ \int_{\sqrt{T}(C_N'-|\lambda_i|)/\sigma}^\infty \phi(\tilde{\lambda}_i) d\tilde{\lambda}_i \\   
	&\le& 2 \int_{\sqrt{T}(C_N'-|\lambda_i|)/\sigma}^\infty \phi(\tilde{\lambda}_i) d\tilde{\lambda}_i \\
	&\le& 2 \frac{ \phi \big(\sqrt{T}(C_N'-|\lambda_i|)/\sigma \big)}{\sqrt{T}(C_N'-|\lambda_i|)/\sigma},           
\end{eqnarray*}
where we used the inequality $\int_x^\infty \phi(\lambda)d\lambda \le \phi(x)/x$. 
Assuming that $N$ is sufficiently large such that $$\sqrt{T}(C_N'-|\lambda_i|)/\sigma > 1$$ for $|\lambda_i|<C_N$, 
we obtain
\[
B_1 \le 8 NC_N^2 \int_{|\lambda_i|< C_N} \exp \left( -\frac{T}{2\sigma^2}(C_N'-|\lambda_i|)^2  \right) \pi(\lambda_i)d\lambda_i.  
\]
We can deduce that $B_1 = o(N^{\epsilon})$ for any $\epsilon>0$ (including $\epsilon_0$) if 
\[
\inf_{|\lambda_i|<C_N} \; \frac{T}{2\sigma^2}(C_N'-|\lambda_i|)^2  > \ln N,
\]
which follows if we choose 
\be
 C_N' = (1+k) \left( \sqrt{\ln N} + C_N\right), \quad k > \max\{0,\sqrt{2\sigma^2/T} -1 \}. 
\label{eq.ap.densityestimator.truncation.cprime}
\ee 
This is the rate that appears in Assumption~\ref{as:ratio.optimality.bandwidth}.

For $B_2$, notice that under Assumption \ref{as:tail.intial.condition}(ii) we obtain
\begin{eqnarray*}
B_2 &=& 4 N C_N^2 \int_{|\lambda_i|< C_N}  \left[	
\int_{|y_{i0}| > C_N'} \pi(y_{i0}|\lambda_i)  dy_{i0}  \right]  \pi(\lambda_i) d \lambda_i \\
&\le& 4 N C_N^2 \int_{|\lambda_i|< C_N}  	2 \exp \big(- m(C_N',\lambda_i) \big)   \pi(\lambda_i) d \lambda_i \\
&\le& 8 C_N^2 \left[ \sup_{|\lambda_i| \leq C_N} \exp \big(- m(C_N',\lambda_i) + \ln N \big) \right] \int_{|\lambda_i|< C_N} \pi(\lambda_i) d \lambda_i \\
& \le & o(N^\epsilon)
\end{eqnarray*}
for any $\epsilon$.
This leads to the desired bound in (\ref{eq.Step 2.desired.2}).
	
\noindent {\bf Step 2.} It remains to be shown that
\be
   N \mathbb{E}_\theta^{{\cal Y}^N} \big[A_{1i}^2 \mathbb{I}(\mathcal{T}_1) \mathbb{I}(\mathcal{T}_2) \mathbb{I}(\mathcal{T}_3) \mathbb{I}(\mathcal{T}_4) \mathbb{I}(\mathcal{T}_{5i}) \mathbb{I}(\mathcal{T}_{6i})\big] \le o(N^{\epsilon_0}).
   \label{eq.bound.A1i2.T1T6}
\ee
We introduce the following notation:
\begin{eqnarray}
	\widetilde{p}^{(-i)}_i &=& \hat{p}^{(-i)} (\hat{\lambda}_i(\hat{\rho}),Y_{i0}) \label{eq.density.abreviated.notation}\\
	d\widetilde{p}^{(-i)}_i &=& \frac{1}{\partial \hat{\lambda}_i(\hat{\rho})} \partial \hat{p}^{(-i)} (\hat{\lambda}_i(\hat{\rho}),Y_{i0}) \nonumber \\
	\hat{p}^{(-i)}_i &=& \hat{p}^{(-i)} (\hat{\lambda}_i(\rho),Y_{i0}) \nonumber \\
	d \hat{p}^{(-i)}_i &=& \frac{1}{\partial \hat{\lambda}_i(\rho)} \partial \hat{p}^{-i} (\hat{\lambda}_i(\rho),Y_{i0}) \nonumber \\
	p_{i} &=& p(\hat{\lambda}_i(\rho),Y_{i0}) \nonumber \\
	p_{*i} &=& p_{*}(\hat{\lambda}_i(\rho),Y_{i0}) \nonumber \\
	d p_{*i} &=& \frac{1}{\partial \hat{\lambda}_i(\rho)} \partial p_{*}(\hat{\lambda}_i(\rho),Y_{i0}). \nonumber
\end{eqnarray}

Using the fact that $| \mu \big( \hat{\lambda}_i(\rho),Y_{i0},\sigma^2/T+B_N^2, p_*(\hat{\lambda}_i(\rho),Y_{i0}) \big) | \leq C_N$ and the triangle inequality, we obtain 
	\begin{eqnarray*}
		| A_{1i} | 
		&=& \bigg| \left[ \mu \big( \hat{\lambda}_i(\hat{\rho}),Y_{i0}, \hat{\sigma}^2/T+B_N^2, \hat{p}^{(-i)}(\hat{\lambda}_i(\hat{\rho}),Y_{i0}) \big) \right]^{C_N}
		- \mu \big( \hat{\lambda}_i(\rho),Y_{i0},\sigma^2/T+B_N^2, p_*(\hat{\lambda}_i(\rho),Y_{i0}) \big) \bigg|  \\
		&\leq&  \bigg| \mu \big( \hat{\lambda}_i(\hat{\rho}),Y_{i0}, \hat{\sigma}^2/T+B_N^2, \hat{p}^{(-i)}(\hat{\lambda}_i(\hat{\rho}),Y_{i0}) \big)
		- \mu \big( \hat{\lambda}_i(\rho),Y_{i0},\sigma^2/T+B_N^2, p_*(\hat{\lambda}_i(\rho),Y_{i0}) \big) \bigg| \\
		&=& \bigg| 
		\hat{\lambda}_i (\hat{\rho}) - \lambda_i(\rho) 
		+ \left( \frac{\hat{\sigma}^2}{T} - \frac{\sigma^2}{T} \right) \frac{d p_{*i}}{p_{*i}}
		+ \left(\frac{\hat{\sigma}^2}{T} + B_N^2 \right)  
		\left( \frac{d\widetilde{p}^{(-i)}_i}{\widetilde{p}^{(-i)}_i} - \frac{d p_{*i}}{p_{*i}} \right)
		\bigg| \\
		&\leq& \big| \hat{\rho} - \rho \big| \big| \bar{Y}_{i,-1} \big| 
		+ \bigg| \frac{\hat{\sigma}^2}{T} - \frac{\sigma^2}{T} \bigg| \bigg| \frac{d p_{*i}}{p_{*i}} \bigg|
		+ \left(\frac{\hat{\sigma}^2}{T} + B_N^2 \right)  \bigg| \frac{d\widetilde{p}^{(-i)}_i}{\widetilde{p}^{(-i)}_i} - \frac{d p_{*i}}{p_{*i}} \bigg|,\\
		&=& A_{11i} + A_{12i} + A_{13i}, \quad \mbox{say}.
	\end{eqnarray*}
	Recall that $\bar{Y}_{i,-1} = \frac{1}{T} \sum_{t=1}^T Y_{it-1}$.
	Using the Cauchy-Schwarz inequality, it suffices to show that 
	\[
   N \mathbb{E}_\theta^{{\cal Y}^N} \big[A_{1ji}^2 \mathbb{I}(\mathcal{T}_1) \mathbb{I}(\mathcal{T}_2) \mathbb{I}(\mathcal{T}_3) \mathbb{I}(\mathcal{T}_4) \mathbb{I}(\mathcal{T}_{5i}) \mathbb{I}(\mathcal{T}_{6i}) \big] \le o(N^{\epsilon_0}), \quad j=1,2,3.
	\] 
	
	First, using a slightly more general argument than the one used in the proof of Lemma~\ref{lemma:ap.ratio.optimality.rhohat}, we can show that 
	\[
	N \mathbb{E}_\theta^{{\cal Y}^N} \big[ A_{11i}^2 \big]
	= \mathbb{E}_\theta^{{\cal Y}^N} \left[ N (\hat{\rho} - \rho)^2 \bar{Y}_{i,-1}  \right] 	
	=o(N^{\epsilon_0}).
	\]
	Second, in the region $\mathcal{T}_{5i}$ we can bound
	\be
	  \left(\frac{\sigma^2}{T}+B_N^2 \right) \left| \frac{d p_{*i}}{p_{*i}} \right| =
	  \bigg| 
	  \hat{\lambda}_i(\rho) - \mathbb{E}_\theta \big[\lambda_i \big| \hat{\lambda}_i(\rho),Y_{i0}; p_*(\hat{\lambda}_i(\rho),Y_{i0}) \big] \bigg| \leq C_N'+ C_N, \label{eq.dpp.bound}
	\ee
	where $\mathbb{E}_\theta[\lambda_i|\cdot]$ is the posterior expectation of $\lambda_i$ conditional on $(\hat{\lambda}_i(\rho),Y_{i0})$ under the prior distribution $p_*(\hat{\lambda}_i(\rho),Y_{i0})$.
    Using Assumption~\ref{as:rhohat.ratio.optimality} we obtain the bound
	\begin{eqnarray*}
		N \mathbb{E}_\theta^{{\cal Y}^N} \big[ A_{12i}^2 \mathbb{I}( \mathcal{T}_{5i}) \big]
		\leq \frac{1}{\left(\sigma^2/T+B_N^2\right)^2} \mathbb{E}_\theta^{{\cal Y}^N}\big[N(\hat{\sigma}^2-\sigma^2)^2\big] (C_N'+ C_N)^2 = o(N^{\epsilon_0}).
	\end{eqnarray*}
	Finally, note that 
	\[
	   A_{13i}^2 \mathbb{I}({\cal T}_1) \le \left( \frac{\sigma^2}{T} + B_N^2 + \frac{1}{L_N} \right)^2 
	   \left( \frac{d\widetilde{p}^{(-i)}_i}{\widetilde{p}^{(-i)}_i} 
	   - \frac{d p_{*i}}{p_{*i}} \right)^2.
	\]
    Thus, the desired result follows if we show
	\begin{eqnarray}
		N \mathbb{E}_\theta^{{\cal Y}^N} \left[ 
		\left( \frac{d\widetilde{p}^{(-i)}_i}{\widetilde{p}^{(-i)}_i} 
		- \frac{d p_{*i}}{p_{*i}} \right)^2  
		\mathbb{I}(\mathcal{T}_{2})\mathbb{I}(\mathcal{T}_{3})\mathbb{I}(\mathcal{T}_{4})\mathbb{I}(\mathcal{T}_{5i})\mathbb{I}(\mathcal{T}_{6i}) \right]= o(N^{\epsilon_0}) \label{eq.step3.desired}
	\end{eqnarray}
	To show (\ref{eq.step3.desired}), we have to control the denominator and consider the following truncation region:
	\begin{equation}
		\mathcal{T}_{7i} = \left\{ (\hat{\lambda}_i,Y_{i0}) \, \bigg| \, \widetilde{p}^{(-i)}_i > \frac{ p_{*i}}{2}    \right\}. \label{eq.truncation.tilde.p}
	\end{equation}
	We first analyze (\ref{eq.step3.desired}) on $\mathcal{T}_{7i}$ (Step 2.1) and then on $\mathcal{T}_{7i}^c$ (Step 2.2). We will use the following decomposition:
	\[
	\frac{d\widetilde{p}^{(-i)}_i}{\widetilde{p}^{(-i)}_i} 
	- \frac{d p_{*i}}{p_{*i}}
	= \frac{d\widetilde{p}^{(-i)}_i - d p_{*i}}{\widetilde{p}^{(-i)}_i - p_{*i} + p_{*i}}
	- \frac{d p_{*i}}{p_{*i}}
	\left( 
	\frac{\widetilde{p}^{(-i)}_i - p_{*i} }{\widetilde{p}^{(-i)}_i - p_{*i} + p_{*i}}
	\right).
	\]
	We also will abbreviate $\mathbb{I}({\cal T}_l) \mathbb{I}({\cal T}_k) = \mathbb{I}({\cal T}_l{\cal T}_k)$.

\noindent {\bf Step 2.1.} For the region $\mathcal{T}_{7i}$ we have
	\begin{eqnarray*}
		\lefteqn{ N \mathbb{E}_\theta^{{\cal Y}^N} \left[ \left( \frac{d\widetilde{p}^{(-i)}_i}{\widetilde{p}^{(-i)}_i} 
		- \frac{d p_{*i}}{p_{*i}} \right)^2 \mathbb{I}(\mathcal{T}_{2}\mathcal{T}_{3}\mathcal{T}_{4}\mathcal{T}_{5i}\mathcal{T}_{6i}\mathcal{T}_{7i}) \right]} \\
		&\leq& 2N \mathbb{E}_\theta^{{\cal Y}^N}  \left[ \left( 
		\frac{d\widetilde{p}^{(-i)}_i - d p_{*i}}{\widetilde{p}^{(-i)}_i - p_{*i} + p_{*i}}
		\right)^2 \mathbb{I}(\mathcal{T}_{2}\mathcal{T}_{3}\mathcal{T}_{4}\mathcal{T}_{5i}\mathcal{T}_{6i}\mathcal{T}_{7i}) \right] \\
		&&+ 2 o(N^{\epsilon_0}) N \mathbb{E}_\theta^{{\cal Y}^N}  \left[ \left( \frac{\widetilde{p}^{(-i)}_i - p_{*i} }{\widetilde{p}^{(-i)}_i - p_{*i} + p_{*i}} \right)^2 
		\mathbb{I}(\mathcal{T}_{2}\mathcal{T}_{3}\mathcal{T}_{4}\mathcal{T}_{5i}\mathcal{T}_{6i}\mathcal{T}_{7i}) \right] \\
		&=& 2 B_{1i} + 2o(N^{\epsilon_0}) B_{2i},
	\end{eqnarray*}
	say. The $o(N^{\epsilon_0})$ bound follows from (\ref{eq.dpp.bound}). Using the mean-value theorem, we can express
	\begin{eqnarray*}
		\sqrt{N}(d\widetilde{p}^{(-i)}_i - dp_{*i}) &=& \sqrt{N}(d\hat{p}^{(-i)}_i - dp_{*i}) + \sqrt{N}(\hat{\rho}-\rho) R_{1i}(\widetilde{\rho}) \\		
		\sqrt{N}(\widetilde{p}^{(-i)}_i - p_{*i}) &=& \sqrt{N}(\hat{p}^{(-i)}_i - p_{*i}) + \sqrt{N}(\hat{\rho}-\rho) R_{2i}(\widetilde{\rho}),
	\end{eqnarray*}
	where 
	\begin{eqnarray*}
	R_{1i}(\rho) &=& -\frac{1}{N-1} \sum_{j \neq i}^N 
	\frac{1}{B_N^2}  \phi \left(\frac{\hat{\lambda}_j(\rho) - \hat{\lambda}_i(\rho) }{B_N}  \right)
	\left( \frac{\hat{\lambda}_j(\rho) - \hat{\lambda}_i(\rho)}{B_N} \right)^2
	\big(\bar{Y}_{j,-1} - \bar{Y}_{i,-1} \big)
	\frac{1}{B_N}\phi \left(\frac{Y_{j0}-Y_{i0}}{B_N} \right) \\
	&& + \frac{1}{N-1} \sum_{j \neq i}^N 
	\frac{1}{B_N^3}  \phi \left(\frac{\hat{\lambda}_j(\rho) - \hat{\lambda}_i(\rho) }{B_N}  \right)
	\big(\bar{Y}_{j,-1} - \bar{Y}_{i,-1} \big)
	\frac{1}{B_N}\phi \left(\frac{Y_{j0}-Y_{i0}}{B_N} \right), \\
	R_{2i}(\rho) &=& \frac{1}{N-1} \sum_{j \neq i}^N 
	\frac{1}{B_N}  \phi \left(\frac{\hat{\lambda}_j(\rho) - \hat{\lambda}_i(\rho) }{B_N}  \right)
	\left(\frac{\hat{\lambda}_j(\rho) - \hat{\lambda}_i(\rho)}{B_N}\right)
	\big(\bar{Y}_{j,-1} - \bar{Y}_{i,-1} \big)
	\frac{1}{B_N}\phi \left(\frac{Y_{j0}-Y_{i0}}{B_N} \right),
	\end{eqnarray*}
	and $\widetilde{\rho}$ is located between $\hat{\rho}$ and $\rho$. 
	
	We proceed with the analysis of $B_2$. Using the lower bound for $\widetilde{p}^{(-i)}_i$ over the region $\mathcal{T}_{7i}$, the $C_r$ inequality, and the law of iterated expectations, 
	we obtain
	\begin{eqnarray*}
      B_{2i} &\le& 8 \mathbb{E}_\theta^{{\cal Y}^i} \left[ \frac{1}{p_{*i}^2} \mathbb{E}_{\theta,{\cal Y}^i}^{{\cal Y}^{(-i)}} \big[ N (\hat{p}^{(-i)}_i - p_{*i})^2 \mathbb{I}(\mathcal{T}_{1}\mathcal{T}_{2}\mathcal{T}_{3}\mathcal{T}_{4}\mathcal{T}_{5i}\mathcal{T}_{6i}\mathcal{T}_{7i})  \big] \right] \\
      && + 8 \mathbb{E}_\theta^{{\cal Y}^i} \left[ \frac{1}{p_{*i}^2} \mathbb{E}_{\theta,{\cal Y}^i}^{{\cal Y}^{(-i)}} \big[ N (\hat{\rho}-\rho)^2 R_{2i}^2(\tilde{\rho}) \mathbb{I}(\mathcal{T}_{1}\mathcal{T}_{2}\mathcal{T}_{3}\mathcal{T}_{4}\mathcal{T}_{5i}\mathcal{T}_{6i}\mathcal{T}_{7i})  \big] \right] \\
      &=& 8 \mathbb{E}_\theta^{{\cal Y}^i}[B_{21i} + B_{22i}],      
    \end{eqnarray*}	 
	say. 
	
	According to Lemma~\ref{lemma:technical.lemma.1}(c) (see Section~\ref{appsubsubsection.further.details})
	\[
	     \mathbb{E}_{\theta,{\cal Y}^i}^{{\cal Y}^{(-i)}} \big[ N (\hat{p}^{(-i)}_i - p_{*i})^2 \mathbb{I}(\mathcal{T}_{1}\mathcal{T}_{2}\mathcal{T}_{3}\mathcal{T}_{4}\mathcal{T}_{5i}\mathcal{T}_{6i}\mathcal{T}_{7i})  \big] 
	     \le \frac{M}{B_N^2} p_i \mathbb{I}({\cal T}_{5i} {\cal T}_{6i}) .
	\]
	This leads to
	\[
	  \mathbb{E}_\theta^{{\cal Y}^i}[B_{21i}]
	   \le \frac{M}{B_N^2} \mathbb{E}_\theta^{{\cal Y}^i}\left[\frac{p_i}{p_{*i}^2} \mathbb{I}({\cal T}_{5i} {\cal T}_{6i}) \right]
	   = \frac{M}{B_N^2}\int_{{\cal T}_{5i} \cap {\cal T}_{6i}} \frac{p_i^2}{p_{*i}^2} d\hat{\lambda}_i d y_{i0}. 
	\]
	According to Lemma~\ref{lemma:technical.lemma.1}(e) (see Section~\ref{appsubsubsection.further.details})
	\[
	  \int_{{\cal T}_{5i} \cap {\cal T}_{6i}} \frac{p_i^2}{p_{*i}^2} d\hat{\lambda}_i d y_{i0} =o(N^\epsilon).
	\]
	Because $1/B_N^2 = o(N^\epsilon)$ according to Assumption~\ref{as:ratio.optimality.bandwidth}, we can deduce that
	\[
	   \mathbb{E}_\theta^{{\cal Y}^i}[B_{21i}] \le o(N^{\epsilon_0}).
	\]
	
	Using the Cauchy-Schwarz Inequality, we obtain
	\[
	   B_{22i} \le \frac{1}{p^2_{*i}} \sqrt{\mathbb{E}_{\theta,{\cal Y}^i}^{{\cal Y}^{(-i)}} \big[ N^2 (\hat{\rho}-\rho)^4 \big]}  \sqrt{\mathbb{E}_{\theta,{\cal Y}^i}^{{\cal Y}^{(-i)}} \big[ R_{2i}^4(\tilde{\rho}) \mathbb{I}(\mathcal{T}_{1}\mathcal{T}_{2}\mathcal{T}_{3}\mathcal{T}_{4}\mathcal{T}_{5i}\mathcal{T}_{6i}\mathcal{T}_{7i})  \big]}. 
	\]
	Using the inequality once more leads to
	\begin{eqnarray*}
		\mathbb{E}_\theta^{{\cal Y}^i}[B_{22i}]
		&\le& \sqrt{\mathbb{E}_\theta^{{\cal Y}^N}\big[ N^2 (\hat{\rho}-\rho)^4 \big] }
		\sqrt{\mathbb{E}_\theta^{{\cal Y}^i} \left[ \frac{1}{p_{*i}^4}\mathbb{E}_{\theta,{\cal Y}^i}^{{\cal Y}^{(-i)}} \big[ R_{2i}^4(\tilde{\rho}) \mathbb{I}(\mathcal{T}_{1}\mathcal{T}_{2}\mathcal{T}_{3}\mathcal{T}_{4}\mathcal{T}_{5i}\mathcal{T}_{6i}\mathcal{T}_{7i})  \big]  \right]} \\
		&\le& M 		\sqrt{\mathbb{E}_\theta^{{\cal Y}^i} \left[ \frac{1}{p_{*i}^4}\mathbb{E}_{\theta,{\cal Y}^i}^{{\cal Y}^{(-i)}} \big[ R_{2i}^4(\tilde{\rho}) \mathbb{I}(\mathcal{T}_{1}\mathcal{T}_{2}\mathcal{T}_{3}\mathcal{T}_{4}\mathcal{T}_{5i}\mathcal{T}_{6i}\mathcal{T}_{7i})  \big]  \right]}.
	\end{eqnarray*}
	The second inequality follows from Assumption~\ref{as:rhohat.ratio.optimality}. 
	According to Lemma~\ref{lemma:technical.lemma.1}(a) (see Section~\ref{appsubsubsection.further.details}) 
	\[
	\mathbb{E}_{\theta,{\cal Y}^i}^{{\cal Y}^{(-i)}} \big[ R_{2i}^4(\tilde{\rho}) \mathbb{I}(\mathcal{T}_{1}\mathcal{T}_{2}\mathcal{T}_{3}\mathcal{T}_{4}\mathcal{T}_{5i}\mathcal{T}_{6i}\mathcal{T}_{7i})  \big] \le M L_N^4 p_i^4 \mathbb{I}(\mathcal{T}_{5i}\mathcal{T}_{6i}),
	\]
	where $L_N = o(N^{\epsilon_0})$ was defined in (\ref{eq.choosing.LN}).
	This leads to the bound
	\begin{eqnarray*}
	  \mathbb{E}_\theta^{{\cal Y}^i}[B_{22i}]
	  & \le & 	M L_N^2 \sqrt{ \mathbb{E}_\theta^{{\cal Y}^i}\left[ \left(\frac{p_i}{p_{*i}} \right)^4  \mathbb{I}(\mathcal{T}_{5i}\mathcal{T}_{6i}) \right] } \\
	  & = & 	M L_N^2 \sqrt{ \int_{\mathcal{T}_{5i} \cap \mathcal{T}_{6i}} \left( \frac{p_i}{p_{*i}} \right)^4 p_i d \hat{\lambda}_i dy_{i0} }	\\
	  & \le & M_* L_N^2 \sqrt{ \int_{\mathcal{T}_{5i} \cap \mathcal{T}_{6i}} \left( \frac{p_i}{p_{*i}} \right)^4 d \hat{\lambda}_i dy_{i0} } \\
	  & \le & o(N^{\epsilon_0}).
	\end{eqnarray*}
	The second inequality holds because the density $p_i$ is bounded from above. The last inequality is proved in Lemma~\ref{lemma:technical.lemma.1}(e) (see Section~\ref{appsubsubsection.further.details}).
	
	We deduce that $B_{2i} = o(N^{\epsilon_0})$. A similar argument can be used to establish that $B_{1i} = o(N^{\epsilon_0})$.

\noindent {\bf Step 2.2.} 	
Over the set $\mathcal{T}_{7i}^c$, since $|A_{1i}| \leq o(N^{{\epsilon_0}})$, we have
	\begin{eqnarray*}
		N \mathbb{E}_\theta^{{\cal Y}^N} \left[ \left( \frac{d\widetilde{p}^{(-i)}_i}{\widetilde{p}^{(-i)}_i} 
		- \frac{d p_{*i}}{p_{*i}} \right)^2
		\mathbb{I}( \mathcal{T}_1 \mathcal{T}_{2}\mathcal{T}_{3}\mathcal{T}_{4}\mathcal{T}_{5i}\mathcal{T}_{6i} \mathcal{T}_{7i}^c) \right] 
		\leq o(N^{\epsilon_0}) N \mathbb{P}_\theta^{{\cal Y}^N} (\mathcal{T}_{1}\mathcal{T}_{2}\mathcal{T}_{3}\mathcal{T}_{4}\mathcal{T}_{5i}\mathcal{T}_{6i}\mathcal{T}_{7i}^c). 
	\end{eqnarray*}
	Notice that 
	\begin{eqnarray*}
		\mathcal{T}_{7i}^c &=&  \left\{ \hat{p}^{(-i)}_i - p_{*i} + (\hat{\rho} - \rho) R_{1i}(\widetilde{\rho}) < -\frac{p_{*i}}{2} \right\} \\
		&\subset& \left\{ \hat{p}^{(-i)}_i - p_{*i} - |\hat{\rho} - \rho| |R_{1i}(\widetilde{\rho})| < -\frac{p_{*i}}{2} \right\} \\
		&\subset& \left\{ \hat{p}^{(-i)}_i - p_{*i}  < -\frac{p_{*i}}{4} \right\} \cup \left\{ |\hat{\rho} - \rho| |R_{1i}(\widetilde{\rho})| > \frac{p_{*i}}{4} \right\}.
	\end{eqnarray*}
	Then, 
	\begin{eqnarray*}
		\lefteqn{ N \mathbb{P}_{\theta,{\cal Y}^i}^{{\cal Y}^{(-i)}} (\mathcal{T}_{1}\mathcal{T}_{2}\mathcal{T}_{3}\mathcal{T}_{4}\mathcal{T}_{5i}\mathcal{T}_{6i}\mathcal{T}_{7i}^c) } \\ 
		&\leq& N \mathbb{P}_{\theta,{\cal Y}^i}^{{\cal Y}^{(-i)}} \left\{ \hat{p}^{(-i)}_i - p_{*i}  < -\frac{p_{*i}}{4} \right\}
		+ N \mathbb{P}_{\theta,{\cal Y}^i}^{{\cal Y}^{(-i)}} \left[ 
		\left\{ |\hat{\rho} - \rho| |R_{2i}(\widetilde{\rho})| > \frac{p_{*i}}{4} \right\}\mathbb{I}(\mathcal{T}_{1}\mathcal{T}_{2}\mathcal{T}_{3}\mathcal{T}_{4}\mathcal{T}_{5i}\mathcal{T}_{6i})
		\right] \\
		&\leq& N \mathbb{P}_{\theta,{\cal Y}^i}^{{\cal Y}^{(-i)}} \left\{ \hat{p}^{(-i)}_i - p_{*i}  < -\frac{p_{*i}}{4} \right\}
		+  \frac{16 L_N^4 }{p_{*i}^2}  
		\mathbb{E}_{\theta,{\cal Y}^i}^{{\cal Y}^{(-i)}} \left[ 
		R_{2i}(\widetilde{\rho})^2  \mathbb{I}(\mathcal{T}_{2}\mathcal{T}_{3}\mathcal{T}_{4}\mathcal{T}_{5i}\mathcal{T}_{6i}\mathcal{T}_{7i}) 
		\right] \\
		&\leq& N \mathbb{P}_{\theta,{\cal Y}^i}^{{\cal Y}^{(-i)}} \left\{ \hat{p}^{(-i)}_i - p_{*i}  < -\frac{p_{*i}}{4} \right\}
		+  \frac{M L_N^4 }{p_{*i}^2} p_i \mathbb{I}(\mathcal{T}_{5i}\mathcal{T}_{6i}).
	\end{eqnarray*}
	The first inequality is based on the superset of ${\cal T}_{7i}^c$ from above. The second inequality is based on Chebychev's inequality and trucation $\mathcal{T}_2$. The third inequality uses a version of the result in Lemma~\ref{lemma:technical.lemma.1}(a) in which the remainder is raised to the power of two instead of to the power of four. Moreover, we use the fact that $p_i$ is bounded from above to absorb one of the $p_i$ terms in the constant $M$.
	
	In Lemma~\ref{lemma:technical.lemma.1}(f) (see Section~\ref{appsubsubsection.further.details}) we  apply Bernstein's inequality to bound the probability $\mathbb{P}_{\theta,{\cal Y}^i}^{{\cal Y}^{(-i)}} \left\{ \hat{p}^{(-i)}_i - p_{*i}  < -\frac{p_{*i}}{4} \right\}$ uniformly over $(\hat{\lambda}_i,Y_{i0})$ in the region $\mathcal{T}_{5i}$, showing that
	\[
		N \mathbb{E}_\theta^{{\cal Y}^i} \left[ \mathbb{P}_{\theta,{\cal Y}^i}^{{\cal Y}^{(-i)}} \left\{ \hat{p}^{(-i)}_i - p_{*i}  < -\frac{p_{*i}}{4} \right\} \mathbb{I}(\mathcal{T}_{5i}\mathcal{T}_{6i}) \right]
		= o(N^{\epsilon_0}),
	\]
	as desired. Moreover, according to Lemma~\ref{lemma:technical.lemma.1}(f) (see Section~\ref{appsubsubsection.further.details})
	\[
	\mathbb{E}_\theta^{{\cal Y}^i} \left[ \frac{p_i}{p_{*i}^2} \mathbb{I}(\mathcal{T}_{5i}\mathcal{T}_{6i}) \right]
	= \int_{{\cal T}_{5i} \cap {\cal T}_{6i}} \left(\frac{p_i}{p_{*i}} \right)^2d\hat{\lambda}_i dy_{i0} = o(N^{\epsilon_0}),
	\]
	which gives us the required result for Step 2.2. Combining the results from Steps 2.1 and 2.2 yields (\ref{eq.bound.A1i2.T1T6}).

The bound in (\ref{eq.step1.bound}) now follows from (\ref{eq.Step 2.desired.1}), (\ref{eq.Step 2.desired.2}), and (\ref{eq.bound.A1i2.T1T6}), which completes the proof of the lemma. $\blacksquare$

\paragraph{Term $A_{2i}$}
\begin{lemma} \label{lemma:ap.ratio.optimality.pstarlambda} Suppose the assumptions in Theorem \ref{thm:ratio.optimality} hold. Then,
\[
 \limsup_{N \rightarrow \infty}
 \;  \frac{ N \mathbb{E}_\theta^{{\cal Y}^i,\lambda_i} \big[ \big(\mu \big( \hat{\lambda}_i(\rho),\sigma^2/T+B_N^2, p_*(\hat{\lambda}_i(\rho),Y_{i0}) \big) -
 	\lambda_i \big)^2 \big] }{N \mathbb{E}_\theta^{{\cal Y}^i,\lambda_i}\big[ (\lambda_i - \mathbb{E}_{\theta,{\cal Y}^i}^{\lambda_i}[\lambda_i] )^2\big]  + N^{\epsilon_0}} \leq  1
\]
\end{lemma}

\noindent {\bf Proof of Lemma~\ref{lemma:ap.ratio.optimality.pstarlambda}}.
Notice that 
$\mu \big( \hat{\lambda}_i(\rho),Y_{i0},\sigma^2/T+B_N^2, p_*(\hat{\lambda}_i(\rho),Y_{i0}) \big)$ can be interpreted $\mu(\cdot)$ as the posterior mean of $\lambda_i$ under the $p_*(\cdot)$ measure. We use $\mathbb{E}_{*,\theta}^{{\cal Y}^i,\lambda_i}[\cdot]$ to denote
the joint distribution of ${\cal Y}^i$ and $\lambda_i$ under the  $p_*(\cdot)$ measure. Let $\{\tau_N\}$ be a non-negative sequence such that $\tau_N = o(N^{\epsilon_0})$. The desired result follows if we 
can show that
\begin{eqnarray*}
	(i)  && \limsup_{N \rightarrow \infty} \; \frac{ N \mathbb{E}_{*,\theta}^{{\cal Y}^i,\lambda_i} 
		\left[ \left( \mu \big( \hat{\lambda}_i(\rho),Y_{i0},\sigma^2/T + B_N^2, p_*(\hat{\lambda}_i(\rho),Y_{i0}) \big) -
		\lambda_i \right)^2 \right] + \tau_N }{N \mathbb{E}_\theta^{{\cal Y}^i,\lambda_i}\big[ (\lambda_i - \mathbb{E}_{\theta,{\cal Y}^i}^{\lambda_i}[\lambda_i] )^2\big]
  + N^{\epsilon_0}} \leq  1 \\
	(ii) && \limsup_{N \rightarrow \infty} \; \frac{ N \mathbb{E}_{\theta}^{{\cal Y}^i,\lambda_i} 
		\left[ \left( \mu \big( \hat{\lambda}_i(\rho),Y_{i0},\sigma^2/T +B_N^2 , p_*(\hat{\lambda}_i(\rho),Y_{i0}) \big) -
		\lambda_i \right)^2 \right] }
	{ N \mathbb{E}_{*,\theta}^{{\cal Y}^i,\lambda_i}  
		\left[ \left( \mu \big( \hat{\lambda}_i(\rho),Y_{i0},\sigma^2/T + B_N^2, p_*(\hat{\lambda}_i(\rho),Y_{i0}) \big) -
		\lambda_i \right)^2 \right] + \tau_N } \leq  1,
\end{eqnarray*}
where
\[
 \mathbb{E}_\theta^{{\cal Y}^i,\lambda_i}\big[ (\lambda_i - \mathbb{E}_{\theta,{\cal Y}^i}^{\lambda_i}[\lambda_i] )^2\big] =
 	\mathbb{E}_\theta^{{\cal Y}^i,\lambda_i} \left[ \left( \mu \big( \hat{\lambda}_i(\rho),Y_{i0},\sigma^2/T, p(\hat{\lambda}_i(\rho),Y_{i0}) \big) -
 	\lambda_i \right)^2 \right].
\]

\noindent {\bf Part (i):} We will construct an upper bound for the numerator. Using the fact that the posterior mean minimizes the integrated risk, we obtain
\begin{eqnarray*}
	\lefteqn{N \mathbb{E}_{*,\theta}^{{\cal Y}_i,\lambda_i} 
		\left[ \left( \mu \big( \hat{\lambda}_i(\rho),Y_{i0},\sigma^2/T + B_N^2, p_*(\hat{\lambda}_i(\rho),Y_{i0}) \big) -
		\lambda_i \right)^2 \right]} \\
	&\le& N \mathbb{E}_{*,\theta}^{{\cal Y}_i,\lambda_i} 
	\left[ \left( \mu \big( \hat{\lambda}_i(\rho),Y_{i0},\sigma^2/T, p(\hat{\lambda}_i(\rho),Y_{i0}) \big) -
	\lambda_i \right)^2 \right] \\
	&=&  N \int \int p_*(\hat{\lambda}_i,y_{i0})  
	\left( \mu \big( \hat{\lambda}_i(\rho),y_{i0},\sigma^2/T, p(\hat{\lambda}_i(\rho),y_{i0}) \big) -
	\lambda_i \right)^2 d \hat{\lambda}_i dy_{i0}  \\
	&\leq& N \int \int p_*(\hat{\lambda}_i,y_{i0}) 
	\left( \mu \big( \hat{\lambda}_i(\rho),y_{i0},\sigma^2/T, p(\hat{\lambda}_i(\rho),y_{i0}) \big) -
	\lambda_i \right)^2 \mathbb{I}(\mathcal{T}_{5i} \mathcal{T}_{6i}) d \hat{\lambda}_i dy_{i0}  \\
	&& + N 4C_N^2 \mathbb{P} (\mathcal{T}_{5i}^c \cup \mathcal{T}_{6i}^c) \\
	&=& N \int \int p_*(\hat{\lambda}_i,y_{i0}) 
	\left( \mu \big( \hat{\lambda}_i(\rho),y_{i0},\sigma^2/T, p(\hat{\lambda}_i(\rho),y_{i0}) \big) -
	\lambda_i \right)^2 \mathbb{I}(\mathcal{T}_{5i} \mathcal{T}_{6i}) d \hat{\lambda}_i dy_{i0} + o(N^{\epsilon_0 }).
\end{eqnarray*}
The second inequality uses the fact that $|\lambda_i| \le C_N$ and therefore the posterior mean has to be bounded in absolute value by $C_N$ as well. The last line follows from an argument similar to that used in Step~1 of the proof of Lemma~\ref{lemma:ap.ratio.optimality.phatpstar}.

According to Lemma~\ref{lemma.pi/p*.limit}, we obtain the following uniform bound over the region $\mathcal{T}_{5i} \cap \mathcal{T}_{6i}$:
\[
p_*(\hat{\lambda}_i,y_{i0}) \leq (1+o(1))  p(\hat{\lambda}_i,y_{i0}).
\]
Therefore,
\begin{eqnarray*}
	\lefteqn{ \int \int p_*(\hat{\lambda}_i,y_{i0}) 
	\left( \mu \big( \hat{\lambda}_i(\rho),y_{i0},\sigma^2/T, p(\hat{\lambda}_i(\rho),y_{i0}) \big) -
	\lambda_i \right)^2 \mathbb{I}(\mathcal{T}_{5i} \mathcal{T}_{6i}) d \hat{\lambda}_i dy_{i0} } \\
	&=& (1+o(1)) \int \int p(\hat{\lambda}_i,y_{i0}) 
	\left( \mu \big( \hat{\lambda}_i(\rho),y_{i0},\sigma^2/T, p(\hat{\lambda}_i(\rho),y_{i0}) \big) -
	\lambda_i \right)^2 \mathbb{I}(\mathcal{T}_{5i} \mathcal{T}_{6i}) d \hat{\lambda}_i dy_{i0}.
\end{eqnarray*}
In turn, we obtain the following bound:
\begin{eqnarray*}
	\lefteqn{N \mathbb{E}_{*,\theta}^{{\cal Y}_i,\lambda_i} 
		\left[ \left( \mu \big( \hat{\lambda}_i(\rho),Y_{i0},\sigma^2/T + B_N^2, p_*(\hat{\lambda}_i(\rho),Y_{i0}) \big) -
		\lambda_i \right)^2 \right] + \tau_N}  \\
	&\leq& (1+o(1)) N \int \int p(\hat{\lambda}_i,y_{i0}) 
	\left( \mu \big( \hat{\lambda}_i(\rho),y_{i0},\sigma^2/T, p(\hat{\lambda}_i(\rho),y_{i0}) \big) -
	\lambda_i \right)^2 \mathbb{I}(\mathcal{T}_{5i} \mathcal{T}_{6i}) d \hat{\lambda}_i dy_{i0} + o(N^{\epsilon_0}) \\
	&\leq& (1+o(1)) N \mathbb{E}_\theta^{{\cal Y}^i,\lambda_i}\big[ (\lambda_i - \mathbb{E}_{\theta,{\cal Y}^i}^{\lambda_i}[\lambda_i] )^2\big] + o(N^{\epsilon_0})  \\
	&\leq& (1+o(1)) N \mathbb{E}_\theta^{{\cal Y}^i,\lambda_i}\big[ (\lambda_i - \mathbb{E}_{\theta,{\cal Y}^i}^{\lambda_i}[\lambda_i] )^2\big] + N^{\epsilon_0}, 
\end{eqnarray*}
which yields the required result for Part (i).

\noindent {\bf Part (ii):} Similar to the proof of Part (i), we construct an upper bound for the numerator as follows
\begin{eqnarray*}
	\lefteqn{N \mathbb{E}_{\theta}^{{\cal Y}_i,\lambda_i} 
		\left[ \left( \mu \big( \hat{\lambda}_i(\rho),Y_{i0},\sigma^2/T  + B_N^2, p_*(\hat{\lambda}_i(\rho),Y_{i0}) \big) -
		\lambda_i \right)^2 \right]} \\
	&=& N \int \int p(\hat{\lambda}_i,y_{i0})  
	\left( \mu \big( \hat{\lambda}_i(\rho),y_{i0},\sigma^2/T+ B_N^2, p_{*}(\hat{\lambda}_i(\rho),y_{i0}) \big) -
	\lambda_i \right)^2 d \hat{\lambda}_i dy_{i0}  \\
	&\leq& \int \int p_*(\hat{\lambda}_i,y_{i0}) \frac{p(\hat{\lambda}_i,y_{i0})}{p_*(\hat{\lambda}_i,y_{i0})}
	\left( \mu \big( \hat{\lambda}_i(\rho),y_{i0},\sigma^2/T+ B_N^2, p_{*}(\hat{\lambda}_i(\rho),y_{i0}) \big) -
	\lambda_i \right)^2 \mathbb{I}(\mathcal{T}_{5i} \mathcal{T}_{6i}) d \hat{\lambda}_i dy_{i0}  \\
	&& + N 4 C_N^2 \mathbb{P} (\mathcal{T}_{5i}^c \cup \mathcal{T}_{6i}^c) \\
	&=&(1+o(1)) N  \int \int p_*(\hat{\lambda}_i,y_{i0})
	\left( \mu \big( \hat{\lambda}_i(\rho),y_{i0},\sigma^2/T  + B_N^2, p_{*}(\hat{\lambda}_i(\rho),y_{i0}) \big) -
	\lambda_i \right)^2 \\
	&& \times \mathbb{I}(\mathcal{T}_{5i} \mathcal{T}_{6i}) d \hat{\lambda}_i dy_{i0} + o(N^{\epsilon}), \quad \mbox{any } \epsilon>0 \\
	&\leq& (1+o(1)) N \mathbb{E}_{*,\theta}^{{\cal Y}_i,\lambda_i} 
	\left[ \left( \mu \big( \hat{\lambda}_i(\rho),Y_{i0},\sigma^2/T + B_N^2, p_*(\hat{\lambda}_i(\rho),Y_{i0}) \big) -
	\lambda_i \right)^2 \right] + \tau_N.
\end{eqnarray*}
For the last line we used the fact that $\tau_N = o(N^{\epsilon_0})$.
We now have the required result for Part (ii). 

\paragraph{Term $A_{3i}$}
\begin{lemma} \label{lemma:ap.ratio.optimality.rhohat} Suppose the assumptions in Theorem \ref{thm:ratio.optimality} hold. Then, for any $\epsilon > 0$: 
\[
   N \mathbb{E}_\theta^{{\cal Y}^N} \big[ \big( \hat{\rho} - \rho \big)^2 Y_{iT}^2 \big] = o(N^\epsilon).
\]
\end{lemma}

\noindent {\bf Proof of Lemma~\ref{lemma:ap.ratio.optimality.rhohat}.}
Using the Cauchy-Schwarz inequality, we can bound
\begin{eqnarray*}
	\mathbb{E}_\theta^{{\cal Y}^N} \left[ \big( \sqrt{N} (\hat{\rho}-\rho) \big)^2 Y_{iT}^2 \right] 
	&\le& \sqrt{ \mathbb{E}_\theta^{{\cal Y}^N} \left[ \big( \sqrt{N} (\hat{\rho}-\rho) \big)^4 \right]
		\mathbb{E}_\theta^{{\cal Y}^N} \left[ Y_{iT}^4 \right]}.
\end{eqnarray*}
By Assumption \ref{as:rhohat.ratio.optimality}, we have 
\[
\mathbb{E}_\theta^{{\cal Y}^N} \left[ \big( \sqrt{N} (\hat{\rho}-\rho) \big)^4 \right] \le o(N^\epsilon)
\]
for any $\epsilon>0$.

For the second term, write
\[
   Y_{iT} = \rho^T Y_{i0} + \sum_{\tau=0}^{T-1} \rho^\tau(\lambda_i + U_{iT-\tau} ).
\]
Using the $C_r$ inequality and the assumptions that $|\rho|<1$ and $U_{it} \sim iid N(0,\sigma^2)$, we deduce that there are finite constants $M_1$, $M_2$, $M_3$ such that 
\begin{eqnarray*}
  \mathbb{E}_\theta^{{\cal Y}^N}\left[  Y_{iT}^4 \right]
  &\le&  M_1 \mathbb{E}_\theta^{{\cal Y}^N}\left[  Y_{i0}^4 \right] 
  + M_2  \mathbb{E}_\theta^{{\cal Y}^N}\left[ \lambda_i^4 \right] 
  + M_3  \mathbb{E}_\theta^{{\cal Y}^N}\left[ U_{i1}^4 \right] \\
  &=& M_1 \mathbb{E}_\theta^{{\cal Y}^N}\left[  Y_{i0}^4 \right] +o(N^{\epsilon_0}) + o(N^{\epsilon})
\end{eqnarray*}
for any $\epsilon$, where the last line holds because $|\lambda_i| \le C_N$ according to Assumption~\ref{as:ratio.optimality.marginal.dist.lambda} and $U_{i1}$ is normally distributed and therefore all its moments are finite. 

The desired $o(N^{\epsilon})$ bound  for the fourth moment of $Y_{i0}$ can be obtained as follows (we are dropping subscripts and superscripts from expectation and probability operators):
\begin{eqnarray*}
\mathbb{E}\big[|Y_{i0}|^4\big] &=& 4 \mathbb{E} \left[ \int_0^\infty \mathbb{I}\{|Y_{i0}| \ge \tau \} \tau^3 d \tau \right] \\
&=& 4 \mathbb{E} \left[\int_0^{\infty} \mathbb{P} \{ |Y_{i0}| \ge \tau | \lambda_i\} \tau^3 d \tau \right] \\
&=& 4 \mathbb{E} \left[\int_0^{\bar{C}} \mathbb{P} \{ |Y_{i0}| \ge \tau | \lambda_i\} \tau^3 d \tau \right] 
+ \mathbb{E} \left[\int_{\bar{C}}^{\infty} \mathbb{P} \{ |Y_{i0}| \ge \tau | \lambda_i\} \tau^3 d \tau \right]\\
&\leq& M + \int \left[ 
\int_{\bar{C}}^{\infty}  \exp\left( -m( \tau,\lambda)\right) \tau^3 d\tau 
\right]\pi_{\lambda}(\lambda) d \lambda
\end{eqnarray*}
for some finite constant $M$, where $\bar{C}$ is the constant in Assumption \ref{as:tail.intial.condition}(ii).

Notice that on the domain $[\bar{C},\infty)$, the function $\exp\left( -m(\tau,\lambda)\right)$ in decreasing in $\tau$, while the function $\tau^3$ is increasing in $\tau$. W.l.o.g, suppose that $\bar{C} = (1+k) (\sqrt{\ln N^*} + C_{N^*})$ and  $(1+k) (\sqrt{\ln N} + C_{N}) > 2 \ln N$ for all $N \geq N^*$.  
Now, let $\tau_N = (1+k) (\sqrt{\ln N} + C_{N})$ and bound the integral with a Riemann sum:
\begin{eqnarray*}
\int_{\bar{C}}^{\infty}  \exp\left( -m( \tau,\lambda)\right) \tau^3 d\tau 
&\leq& \sum_{N = N^*}^{\infty} \exp \left( -m(\tau_N,\lambda) \right) \tau_{N+1}^3 (\tau_{N+1} - \tau_N) \\
&\leq& 	\sum_{N = N^*}^{\infty} \exp \left( -m(\tau_N,\lambda) \right) \tau_{N+1}^4 \\
&=& \sum_{N = N^*}^{\infty} \exp \left( -m(\tau_N,\lambda) + 4 \ln \tau_{N+1} \right) \\
&\leq & \sum_{N = N^*}^{\infty} \exp \left( - (2+\epsilon) \ln N + 4 \ln \tau_{N+1} \right) \\
&=& \sum_{N = N^*}^{\infty} \frac{\tau_{N+1}^4}{N^{2+\epsilon}},
\end{eqnarray*}
for some constant $\epsilon \geq 0$. The last inequality holds by Assumption \ref{as:tail.intial.condition}(ii).
Because $\tau_N^4 = o(N^{\epsilon})$, there exists a finite constant $M$ such that
\begin{eqnarray*}
\sum_{N = N^*}^{\infty} \frac{\tau_{N+1}^4}{N^{2+\epsilon}} 
\leq M \sum_{N = N^*}^{\infty} \frac{1}{N^{2}} < \infty. 
\end{eqnarray*}
This leads to the desired result
\[
\mathbb{E}\big[|Y_{i0}|^4\big] < \infty. \quad \blacksquare
\]

\subsubsection{Further Details}
\label{appsubsubsection.further.details}

We now provide more detailed derivations for some of the bounds used in Section~\ref{appsubsubsec:three.lemmas}. 
Recall that
\begin{eqnarray*}
	R_{1i}(\rho) &=& -\frac{1}{N-1} \sum_{j \neq i}^N 
	\frac{1}{B_N^2}  \phi \left(\frac{\hat{\lambda}_j(\rho) - \hat{\lambda}_i(\rho) }{B_N}  \right)
	\left( \frac{\hat{\lambda}_j(\rho) - \hat{\lambda}_i(\rho)}{B_N} \right)^2
	\big(\bar{Y}_{j,-1} - \bar{Y}_{i,-1} \big)
	\frac{1}{B_N}\phi \left(\frac{Y_{j0}-Y_{i0}}{B_N} \right) \\
	&& + \frac{1}{N-1} \sum_{j \neq i}^N 
	\frac{1}{B_N^3}  \phi \left(\frac{\hat{\lambda}_j(\rho) - \hat{\lambda}_i(\rho) }{B_N}  \right)
	\big(\bar{Y}_{j,-1} - \bar{Y}_{i,-1} \big)
	\frac{1}{B_N}\phi \left(\frac{Y_{j0}-Y_{i0}}{B_N} \right) \\	
	R_{2i}(\rho) &=& \frac{1}{N-1} \sum_{j \neq i}^N 
	\frac{1}{B_N}  \phi \left(\frac{\hat{\lambda}_j(\rho) - \hat{\lambda}_i(\rho) }{B_N}  \right)
	\left(\frac{\hat{\lambda}_j(\rho) - \hat{\lambda}_i(\rho)}{B_N}\right)
	\big(\bar{Y}_{j,-1} - \bar{Y}_{i,-1} \big)
	\frac{1}{B_N}\phi \left(\frac{Y_{j0}-Y_{i0}}{B_N} \right) 
\end{eqnarray*}
For expositional purposes, our analysis focuses on the slightly simpler term $R_{2i}(\widetilde{\rho})$. The extension to $R_{1i}(\widetilde{\rho})$ is fairly straightforward. By definition,
\[
\hat{\lambda}_j(\widetilde{\rho}) - \hat{\lambda}_i(\widetilde{\rho}) 
= \hat{\lambda}_j(\rho) - \hat{\lambda}_i(\rho) - (\widetilde{\rho}-\rho) (\bar{Y}_{j,-1} - \bar{Y}_{i,-1}). 
\]
Therefore,
\begin{eqnarray*}
	R_{2i}(\widetilde{\rho}) &=& \frac{1}{N-1} \sum_{j \neq i}^N 
	\frac{1}{B_N}  \phi \left(\frac{\hat{\lambda}_j(\rho) - \hat{\lambda}_i(\rho) }{B_N} - (\widetilde{\rho} - \rho) \left( \frac{\bar{Y}_{j,-1} - \bar{Y}_{i,-1}}{B_N} \right) \right) \\
	&& \times \left(\frac{\hat{\lambda}_j(\rho) - \hat{\lambda}_i(\rho)}{B_N}
	- (\widetilde{\rho} - \rho) \left( \frac{\bar{Y}_{j,-1} - \bar{Y}_{i,-1}}{B_N} \right) \right) \\
	&& \times
	\big(\bar{Y}_{j,-1} - \bar{Y}_{i,-1} \big)
	\frac{1}{B_N}\phi \left(\frac{Y_{j0}-Y_{i0}}{B_N} \right). \\
\end{eqnarray*}

Consider the region ${\cal T}_2 \cap {\cal T}_3 \cap {\cal T}_4$. First, using   (\ref{eq.truncation.rhohat-rho.Ybarj-Ybari}) we can bound
\[
\max_{1\leq i, i \leq N} \left| (\hat{\rho} - \rho ) (\bar{Y}_{j,-1} - \bar{Y}_{i,-1}) \right| \leq \frac{M}{L_N} .
\]
Thus,  
\begin{eqnarray*}
	\lefteqn{ \phi \left(\frac{\hat{\lambda}_j(\rho) - \hat{\lambda}_i(\rho) }{B_N} - (\widetilde{\rho} - \rho) \left( \frac{\bar{Y}_{j,-1} - \bar{Y}_{i,-1}}{B_N} \right) \right) \mathbb{I}({\cal T}_2 {\cal T}_3 {\cal T}_4 ) }\\
	&\leq& \phi \left(\frac{\hat{\lambda}_j(\rho) - \hat{\lambda}_i(\rho) }{B_N} 
	+ \left( \frac{M}{ L_N B_N}  \right) \right) 
	\mathbb{I}\left\{ \frac{\hat{\lambda}_j(\rho) - \hat{\lambda}_i(\rho) }{B_N} 
	\leq - \frac{M}{ L_N B_N}  \right\} \\
	&& + \phi(0) \mathbb{I}\left\{ \left| 
	\frac{\hat{\lambda}_j(\rho) - \hat{\lambda}_i(\rho) }{B_N} \right| 
	\leq  \frac{M}{ L_N B_N}   \right\} \\
	&& + \phi \left(\frac{\hat{\lambda}_j(\rho) - \hat{\lambda}_i(\rho) }{B_N} - \left(  \frac{M}{ L_N B_N}   \right) \right) 
	\mathbb{I}\left\{ \frac{\hat{\lambda}_j(\rho) - \hat{\lambda}_i(\rho) }{B_N} \geq   \frac{M}{ L_N B_N}   \right\} \\
	&=& \bar{\phi} \left(\frac{\hat{\lambda}_j(\rho) - \hat{\lambda}_i(\rho) }{B_N} \right),
\end{eqnarray*}
say. The function $\bar{\phi}(x)$ is flat for $|x|< M/L_NB_N$ and is proportional to a Gaussian density outside of this region. 

Second, we can use the bound
\[
\left|\frac{\hat{\lambda}_j(\rho) - \hat{\lambda}_i(\rho)}{B_N}
- (\widetilde{\rho} - \rho) \left( \frac{\bar{Y}_{j,-1} - \bar{Y}_{i,-1}}{B_N} \right) \right|
 \le \left|\frac{\hat{\lambda}_j(\rho) - \hat{\lambda}_i(\rho)}{B_N} \right| + \frac{M}{L_NB_N}.
\]
Third, for the region ${\cal T}_3 \cap {\cal T}_4$ we can deduce from (\ref{eq.truncation.Ybar_j - Ybar_i}) that
\[
\max_{1 \leq i,j \leq N} | \bar{Y}_{j,-1} - \bar{Y}_{i,-1}| 
\leq M L_N.
\] 
Therefore,
\[
  	\big| \bar{Y}_{j,-1} - \bar{Y}_{i,-1} \big|
  	\frac{1}{B_N}\phi \left(\frac{Y_{j0}-Y_{i0}}{B_N} \right)
  	\le \frac{M L_N}{B_N} \phi \left(\frac{Y_{j0}-Y_{i0}}{B_N} \right).
\]

Now, define the function
\[
\bar{\phi}_{*}(x) = \bar{\phi} \left(x\right) 
\left( \left| x \right| + \frac{M}{ L_N B_N}   \right).
\]
Because for random variables with bounded densities and Gaussian tails all moments exist and because $L_NB_N>1$ by definition of $L_N$ in (\ref{eq.choosing.LN}), the function $\bar{\phi}_{*}(x)$ has the property that for any finite positive integer $m$ there is a finite constant $M$ such that 
\[
	\int \bar{\phi}_{*}(x)^m dx \leq M.
\]

Combining the previous results we obtain the following bound for $R_{2i}(\widetilde{\rho})$:
\be
\big|R_{2i}(\widetilde{\rho}) \mathbb{I}({\cal T}_2 {\cal T}_3 {\cal T}_4) \big|
\le \frac{M L_N}{N-1} \sum_{j \neq i}^N \frac{1}{B_N} 
\bar{\phi}_{*} \left(\frac{\hat{\lambda}_j(\rho) - \hat{\lambda}_i(\rho) }{B_N} \right) 
\frac{1}{B_N} \phi\left( \frac{Y_{j0}-Y_{i0}}{B_N}\right) \label{bound.R_2rhobar}.
\ee
For the subsequent analysis it is convenient define the function
\be
  f(\hat{\lambda}_j - \hat{\lambda}_i,Y_{j0}-Y_{i0})
  = \frac{1}{B_N^2} \bar{\phi}_{*} \left(\frac{\hat{\lambda}_j(\rho) - \hat{\lambda}_i(\rho) }{B_N} \right)\phi\left( \frac{Y_{j0}-Y_{i0}}{B_N}\right).
  \label{bound.R_2rhobar.fdef}
\ee
In the remainder of this section we will state and prove three technical lemmas that establish moment bounds for $R_{1i}(\widetilde{\rho})$ and $R_{2i}(\widetilde{\rho})$. The bounds are used in Section~\ref{appsubsubsec:three.lemmas}. We will abbreviate $\mathbb{E}_{\theta,{\cal Y}^i}^{{\cal Y}^{(-i)}}[\cdot] = \mathbb{E}_i[\cdot]$ and simply use $\mathbb{E}[\cdot]$ to denote $\mathbb{E}_\theta^{{\cal Y}^N}[\cdot]$. 

\begin{lemma}\label{lemma.E_i(f)^m}
	Suppose the assumptions required for Theorem~\ref{thm:ratio.optimality} are satisfied. Then, for a finite positive integer $m$, over the region $\mathcal{T}_{5i}$, we have
	\[
	  \mathbb{E}_i \big[ f^m(\hat{\lambda}_j - \hat{\lambda}_i,Y_{j0}-Y_{i0})  \big]
	\leq \frac{M}{B_N^{2(m-1)}}p_i.
	\]
\end{lemma}

\noindent {\bf Proof of Lemma	\ref{lemma.E_i(f)^m}.} 
We have
\begin{eqnarray*}
	\lefteqn{\mathbb{E}_i \big[ f^m(\hat{\lambda}_j - \hat{\lambda}_i,Y_{j0}-Y_{i0})  \big]} \\ 
	&=& \int \left( 
	\frac{1}{B_N} \bar{\phi}_{*} \left(\frac{\hat{\lambda} - \hat{\lambda}_i }{B_N}  \right)
	\frac{1}{B_N}\phi \left(\frac{y_{0}-Y_{i0}}{B_N} \right) 
	\right)^m p(\hat{\lambda},y_0) d (\hat{\lambda},y_0) \\
	&=& \frac{1}{B_N^{2(m-1)}} \int \left\{ \int  
	\frac{1}{B_N} \bar{\phi}_{*} \left(\frac{\hat{\lambda} - \hat{\lambda}_i }{B_N}  \right)^m 
	\frac{1}{B_N}\phi \left(\frac{y_{0}-Y_{i0}}{B_N} \right)^m 
	p(\hat{\lambda},y_0|\lambda) 
	d (\hat{\lambda},y_0) \right\} \pi(\lambda) d \lambda.
\end{eqnarray*}
The inner integral is 
\begin{eqnarray*}
	\lefteqn{ \int  
	\frac{1}{B_N} \bar{\phi}_{*} \left(\frac{\hat{\lambda} - \hat{\lambda}_i }{B_N}  \right)^m 
	\frac{1}{B_N}\phi \left(\frac{y_{0}-Y_{i0}}{B_N} \right)^m 
	p(\hat{\lambda},y_0|\lambda) 
	d (\hat{\lambda},y_0) } \\
	&=& \int 
	\frac{1}{B_N} \bar{\phi}_{*} \left(\frac{\hat{\lambda} - \hat{\lambda}_i }{B_N}  \right)^m
	\frac{1}{\sigma/\sqrt{T}} \exp\left( -\frac{1}{2} \left( \frac{\hat{\lambda} - \lambda_i }{\sigma/\sqrt{T}} \right)^2 \right) d \hat{\lambda} \\
	&& \times \int 
	\frac{1}{B_N}\phi \left(\frac{y_{0}-Y_{i0}}{B_N} \right)^m 
	\pi(y_0|\lambda) dy_0 \\
	&=& I_1 \times I_2, 
\end{eqnarray*}
say.

Notice that 
\begin{eqnarray*}
	I_1&=& 	\int 
	\frac{1}{B_N} \bar{\phi}_{*} \left(\frac{\hat{\lambda} - \hat{\lambda}_i }{B_N}  \right)^m
	\frac{1}{\sigma/\sqrt{T}} \exp\left( -\frac{1}{2} \left( \frac{\hat{\lambda} - \lambda_i }{\sigma/\sqrt{T}} \right)^2 \right) d \hat{\lambda} 
	\\
	&=& \int \bar{\phi}_{*}(\lambda^*)^m 
	\frac{1}{\sigma/\sqrt{T}} \exp\left( -\frac{1}{2} \left( \frac{\hat{\lambda}_i - \lambda_i + B_N \lambda^* }{\sigma/\sqrt{T}} \right)^2 \right) d \lambda^{*} \\
	&=& \int \bar{\phi}_{*}(\lambda^*)^m \exp\left( -\left( (\hat{\lambda}_i - \lambda_i)B_N \lambda^* \right) \frac{1}{\sigma^2/T} \right)
	\exp\left( -\frac{1}{2} \left( \frac{ B_N \lambda^* }{\sigma/\sqrt{T}} \right)^2 \right)   d \lambda^{*} \\
	&& \times \left[  \frac{1}{\sigma/\sqrt{T}} \exp\left( -\frac{1}{2} \left( \frac{\hat{\lambda}_i - \lambda_i }{\sigma/\sqrt{T}} \right)^2 \right)\right] \\
	&\leq& M \left(\int \bar{\phi}_{*}(\lambda^*)^m \exp\left( v_N \lambda^* \right) d \lambda^*\right) 
	\left[  \frac{1}{\sigma/\sqrt{T}} \exp\left( -\frac{1}{2} \left( \frac{\hat{\lambda}_i - \lambda_i }{\sigma/\sqrt{T}} \right)^2 \right)\right] \\
	&\leq& M \left[  \frac{1}{\sigma/\sqrt{T}} \exp\left( -\frac{1}{2} \left( \frac{\hat{\lambda}_i - \lambda_i }{\sigma/\sqrt{T}} \right)^2 \right)\right] = M p(\hat{\lambda}_i|\lambda_i,Y_{i0}).
\end{eqnarray*}	
We used the change-of-variable $\lambda_* = (\hat{\lambda}-\hat{\lambda}_i)/B_N$ to replace $\hat{\lambda}$.
Here the second inequality holds because the exponential function $\exp\left( -\frac{1}{2} \left( \frac{ B_N \lambda^* }{\sigma/\sqrt{T}} \right)^2 \right)$ is bounded by a constant. Moreover,
under truncation $\mathcal{T}_{5i}$, $| \hat{\lambda}_i | \leq C_N'$ and the support of $\lambda_i$ is bounded by $[-C_N,C_N]$ (under Assumption \ref{as:ratio.optimality.marginal.dist.lambda}). Thus, $v_N= B_N (C_N'+2C_N)$. According to Assumption~\ref{as:ratio.optimality.bandwidth} $v_N= B_N (C_N'+2C_N) = o(1)$. Thus, the last inequality holds because $\int \bar{\phi}_{*}(\lambda^*)^m \exp\left( v_N \lambda^* \right) d \lambda^*$ is finite. Finally, note that $p(\hat{\lambda}_i|\lambda_i,Y_{i0}) = p(\hat{\lambda}_i|\lambda_i)$.

We now proceed with a bound for the second integral, $I_2$. Using the fact that the Gaussian pdf $\phi(x)$ is bounded, we can write
\begin{eqnarray*}
 I_2 &=& \int 
 \frac{1}{B_N}\phi \left(\frac{y_{0}-Y_{i0}}{B_N} \right)^m 
 \pi(y_0|\lambda) dy_0 \\
     &\le& M  \int 
     \frac{1}{B_N}\phi \left(\frac{y_{0}-Y_{i0}}{B_N} \right) 
     \pi(y_0|\lambda) dy_0 \\
     &=& M \big(1+o(1) \big) \pi(Y_{i0}|\lambda), 
\end{eqnarray*}
uniformly in $|y_0| \leq C_N' $ and $| \lambda |  \leq C_N$.
Here the last equality follows from Assumption~\ref{as:tail.intial.condition}(iii).
Combining the bounds for $I_1$ and $I_2$ and integrating over $\lambda$, we obtain
\begin{eqnarray*}
 \mathbb{E}_i \big[ f^m(\hat{\lambda}_j - \hat{\lambda}_i,Y_{j0}-Y_{i0})  \big]
 & = & \frac{1}{B_N^{2(m-1)}} \int I_1 \times I_2 \pi(\lambda_i) d\lambda_i \\
 & \le & \frac{1}{B_N^{2(m-1)}} M\big(1+o(1)\big) \int p(\hat{\lambda}_i|\lambda_i,Y_{i0}) p(Y_{i0}|\lambda_i) \pi(\lambda_i) d \lambda_i \\
 & = & \frac{1}{B_N^{2(m-1)}} M\big(1+o(1)\big) p_i,
\end{eqnarray*}
as required.

\begin{lemma} \label{lemma.pi/p*.limit}
	Suppose the assumptions required for Theorem~\ref{thm:ratio.optimality} are satisfied. Then, 
	\begin{eqnarray}
	\sup_{(\hat{\lambda}_i,Y_{i0}) \in \mathcal{T}_{5i} \cap \mathcal{T}_{6i} }  \frac{p_i}{p_{*i}} &=& 1 + o(1) \label{eq.lemma.pi/p*.limit} \\
\sup_{(\hat{\lambda}_i,Y_{i0}) \in \mathcal{T}_{5i} \cap \mathcal{T}_{6i} }  \frac{p_{*i}}{p_{i}} &=& 1 + o(1). \label{eq.lemma.p*/pi.limit}	
	\end{eqnarray}
\end{lemma}

\noindent {\bf Proof of Lemma~\ref{lemma.pi/p*.limit}.}
We begin by verifying (\ref{eq.lemma.pi/p*.limit}). Let
\begin{eqnarray*}
	p(\hat{\lambda}_i,y_{i0}|\lambda_i) &=& \frac{1}{\sqrt{\sigma^2/T}} \phi\left(\frac{\hat{\lambda}_i-\lambda_i}{\sqrt{\sigma^2/T}}\right) \pi(y_{i0}|\lambda_i) \\
	p_*(\hat{\lambda}_i,y_{i0}|\lambda_i) &=& 
	\frac{1}{\sqrt{B_N^2+\sigma^2/T}} \phi\left( \frac{\hat{\lambda}_i - \lambda_i }{\sqrt{B_N^2+\sigma^2/T}}\right) 
	\left[ \int \frac{1}{B_N} \phi\left( \frac{y_{i0} - \tilde{y}_{i0}}{B_N}\right)  \pi(\tilde{y}_{i0}|\lambda_i)  d \tilde{y}_{i0} \right]
\end{eqnarray*}
such that
\[
  p_i = \int p(\hat{\lambda}_i,y_{i0}|\lambda_i) \pi(\lambda_i) d\lambda_i, \quad
  p_{*i} = \int p_*(\hat{\lambda}_i,y_{i0}|\lambda_i) \pi(\lambda_i) d\lambda_i.
\]

Because $|\lambda_i| \le C_N$ by Assumption~\ref{as:ratio.optimality.marginal.dist.lambda} and $|\hat{\lambda}_i| \leq C_N'$ in the region $\mathcal{T}_{5i}$, for some finite constant $M$ we have
\begin{eqnarray}
\frac{1}{\sqrt{\sigma^2/T}} \phi\left(\frac{\hat{\lambda}_i-\lambda_i}{\sqrt{\sigma^2/T}}\right) 
&=& \frac{1}{\sqrt{B_N^2+\sigma^2/T}} \phi\left( \frac{\hat{\lambda}_i - \lambda_i }{\sqrt{B_N^2+\sigma^2/T}}\right) \nonumber \\
&& \times \frac{\sqrt{B_N^2+\sigma^2/T}}{\sqrt{\sigma^2/T}}
\exp \left\{ -\frac{1}{2}\left(\frac{\hat{\lambda}_i-\lambda_i}{\sqrt{B_N^2+\sigma^2/T}} \right)^2
\frac{B_N^2}{\sigma^2/T}  \right\} \nonumber \\
&\leq& \frac{1}{\sqrt{B_N^2+\sigma^2/T}} \phi\left( \frac{\hat{\lambda}_i - \lambda_i }{\sqrt{B_N^2+\sigma^2/T}}\right) \nonumber \\
&& \times \sqrt{1+M B_N^2} \exp(-M (C_N'+C_N)^2B_N^2)  \nonumber \\
&=& (1+o(1)) \frac{1}{\sqrt{B_N^2+\sigma^2/T}} \phi\left( \frac{\hat{\lambda}_i - \lambda_i }{\sqrt{B_N^2+\sigma^2/T}}\right),
\label{eq.p/p*.bound.a1}
\end{eqnarray}
where $o(1)$ is uniform in $(\hat{\lambda}_i,Y_{i0}) \in  \mathcal{T}_{5i} \cap \mathcal{T}_{6i}$.
Here we used Assumption~\ref{as:ratio.optimality.bandwidth} which implies that $v_N=(C_N'+C_N)B_N = o(1)$.

According to Assumption~\ref{as:tail.intial.condition}(iii), 
\[
\int \frac{1}{B_N} \phi\left( \frac{y_{i0} - \tilde{y}_{i0}}{B_N}\right)  \pi(\tilde{y}_{i0}|\lambda_i)  d \tilde{y}_{i0}
= (1+o(1)) \pi(y_{i0}|\lambda_i) 
\]
uniformly in $|y_{i0}| \leq C_N'$ and $|\lambda_i| \leq C_N$. This implies that
\begin{eqnarray}
\pi(y_{i0}|\lambda_i) 
\leq (1+ o(1)) \int \frac{1}{B_N} \phi\left( \frac{y_{i0} - \tilde{y}_{i0}}{B_N}\right)  \pi(\tilde{y}_{i0}|\lambda_i)  d \tilde{y}_{i0}. \label{eq.p/p*.bound.a2}
\end{eqnarray}
uniformly in $|y_{i0}| \leq C_N'$ and $|\lambda_i| \leq C_N$.

Then, by combining the bounds in (\ref{eq.p/p*.bound.a1}) and ({\ref{eq.p/p*.bound.a2}}) we deduce
	\begin{eqnarray*}
		\lefteqn{  p(\hat{\lambda}_i,y_{i0}|\lambda_i) - p_*(\hat{\lambda}_i,y_{i0}|\lambda_i) }  \\
		&=& \frac{1}{\sqrt{\sigma^2/T}} \phi\left(\frac{\hat{\lambda}_i-\lambda_i}{\sqrt{\sigma^2/T}}\right) 
		\pi(y_{i0}|\lambda_i) \\
		&&
		- \frac{1}{\sqrt{B_N^2+\sigma^2/T}} \phi\left( \frac{\hat{\lambda}_i - \lambda_i }{\sqrt{B_N^2+\sigma^2/T}}\right) 
		\int \frac{1}{B_N} \phi\left( \frac{y_{i0} - \tilde{y}_{i0}}{B_N}\right)  \pi(\tilde{y}_{i0}|\lambda_i)  d \tilde{y}_{i0}  \\
		&\leq& \big[(1+o(1))^2-1\big] 
		\frac{1}{\sqrt{B_N^2+\sigma^2/T}} \phi\left( \frac{\hat{\lambda}_i - \lambda_i }{\sqrt{B_N^2+\sigma^2/T}}\right) 
		\int \frac{1}{B_N} \phi\left( \frac{y_{i0} - \tilde{y}_{i0}}{B_N}\right)  \pi(\tilde{y}_{i0}|\lambda_i)  d \tilde{y}_{i0} \\
		&=& o(1)\cdot p_*(\hat{\lambda}_i,y_{i0}|\lambda_i).
	\end{eqnarray*}
Note that the $o(1)$ term does not depend on $(\hat{\lambda}_i,Y_{i0}) \in  \mathcal{T}_{5i} \cap \mathcal{T}_{6i}$.

We deduce that 
\begin{eqnarray*}
  \sup_{(\hat{\lambda}_i,Y_{i0}) \in \mathcal{T}_{5i} \cap \mathcal{T}_{6i} } \; \frac{p_i}{p_{*i}} 
  &=& 1 + \sup_{(\hat{\lambda}_i,Y_{i0}) \in \mathcal{T}_{5i} \cap \mathcal{T}_{6i} } \; \frac{p_i - p_{*i}}{p_{*i}} \\
  &=& 1+ \sup_{(\hat{\lambda}_i,Y_{i0}) \in \mathcal{T}_{5i} \cap \mathcal{T}_{6i} } \;
  \frac{\int \left[p(\hat{\lambda}_i,y_{i0}|\lambda_i)-p_*(\hat{\lambda}_i,y_{i0}|\lambda_i) \right] \pi(\lambda_i) d \lambda_i}{p_{*i}} \\
  &=& 1+o(1).
\end{eqnarray*}
This proves (\ref{eq.lemma.pi/p*.limit}). A similar argument can be used to establish (\ref{eq.lemma.p*/pi.limit}). 
$\blacksquare$

\begin{lemma}\label{lemma:technical.lemma.1}
	Under the assumptions required for Theorem~\ref{thm:ratio.optimality}, we obtain the following bounds:
    \begin{itemize}
		\item[(a)] $\mathbb{E}_i\big[R_{2i}^4(\widetilde{\rho})
		\mathbb{I}(\mathcal{T}_{2}\mathcal{T}_{3}\mathcal{T}_{4}\mathcal{T}_{5i}\mathcal{T}_{6i}\mathcal{T}_{7i})\big] 
		\leq M L_N^4 p_{i}^4 \mathbb{I}(\mathcal{T}_{5i}\mathcal{T}_{6i})$ 
		
		\item[(b)]  $\mathbb{E}_i \big[R_{1i}^4\mathbb{I}(\mathcal{T}_{2}\mathcal{T}_{3}\mathcal{T}_{4}\mathcal{T}_{5i}\mathcal{T}_{6i}\mathcal{T}_{7i})\big] \leq M \frac{L_N^4}{B_N^4}p_{i}^4\mathbb{I}(\mathcal{T}_{5i}\mathcal{T}_{6i})$ 
		
		\item[(c)]  $\mathbb{E}_i \left[N (\hat{p}^{(-i)}_i - p_{*i})^2 \mathbb{I}(\mathcal{T}_{2}\mathcal{T}_{3}\mathcal{T}_{4}\mathcal{T}_{5i}\mathcal{T}_{6i}\mathcal{T}_{7i}) \right] 
	     \leq \frac{M}{B_N^2}  p_{i} \mathbb{I}(\mathcal{T}_{5i}\mathcal{T}_{6i})$
	
         \item[(d)] $\mathbb{E}_i \left[N(d\hat{p}^{(-i)}_i - d p_{*i})^2\mathbb{I}(\mathcal{T}_{2}\mathcal{T}_{3}\mathcal{T}_{4}\mathcal{T}_{5i}\mathcal{T}_{6i}\mathcal{T}_{7i}) \right] 
	\leq  \frac{M}{B_N^2}  p_{i} \mathbb{I}(\mathcal{T}_{5i}\mathcal{T}_{6i})$
	
	\item[(e)] $ 
	 \int_{\mathcal{T}_{5i} \cap \mathcal{T}_{6i}} \left(\frac{p_i}{p_{*i}} \right)^m d \hat{\lambda}_i d y_{i0} = o(N^{\epsilon})$, $m>1$. 	
	
	\item[(f)] $N \mathbb{E} \big[\mathbb{P}_i \big\{\hat{p}^{(-i)}_i - p_{*i}  < - p_{*i}/4 \big\} \mathbb{I}(\mathcal{T}_{5i}\mathcal{T}_{6i}) \big] = o(N^{\epsilon})$
	
\end{itemize}
\end{lemma}

\noindent {\bf Proof of Lemma~\ref{lemma:technical.lemma.1}}. 
{\bf Part (a).}   
Recall the following definitions
\begin{eqnarray*}
	\bar{\phi}(x) &=&  
	\phi \left( x +  \frac{M}{ L_N B_N} \right) 
	\mathbb{I}\left\{ x \leq - \frac{M}{ L_N B_N}  \right\} 
	 + \phi(0) \mathbb{I}\left\{ \left| x \right| \leq   \frac{M}{ L_N B_N}  \right\} \\
	&& + \phi \left( x -  \frac{M}{ L_N B_N}  \right) 
	\mathbb{I}\left\{ x \geq   \frac{M}{ L_N B_N}  \right\} \\
	\bar{\phi}_{*}(x) &=& \bar{\phi} \left(x\right) 
	\left( \left| x \right| + \frac{M}{ L_N B_N} \right).
\end{eqnarray*}

First, recall that according to (\ref{bound.R_2rhobar}), in the region ${\cal T}_2 \cap {\cal T}_3 \cap {\cal T}_4$
\[
|R_{2i}(\widetilde{\rho})| \leq \frac{M L_N}{N-1} \sum_{j \neq i}^N 
f(\hat{\lambda}_j- \hat{\lambda}_i, Y_{j0}-Y_{i0}).  
\] 
Then, 
\begin{eqnarray*}
	| R_{2i}(\widetilde{\rho}) |^4 
	&\leq& 
	\left[ \frac{M L_N}{N-1} \sum_{j \neq i}^N 
	f(\hat{\lambda}_j- \hat{\lambda}_i, Y_{j0}-Y_{i0}) 
	\right]^4 \\
	&=&  \bigg[ \frac{M L_N}{N-1} \sum_{j \neq i}^N 
	\bigg\{ f(\hat{\lambda}_j- \hat{\lambda}_i, Y_{j0}-Y_{i0}) - \mathbb{E}_i[f(\hat{\lambda}_j- \hat{\lambda}_i, Y_{j0}-Y_{i0})] \\ 
	&& + \mathbb{E}_i [f(\hat{\lambda}_j- \hat{\lambda}_i, Y_{j0}-Y_{i0})]
	\bigg\} \bigg]^4 \\
	&\leq& M L_N^4 \left[ \frac{1}{N-1} \sum_{j \neq i}^N 
	\left( f(\hat{\lambda}_j- \hat{\lambda}_i, Y_{j0}-Y_{i0}) - \mathbb{E}_i[f(\hat{\lambda}_j- \hat{\lambda}_i, Y_{j0}-Y_{i0})] \right) \right]^4 \\
	&&+ M L_N^4 \left[ \mathbb{E}_i[ f(\hat{\lambda}_j- \hat{\lambda}_i, Y_{j0}-Y_{i0})] \right]^4 \\
	&=& M L_N^4 \big( A_1 + A_2 \big),
\end{eqnarray*}
say. The second inequality holds because $|x+y|^4 \le 8(|x|^4 +|y|^4)$.

The term $(N-1)^4 A_1$ takes the form 
\begin{eqnarray*}
   \left( \sum a_j \right)^4
   &=& \left( \sum a_j^2 + 2 \sum_j \sum_{i > j} a_j a_i \right)^2 \\
   &=& \left( \sum a_j^2 \right)^2 + 4 \left( \sum a_j^2 \right) \left( \sum_j \sum_{i >j} a_j a_i\right) 
   + 4 \left( \sum_j \sum_{i > j} a_j a_i \right)^2 \\
   &=&  \sum a_j^4 + 6 \sum_j \sum_{i >j} a_j^2 a_i^2 \\
   &&   + 4 \left( \sum a_j^2 \right) \left( \sum_j \sum_{i > j} a_j a_i\right) + 4 \sum_j \sum_{i>j} \sum_{l \not=j} \sum_{k>l} a_j a_i a_l a_k,
\end{eqnarray*}
where 
\[
  a_j = f(\hat{\lambda}_j- \hat{\lambda}_i, Y_{j0}-Y_{i0}) - \mathbb{E}_i [f(\hat{\lambda}_j- \hat{\lambda}_i, Y_{j0}-Y_{i0})], \quad j \not=i.
\]
Notice that conditional on $(\hat{\lambda}_i(\rho),Y_{i0})$, the random variables $a_j$ have mean zero and are $iid$ across $j \not=i$.
This implies that 
\[
   \mathbb{E}_i \left[\left( \sum a_j \right)^4 \right]
   = \sum \mathbb{E}_i\big[a_j^4\big] + 6 \sum_j \sum_{i >j} \mathbb{E}_i\big[a_j^2 a_i^2 \big].
\]
The remaining terms drop out because they involve at least one term $a_j$ that is raised to the power of one and therefore has mean zero.

Using the $C_R$ inequality, Jensen's inequality, the conditional independence of $a_j^2$ and $a_i^2$ and Lemma~\ref{lemma.E_i(f)^m}, we can bound
\[
 \mathbb{E}_i[a_j^4] \le \frac{M}{B_N^{6}} p_i, \quad
 \mathbb{E}_i[a_j^2 a_i^2] \le \frac{M}{B_N^4} p_i^2.
\]
Thus, in the region ${\cal T}_2 \cap {\cal T}_3 \cap {\cal T}_4 \cap {\cal T}_{5i} \cap \mathcal{T}_{6i}$
\[
  \mathbb{E}_i[A_1] \le \frac{M p_i}{N^3 B_N^6} + \frac{M p_i^2}{N^2 B_N^4} \le M p_i^4.
\]
The second inequality holds because over $\mathcal{T}_{6i}$, $p_i \geq  \frac{N^{\epsilon'}}{N} \geq \frac{M}{N B_N^2}$.
Using a similar argument, we can also deduce that 
\[
   \mathbb{E}_i[A_2] \le M p_i^4,
\]
which proves Part (a) of the lemma.

\noindent {\bf Part (b).} Similar to proof of Part (a).

\noindent {\bf Part (c).} Can be established using existing results for the variance of a kernel density estimator.

\noindent {\bf Part (d).} Similar to proof of Part (c).

\noindent {\bf Part (e).} We have the desired result because by Lemma \ref{lemma.pi/p*.limit} we can choose a constant $c$ such that  
\begin{equation*}
p_{i} - p_{*i} \leq c p_{*i} \label{eq.desired.p/p*.sufficient}
\end{equation*}
over truncations $\mathcal{T}_{5i}$ and $\mathcal{T}_{6i}$. Thus,
\[
    \left( \frac{p_i}{p_{*i}} \right)^m = \left( 1 + \frac{p_i - p_{*i}}{p_{*i}} \right)^m \le (1+c)^m.
\]
We deduce that  
\[
\int_{\mathcal{T}_{5i} \cap \mathcal{T}_{6i}} \left( \frac{p_i}{p_{*i}} \right)^m d \hat{\lambda}_i d y_{i0} 
\leq  (1+c)^m	\int_{\mathcal{T}_{5i} \cap \mathcal{T}_{6i}} d \hat{\lambda}_i d y_{i0} = \big(2C_N'\big)^2 = o(N^{\epsilon}),
\]
as required. 

\noindent {\bf Part (f).} Define
\[
  \psi_i(\hat\lambda_j,Y_{j0}) = \phi\left( \frac{\hat{\lambda}_j - \hat{\lambda}_i}{B_N}\right)
  \phi\left( \frac{Y_{j0} - Y_{i0}}{B_N}\right)
\]
and write 
\begin{eqnarray*}
  \hat{p}^{(-i)}_i - p_{*i} 
	&=&  \frac{1}{N-1} \sum_{j \neq i}^{N} \left\{ 
	\frac{1}{B_N} \phi\left( \frac{\hat{\lambda}_j - \hat{\lambda}_i}{B_N}\right)
	\frac{1}{B_N} \phi\left( \frac{Y_{j0} - Y_{i0}}{B_N}\right) \right. \\ 
	&& - \left.  \mathbb{E}_i \left[ 
	\frac{1}{B_N} \phi\left( \frac{\hat{\lambda}_j - \hat{\lambda}_i}{B_N}\right)
	\frac{1}{B_N} \phi\left( \frac{Y_{j0} - Y_{i0}}{B_N}\right)
	\right]
	\right\} \\
	&=& \frac{1}{B_N^2(N-1)} \sum_{j \neq i}^{N} \left( \psi_i(\hat\lambda_j,Y_{j0}) - \mathbb{E}_i[\psi_i(\hat\lambda_j,Y_{j0})] \right).
\end{eqnarray*}

Notice that for $\psi_i(\lambda_j,Y_{j0}) \sim iid$ across $j \neq i$ with $|\psi_i(\hat\lambda_j,Y_{j0})| \leq M$ for some finite constant $M$. 
Then, by Bernstein's inequality \footnote{ For a bounded function $f$ and a sequence of $iid$ random variables $X_i$,
	\[
	\mathbb{P} \left\{ 
	\left| \frac{1}{\sqrt{N}} \sum_{i=1}^{N} \left( f(X_i) - \mathbb{E}[f(X_i)] \right) \right| > x 
	\right\}
	\leq 2 \exp\left( 
	-\frac{1}{4} \frac{x^2}{\mathbb{E} [ f(X_i)^2] + \frac{1}{\sqrt{N}} x \sup_x|f(x)|}
	\right).
	\]
} (e.g., Lemma 19.32 in van der Vaart (1998)), 
\begin{eqnarray*}
	\lefteqn{ N \mathbb{P}_i \left\{ \hat{p}^{(-i)}_i - p_{*i}  < -\frac{p_{*i}}{4} \right\} \mathbb{I}(\mathcal{T}_{5i}\mathcal{T}_{6i}) } \\
	&=& N  \mathbb{P}_i  \left\{ \frac{1}{B_N^2(N-1)} \sum_{j \neq i}^{N} \left( \psi_i(\hat\lambda_j,Y_{j0}) - \mathbb{E}_i[\psi_i(\hat\lambda_j,Y_{j0})] \right) < -\frac{p_{*i}}{4} \right\} \mathbb{I}(\mathcal{T}_{5i}\mathcal{T}_{6i}) \\
	&\leq&  2N\exp \left( -\frac{1}{4} \frac{B_N^4(N-1) p_{*i}^2/16}{\mathbb{E}_i[ \psi_i(\hat\lambda_j,Y_{j0})^2] + M B_N^2 p_{i*}/4  }  \right) \mathbb{I}(\mathcal{T}_{5i}\mathcal{T}_{6i}).
\end{eqnarray*}
Using an argument similar to the proof of Lemma~\ref{lemma.E_i(f)^m} one can show that
\[
   \mathbb{E}_i[ \psi_i(\lambda_j,Y_{j0})^2/B_N^4] \leq M p_i/B_N^2.
\]	
In turn
\[
N \mathbb{P}_i \left\{ \hat{p}^{(-i)}_i - p_{*i}  < -\frac{p_{*i}}{4} \right\} \mathbb{I}(\mathcal{T}_{5i}\mathcal{T}_{6i})  \le 2 \exp \left( -M N B_N^2 \frac{p_{*i}^2}{p_i + p_{*i}} + \ln N\right) \mathbb{I} (\mathcal{T}_{5i}\mathcal{T}_{6i}).
\]

From Lemma \ref{lemma.pi/p*.limit} we can find a constant $c$ such that $p_i \leq (1+c) p_{*i}$ and $p_{*i} \leq (1+c) p_i$. This leads to
\begin{eqnarray*}
	\frac{p_{*i}^2}{p_i + p_{*i}} \geq \frac{p_i}{(2+c)(1+c)^2}. 
\end{eqnarray*}	
Then, on the region ${\cal T}_{6i}$
\begin{eqnarray*}
	\lefteqn{N \mathbb{E} \left[ \mathbb{P}_i \left\{ \hat{p}^{(-i)}_i - p_{*i}  < -\frac{p_{*i}}{4} \right\}\mathbb{I}(\mathcal{T}_{5i}\mathcal{T}_{6i}) \right]} \\ 
	&\leq& 2 \mathbb{E} \left[  \exp \left( -M N B_N^2\frac{p_{*i}^2}{p_i + p_{*i}} + \ln N \right) \mathbb{I}( \mathcal{T}_{5i}\mathcal{T}_{6i} ) \right] \\
	&\leq& 2 \mathbb{E} \big[  \exp \left( -M N B_N^2 p_i +\ln N \right) \mathbb{I}( \mathcal{T}_{5i}\mathcal{T}_{6i} ) \big] \\
	&\leq& 2 \exp \left( -M B_N^2 N^{\epsilon'} + \ln N \right) \\
	&=& o(N^{\epsilon}),
\end{eqnarray*}
as desired. $\blacksquare$

\newpage

\subsection{Derivations for Section~\ref{sec:monte-carlo-simulations}}

\subsubsection{Consistency of QMLE in Experiments 2 and 3}

We show for the basic dynamic panel data model that even if the Gaussian correlated random effects distribution is misspecified, the pseudo-true value of the QMLE estimator of $\theta$ corresponds to the ``true'' $\theta_0$. We do so, by calculating
\be
(\theta_*,\xi_*) = \mbox{argmax}_{\theta,\xi} \; \mathbb{E}_{\theta_0}^{\cal Y} \left[ \ln p(Y,X_2|H,\theta,\xi) \right],
\ee
and verifying that $\theta_*=\theta_0$.
Here, $p(y,x_2|h,\theta,\xi)$ is given in (\ref{eq.qmleobj.thetahat.xihat}). Because the observations are conditionally independent across $i$ and the likelihood function is symmetric with respect to $i$, we can drop the $i$ subscripts.

We make some adjustment to the notation. The covariance matrix $\Sigma$ only depends on $\gamma$, but not on $(\rho,\alpha)$. Moreover, we will split $\xi$ into the parameters that characterize the conditional mean of $\lambda$, denoted by $\Phi$, and $\omega$, which are the non-redundant elements of the prior covariance matrix $\underline{\Omega}$. Finally, we define
\[
\tilde{Y}(\theta_1) = Y - X \rho - Z \alpha
\]
with the understanding that $\theta_1=(\rho,\alpha)$ and excludes $\gamma$. Moreover, let $\phi = \mbox{vec}(\Phi')$ and $\tilde{h}' = I \otimes h'$, such that we can write $\Phi h = \tilde{h}' \phi$. Using this notation, we can write
\begin{eqnarray}
\lefteqn{\ln p(y,x_2|h,\theta_1,\gamma,\phi,\omega)} \label{eq.full.lh} \\
&=& C -\frac{1}{2} \ln|\Sigma(\gamma)| - \frac{1}{2} \big(\tilde{y}(\theta_1)-w\hat{\lambda}(\theta)\big)'\Sigma^{-1}(\gamma)\big(\tilde{y}(\theta_1)-w\hat{\lambda}(\theta)\big) \nonumber \\
&& - \frac{1}{2} \ln \big|\underline{\Omega} \big|  + \frac{1}{2} \ln \big|\bar{\Omega}(\gamma,\omega)\big| \nonumber \\
&& -\frac{1}{2} \bigg( \hat{\lambda}(\theta)'w'\Sigma^{-1}(\gamma) w \hat{\lambda}(\theta) +
\phi'\tilde{h} \underline{\Omega}^{-1} \tilde{h}'\phi - \bar{\lambda}'(\theta,\xi) \bar{\Omega}^{-1}(\gamma,\omega) \bar \lambda(\theta,\xi) \bigg), \nonumber
\end{eqnarray}
where
\begin{eqnarray*}
	\hat{\lambda}(\theta) &=& (w'\Sigma^{-1}(\gamma)w)^{-1}w'\Sigma^{-1}(\gamma)\tilde{y}(\theta_1) \\
	\bar{\Omega}^{-1}(\gamma,\omega) &=& \underline{\Omega}^{-1} + w'\Sigma^{-1}(\gamma) w, \quad
	\bar{\lambda}(\theta,\xi)     = \bar{\Omega}(\gamma,\omega) \big( \underline{\Omega}^{-1} \tilde{h}'\phi  + w'\Sigma^{-1}(\gamma) w \hat{\lambda}(\theta) \big).
\end{eqnarray*}

In the basic dynamic panel data model $\lambda$ is scalar, $w=\iota$, $\Sigma(\gamma) = \gamma I$, $x_2 = \emptyset$, $z=\emptyset$, $h=[1,y_0]'$, $\underline{\Omega}= \omega^2$. Thus, splitting the $(T-1)(\ln \gamma^2)/2$, we can write
\begin{eqnarray*}
	\ln p(y|h,\rho,\gamma,\phi,\omega)
	&=& C -\frac{T-1}{2} \ln|\gamma^2| - \frac{1}{2\gamma^2} \big(\tilde{y}(\rho)-\iota\hat{\lambda}(\rho)\big)'\big(\tilde{y}(\rho)-\iota\hat{\lambda}(\rho)\big) \nonumber \\
	&& - \frac{1}{2} \ln \big|\omega^2 \big| - \frac{1}{2} \ln \big|\gamma^2/T \big| + \frac{1}{2}\ln (1/T) + \frac{1}{2} \ln \big|\bar{\Omega}(\gamma,\omega)\big| \nonumber \\
	&& -\frac{1}{2} \bigg( \frac{T}{\gamma^2} \hat{\lambda}^2(\rho) +
	\frac{1}{\omega^2} \phi'\tilde{h} \tilde{h}'\phi - \frac{1}{\bar{\Omega}(\gamma,\omega) }\bar{\lambda}^2(\theta,\xi)  \bigg), \nonumber
\end{eqnarray*}
where
\begin{eqnarray*}
	\hat{\lambda}(\rho) &=& \frac{1}{T} \iota'\tilde{y}(\rho) \\
	\bar{\Omega}^{-1}(\gamma,\omega) &=& \frac{1}{\omega^2} + \frac{1}{\gamma^2/T}, \quad
	\bar{\lambda}(\theta,\xi)     = \bar{\Omega}(\gamma,\omega) \left( \frac{1}{\omega^2} \tilde{h}'\phi  + \frac{T}{\gamma^2} \hat{\lambda}(\rho) \right).
\end{eqnarray*}
Note that
\[
- \frac{1}{2} \ln \big|\omega^2 \big| + \frac{1}{2} \ln \big|T/\gamma^2 \big| + \frac{1}{2} \ln \big|\bar{\Omega}(\gamma,\omega)\big| = \frac{1}{2} \ln \left| \frac{ \frac{1}{\omega^2} \frac{T}{\gamma^2}}{\frac{1}{\omega^2}+\frac{T}{\gamma^2}} \right| = - \frac{1}{2} \ln \big| \omega^2 + \gamma^2/T \big|.
\]
In turn, we can write
\begin{eqnarray*}
	\lefteqn{ \ln p(y|h,\rho,\gamma,\phi,\omega) } \\
	&=& C -\frac{T-1}{2} \ln|\gamma^2| - \frac{1}{2\gamma^2} \tilde{y}(\rho)'(I-\iota \iota'/T)\tilde{y}(\rho) 
	- \frac{1}{2} \ln \big| \omega^2 + \gamma^2/T \big|\\
	&& -\frac{1}{2} \bigg( \frac{T}{\gamma^2} \hat{\lambda}^2(\rho) +
	\frac{1}{\omega^2} \phi'\tilde{h} \tilde{h}'\phi - \frac{\omega^2 \gamma^2/T}{\omega^2 + \gamma^2/T} \left( \frac{1}{\omega^2} \tilde{h}'\phi  + \frac{T}{\gamma^2} \hat{\lambda}(\rho) \right)^2  \bigg) \\
	&=& C -\frac{T-1}{2} \ln|\gamma^2| - \frac{1}{2\gamma^2} \tilde{y}(\rho)'(I-\iota \iota'/T)\tilde{y}(\rho) 
	- \frac{1}{2} \ln \big| \omega^2 + \gamma^2/T \big|\\
	&& -\frac{1}{2(\omega^2+\gamma^2/T)} \bigg( \phi'\tilde{h}\tilde{h}'\phi - 2 \hat{\lambda}(\rho) \tilde{h}'\phi +  \hat{\lambda}^2(\rho)\bigg) .
\end{eqnarray*}
Taking expectations (we omit the subscripts from the expectation operator), we can write
\begin{eqnarray}
\lefteqn{ \mathbb{E}\big[ \ln p(Y|H,\rho,\gamma,\phi,\omega) \big] } \label{eq.simple.expectedlh} \\
&=& C -\frac{T-1}{2} \ln|\gamma^2| - \frac{1}{2\gamma^2} \mathbb{E}\big[ \tilde{Y}(\rho)'(I-\iota\iota'/T)\tilde{Y}(\rho) \big]
- \frac{1}{2} \ln \big| \omega^2 + \gamma^2/T \big| \nonumber \\
&& -\frac{1}{2(\omega^2+\gamma^2/T)} \bigg( \big( \phi - \big( \mathbb{E}[\tilde{H}\tilde{H}'] \big)^{-1} \mathbb{E}[\tilde{H}\hat{\lambda}(\rho)] \big)' \mathbb{E}[\tilde{H}\tilde{H}'] \big( \phi - \big( \mathbb{E}[\tilde{H}\tilde{H}'] \big)^{-1} \mathbb{E}[\tilde{H}\hat{\lambda}(\rho)] \big) \nonumber \\
&& - \mathbb{E}[\hat{\lambda}(\rho)\tilde{H}']\big(\mathbb{E}[\tilde{H}\tilde{H}'] \big)^{-1}\mathbb{E}[\tilde{H}\hat{\lambda}(\rho)] + \mathbb{E}[\hat{\lambda}^2(\rho)]\bigg). \nonumber
\end{eqnarray}
We deduce that 
\be
\phi_*(\rho) = \big( \mathbb{E}[\tilde{H}\tilde{H}'] \big)^{-1} \mathbb{E}[\tilde{H}\hat{\lambda}(\rho)].
\label{eq.simple.phistarofrho} 
\ee
To evaluate $\phi_*(\rho_0)$, note that $\hat{\lambda}(\rho_0)= \lambda + \iota'u/T$. Using that fact that the initial observation $Y_{i0}$ is uncorrelated with the shocks $U_{it}$, $t\ge 1$, we deduce that $\mathbb{E}[\tilde{H}\hat{\lambda}(\rho_0)] = \mathbb{E}[\tilde{H}\lambda]$. Thus,
\be
\phi_*(\rho_0) = \big( \mathbb{E}[\tilde{H}\tilde{H}'] \big)^{-1} \mathbb{E}[\tilde{H}\lambda].
\label{eq.simple.phistarofrho0} 
\ee 
The pseudo-true value is obtained through a population regression of $\lambda$ on $H$.

Plugging the pseudo-true value for $\phi$ into (\ref{eq.simple.expectedlh}) yields the concentrated objective function
\begin{eqnarray}
\lefteqn{ \mathbb{E}\big[ \ln p(Y|H,\rho,\gamma,\phi_*(\rho),\omega) \big] } \label{eq.simple.expectedlh.phistar}\\
&=& C -\frac{T-1}{2} \ln|\gamma^2| - \frac{1}{2\gamma^2} \mathbb{E}\big[ \tilde{Y}(\rho)'(I-\iota \iota'/T)\tilde{Y}(\rho) \big] \nonumber \\
&& - \frac{1}{2} \ln \big| \omega^2 + \gamma^2/T \big| - \frac{1}{2(\omega^2+\gamma^2/T)} \big( \mathbb{E}[\hat{\lambda}^2(\rho)] - \mathbb{E}[\hat{\lambda}(\rho)\tilde{H}']\big(\mathbb{E}[\tilde{H}\tilde{H}'] \big)^{-1}\mathbb{E}[\tilde{H}\hat{\lambda}(\rho)] \big). \nonumber	
\end{eqnarray}
Using well-known results for the maximum likelihood estimator of a variance parameter in a Gaussian regression model, we can immediately deduce that
\begin{eqnarray}
\gamma^2_*(\rho) &=& \frac{1}{T-1} \mathbb{E}\big[ \tilde{Y}(\rho)'(I-\iota \iota'/T)\tilde{Y}(\rho) \big] \label{eq.simple.gammastarofrho.omegastarofrho}\\
\omega^2_*(\rho)+\gamma^2_*(\rho)/T &=& \big( \mathbb{E}[\hat{\lambda}^2(\rho)]-\mathbb{E}[\hat{\lambda}(\rho)\tilde{H}']\big(\mathbb{E}[\tilde{H}\tilde{H}'] \big)^{-1}\mathbb{E}[\tilde{H}\hat{\lambda}(\rho)] \big). \nonumber 
\end{eqnarray}
At $\rho=\rho_0$ we obtain $\tilde{Y}(\rho_0) = \iota \lambda + u$. Thus, $\mathbb{E}[\hat{\lambda}^2(\rho_0)] = \gamma_0^2/T + \mathbb{E}[\lambda^2]$ and $\mathbb{E}[\tilde{H}\hat{\lambda}(\rho_0)] = \mathbb{E}[\tilde{H}\lambda]$.
In turn,
\be
\gamma_*^2(\rho_0) = \gamma_0^2, \quad \omega_*^2(\rho_0) =  \mathbb{E}[\lambda^2]-\mathbb{E}[\lambda \tilde{H}']\big(\mathbb{E}[\tilde{H}\tilde{H}'] \big)^{-1}\mathbb{E}[\tilde{H}\lambda].
\label{eq.simple.gammastarofrho.omegastarofrho0}
\ee
Given $\rho=\rho_0$ the pseudo-true value for $\gamma^2$ is the ``true'' $\gamma_0^2$ and the pseudo-true variance of the correlated random-effects distribution is given by the expected value of the squared residual from a projection of $\lambda$ onto $H$. 

Using (\ref{eq.simple.gammastarofrho.omegastarofrho}), we can now concentrate out $\gamma^2$ and $\omega^2$
from the objective function (\ref{eq.simple.expectedlh.phistar}):
\begin{eqnarray}
\lefteqn{ \mathbb{E}\big[ \ln p(Y|H,\rho,\gamma_*(\rho),\phi_*(\rho),\omega_*(\rho) \big] } \label{eq.simple.expectedlh.phistar.gammastar.omegastar}\\
&=& C -\frac{T-1}{2} \ln \big|\mathbb{E}\big[ \tilde{Y}(\rho)'(I-\iota \iota'/T)\tilde{Y}(\rho) \big] \big| \nonumber \\
&&
- \frac{1}{2} \ln \big| \mathbb{E}[\tilde{Y}'(\rho) \iota \iota' \tilde{Y}(\rho)]-\mathbb{E}[\tilde{Y}'(\rho)\iota \tilde{H}']\big(\mathbb{E}[\tilde{H}\tilde{H}'] \big)^{-1}\mathbb{E}[\tilde{H}\iota'\tilde{Y}(\rho)] \big|. \nonumber 	
\end{eqnarray}
To find the maximum of $\mathbb{E}\big[ \ln p(Y|H,\rho,\gamma_*(\rho),\phi_*(\rho),\omega_*(\rho) \big]$ with respect to $\rho$ we will calculate the first-order condition. Differentiating (\ref{eq.simple.expectedlh.phistar.gammastar.omegastar}) with respect to $\rho$ yields
\begin{eqnarray*}
	\mbox{F.O.C.}(\rho) &=& (T-1) \frac{ \mathbb{E}\big[ X'(I-\iota \iota'/T)\tilde{Y}(\rho) \big]}{\mathbb{E}\big[ \tilde{Y}(\rho)'(I-\iota \iota'/T)\tilde{Y}(\rho) \big]} \\
	&& + \frac{\mathbb{E}[X' \iota \iota' \tilde{Y}(\rho)]-\mathbb{E}[X'\iota \tilde{H}']\big(\mathbb{E}[\tilde{H}\tilde{H}'] \big)^{-1}\mathbb{E}[\tilde{H}\iota'\tilde{Y}(\rho)]}{\mathbb{E}[\tilde{Y}'(\rho) \iota \iota' \tilde{Y}(\rho)]-\mathbb{E}[\tilde{Y}'(\rho)\iota \tilde{H}']\big(\mathbb{E}[\tilde{H}\tilde{H}'] \big)^{-1}\mathbb{E}[\tilde{H}\iota'\tilde{Y}(\rho)]}. 
\end{eqnarray*}
We will now verify that $\mbox{F.O.C.}(\rho_0)=0$. 
Because both denominators are strictly positive, we can rewrite the condition as 
\begin{eqnarray}
\mbox{F.O.C.}(\rho_0)&=& (T-1) \mathbb{E}\big[ X'(I-\iota \iota'/T)\tilde{Y}(\rho_0) \big] \label{eq.simple.expectedlh.foc.rho0}\\
&& \times \bigg( \mathbb{E}[\tilde{Y}'(\rho_0) \iota \iota' \tilde{Y}(\rho_0)]-\mathbb{E}[\tilde{Y}'(\rho_0)\iota \tilde{H}']\big(\mathbb{E}[\tilde{H}\tilde{H}'] \big)^{-1}\mathbb{E}[\tilde{H}\iota'\tilde{Y}(\rho_0)] \bigg) \nonumber \\
&& + \mathbb{E}\big[ \tilde{Y}(\rho_0)'(I-\iota \iota'/T)\tilde{Y}(\rho_0) \big] \nonumber \\
&& \times \bigg(\mathbb{E}[X' \iota \iota' \tilde{Y}(\rho_0)]-\mathbb{E}[X'\iota \tilde{H}']\big(\mathbb{E}[\tilde{H}\tilde{H}'] \big)^{-1}\mathbb{E}[\tilde{H}\iota'\tilde{Y}(\rho_0)] \bigg).	\nonumber
\end{eqnarray}
Using again the fact that $\tilde{Y}(\rho_0) = \iota \lambda + U$, we can rewrite the terms appearing in the first-order condition as follows:
\begin{eqnarray*}
	\mathbb{E}\big[ X'(I-\iota \iota'/T)\tilde{Y}(\rho_0) \big] &=& \mathbb{E}\big[ X'(I-\iota \iota'/T)u \big] = \mathbb{E}[X'u] - \mathbb{E}[X'\iota \iota' u]/T= - \mathbb{E}[X'\iota \iota' u]/T\\
	\mathbb{E}[\tilde{Y}'(\rho_0) \iota \iota' \tilde{Y}(\rho)] &=& \mathbb{E}\big[ (\lambda \iota'+u')\iota \iota'(\iota \lambda +u) \big] = T^2 \mathbb{E}[\lambda^2] + \mathbb{E}[u'\iota \iota'u] =  T^2 \mathbb{E}[\lambda^2] + T \gamma_0^2 \\
	\mathbb{E}[\tilde{H}\iota'\tilde{Y}(\rho_0)] &=&  \mathbb{E}[\tilde{H}\iota'(\iota \lambda + u)] = T \mathbb{E}[\tilde{H} \lambda] \\
	\mathbb{E}\big[ \tilde{Y}(\rho_0)'(I-\iota \iota'/T)\tilde{Y}(\rho_0) \big]
	&=& \mathbb{E}\big[ u'(I-\iota \iota'/T)u \big] = (T-1)\gamma^2\\
	\mathbb{E}[X' \iota \iota' \tilde{Y}(\rho_0)] &=& \mathbb{E}[X' \iota \iota' (\iota \lambda + u)] = T \mathbb{E}[X'\iota \lambda] + \mathbb{E}[X'\iota \iota' u] .
\end{eqnarray*}
For the first equality we used the fact that $X_{it}=Y_{it-1}$ is uncorrelated with $U_{it}$. We can now re-state the first-order condition (\ref{eq.simple.expectedlh.foc.rho0}) as follows: 
\begin{eqnarray}
\lefteqn{\mbox{F.O.C.}(\rho_0)} \label{eq.simple.expectedlh.foc.rho0.2} \\
&=& -(T-1) \big( \mathbb{E}[X'\iota \iota' u] \big)\bigg( \gamma_0^2 + T\big( \mathbb{E}[\lambda^2] - \mathbb{E}[\lambda \tilde{H}']\big(\mathbb{E}[\tilde{H}\tilde{H}'] \big)^{-1}\mathbb{E}[\tilde{H} \lambda] \big) \bigg) \nonumber \\
&&+ \bigg(\mathbb{E}[X'\iota \iota' u] + T \big( \mathbb{E}[X'\iota \lambda]  - \mathbb{E}[X'\iota \tilde{H}']\big(\mathbb{E}[\tilde{H}\tilde{H}'] \big)^{-1} \mathbb{E}[\tilde{H} \lambda] \big)\bigg) (T-1) \gamma_0^2 \nonumber \\
&=& T(T-1) \bigg[ \gamma_0^2\bigg( \mathbb{E}[X'\iota \lambda]  - \mathbb{E}[X'\iota \tilde{H}']\big(\mathbb{E}[\tilde{H}\tilde{H}'] \big)^{-1} \mathbb{E}[\tilde{H} \lambda] \bigg) \nonumber \\
&& - \mathbb{E}[X'\iota \iota' u] \bigg(  \mathbb{E}[\lambda^2] - \mathbb{E}[\lambda \tilde{H}']\big(\mathbb{E}[\tilde{H}\tilde{H}'] \big)^{-1}\mathbb{E}[\tilde{H} \lambda]  \bigg) \bigg]. \nonumber 
\end{eqnarray}

We now have to analyze the terms involving $X'\iota$. Note that we can express
\[
Y_t = \rho_0^t Y_0 + \sum_{\tau=0}^{t-1} \rho_0^\tau (\lambda + U_{t-\tau}).
\]
Define $a_t = \sum_{\tau=0}^{t-1} \rho_0^\tau$ and $b = \sum_{t=1}^{T-1} a_t$. Thus, we can write
\[
Y_t = \rho_0^t Y_0 + \lambda a_t + \sum_{\tau=0}^{t-1} \rho_0^\tau U_{t-\tau}, \quad t>0.
\]
Consequently,
\[
X'\iota = \sum_{t=0}^{T-1} Y_{t} = Y_0 \left( \sum_{t=0}^{T-1} \rho_0^t\right) +  \lambda \left(\sum_{t=1}^{T-1} a_t \right) + \sum_{t=1}^{T-1} \sum_{\tau=0}^{t-1} \rho_0^\tau U_{t-\tau} = a_T y_0 + b \lambda + \sum_{t=1}^{T-1} a_t U_{T-t}.
\]
Thus, we obtain
\begin{eqnarray*}
	\mathbb{E}[ X'\iota \iota'u] &=& \mathbb{E} \left[ \left( a_T Y_0 + b \lambda + \sum_{t=1}^{T-1} a_t U_{T-t} \right)\left( \sum_{t=1}^{T} U_t \right) \right] = b \gamma_0^2 \\
	\mathbb{E}[X'\iota \lambda] &=& \mathbb{E} \left[ \left( a_T Y_0 + b \lambda + \sum_{t=1}^{T-1} a_t U_{T-t} \right)\lambda \right] = a_T \mathbb{E}[Y_0\lambda] + b \mathbb{E}[\lambda^2] \\ 
	\mathbb{E}[X'\iota \tilde{H}'] &=& \mathbb{E} \left[ \left( a_T Y_0 + b \lambda + \sum_{t=1}^{T-1} a_t U_{T-t} \right) \tilde{H}' \right] = a_T \mathbb{E}[Y_0 \tilde{H}'] + b \mathbb{E}[\lambda \tilde{H}'].
\end{eqnarray*}
Using these expressions, most terms that appear in (\ref{eq.simple.expectedlh.foc.rho0.2}) cancel out and the condition simplifies to
\be
\mbox{F.O.C.}(\rho_0) = T(T-1) \gamma_0 a_T \bigg( \mathbb{E}[Y_0\lambda] \label{eq.simple.expectedlh.foc.rho0.3}
- \mathbb{E}[Y_0 \tilde{H}']  \big(\mathbb{E}[\tilde{H}\tilde{H}'] \big)^{-1} \mathbb{E}[\tilde{H} \lambda] \bigg).
\ee
Now consider
\begin{eqnarray*}
	\lefteqn{\mathbb{E}[Y_0 \tilde{H}']  \big(\mathbb{E}[\tilde{H}\tilde{H}'] \big)^{-1} \mathbb{E}[\tilde{H} \lambda]} \\
	&=& \frac{1}{\mathbb{E}[Y_0^2] -(\mathbb{E}[Y_0])}
	\left[ \begin{array}{cc} \mathbb{E}[Y_0] & \mathbb{E}[Y_0^2] \end{array} \right]
	\left[ \begin{array}{cc} \mathbb{E}[Y_0^2] & - \mathbb{E}[Y_0] \\ - \mathbb{E}[Y_0] & 1 \end{array} \right]
	\left[ \begin{array}{c}  \mathbb{E}[Y_0] \\ \mathbb{E}[Y_0^2] \end{array} \right] \\
	&=& \mathbb{E}[Y_0 \lambda].
\end{eqnarray*}
Thus, we obtain the desired result that $\mbox{F.O.C.}(\rho_0) = 0$. To summarize, the pseudo-true values are given by
\begin{eqnarray}
\rho_*&=&\rho_0, \quad \gamma_*^2 = \gamma_0, \quad \phi_* = \big( \mathbb{E}[\tilde{H}\tilde{H}'] \big)^{-1} \mathbb{E}[\tilde{H}\lambda], \label{eq.simple.expectedlh.ptv}\\ 
\omega_*^2 &=& \mathbb{E}[\lambda^2]-\mathbb{E}[\lambda \tilde{H}']\big(\mathbb{E}[\tilde{H}\tilde{H}'] \big)^{-1}\mathbb{E}[\tilde{H}\lambda]. \quad \blacksquare \nonumber
\end{eqnarray}

\subsubsection{Computation of the Oracle Predictor in Experiment 3}

We are using a Gibbs sampler to compute the oracle predictor under the mixture distributions for $U_{it}$.

\noindent {\em Scale Mixture.} Let $a_{it}=1$ if $U_{it}$ is generated from the mixture component with variance $\gamma_+^2$ and $a_{it}=0$ if $U_{it}$ is generated from the mixture component with variance $\gamma_-^2$. Omitting $i$ subscripts from now on, define
\[
\tilde{Y}_{t} = Y_{t} - \rho Y_{t-1}, \quad \gamma^2(a_t) = a_{t}\gamma_+^2 + (1-a_t) \gamma_-^2
\]
such that 
\[
\tilde{Y}_{t}|(\lambda,a_{t}) \sim N \big(\lambda, \gamma^2(a_t) \big).
\]
Now let 
\[
\hat{\lambda} = \frac{1}{T} \sum_{t=1}^T \tilde{Y}_{t} \sim N \big( \lambda,\bar{\gamma}^2(a_{1:T})/T \big),
\]
where
\[
\bar{\gamma}^2(a_{1:T}) = \frac{1}{T} \sum_{t=1}^T \gamma^2(a_t).
\]
Under the prior distribution
\[
\lambda|Y_{0} \sim N(\phi_0+\phi_1Y_{0},\underline{\Omega}),
\]
we obtain a posterior distribution of the form
\be
\label{eq.mc3.scale.gibbs.lambda}
\lambda|(a_{1:T},Y_{0:T})
\sim N \big( \bar{\lambda}(a_{1:T}), \bar{\Omega}(a_{1:T}) \big),
\ee
where
\begin{eqnarray*}
	\bar{\Omega}(a_{1:T})   &=& \big( \underline{\Omega}^{-1} + T/\bar{\gamma}^2(a_{1:T}) \big)^{-1} \\	
	\bar{\lambda}(a_{1:T})  &=& \bar{\Omega}(a_{1:T}) \big( (\phi_0+\phi_1Y_{0}) + (T/\bar{\gamma}^2(a_{1:T})) \hat{\lambda} \big).   
\end{eqnarray*}
The posterior probability of $a_{t}=1$ conditional on $(\lambda,Y_{0:T})$ is given by
\begin{eqnarray}
\lefteqn{\mathbb{P}\big( a_t=1|\lambda,Y_{0:T}) } \label{eq.mc3.scale.gibbs.at}\\
&=& \frac{p_u (\gamma_+)^{-1} \exp \left\{ -\frac{1}{2 \gamma_+^2}(Y_t-\rho Y_{t-1} -\lambda)^2\right\}} {p_u(\gamma_+)^{-1} \exp \left\{ -\frac{1}{2 \gamma^2_+}(Y_t-\rho Y_{t-1} -\lambda)^2\right\} + (1-p_u)(\gamma_-)^{-1} \exp \left\{ -\frac{1}{2 \gamma_-^2}(Y_t-\rho Y_{t-1} -\lambda )^2\right\}}. \nonumber
\end{eqnarray}
The posterior mean $\mathbb{E}[\lambda|{\cal Y}_i]$ can be approximated with the following Gibbs sampler. Generate a sequence of draws $\{ \lambda^s,a_{1:T}^s\}_{s=1}^{N_{sim}}$
by iterating over the conditional distributions given in (\ref{eq.mc3.scale.gibbs.lambda}) and (\ref{eq.mc3.scale.gibbs.at}). Then,
\begin{eqnarray}
\widehat{\mathbb{E}}[\lambda|Y_{0:T}] 
&=& \frac{1}{N_{sim}} \sum_{s=1}^{N_{sim}} \bar{\lambda}(a^s_{1:T}), \\
\widehat{\mathbb{V}}[\lambda|Y_{0:T}] &=& \left( \frac{1}{N_{sim}} \sum_{s=1}^{N_{sim}}  \bar{\Omega}(a_{1:T}^s) + \bar{\lambda}^2(a_{1:T}^s) \right) - \left( \frac{1}{N_{sim}} \sum_{s=1}^{N_{sim}} \bar{\lambda}(a_{1:T}^s) \right)^2. \nonumber
\end{eqnarray}

\noindent {\em Location Mixture.} Let $a_{it}=1$ if $U_{it}$ is generated from the mixture component with mean $\mu_+$ and $a_{it}=0$ if $U_{it}$ is generated from the mixture component with mean $-\mu_-$. Omitting $i$ subscripts from now on, define
\[
   \tilde{Y}_{t}(a_{t}) = Y_{t} - \rho Y_{t-1} - (a_{t} \mu_+ - (1-a_{t}) \mu_-),
\]
such that 
\[
   \tilde{Y}_{t}(a_{t})|(\lambda,a_{t}) \sim N \big(\lambda ,\gamma^2 \big).
\]
Now let 
\[
   \hat{\lambda}(a_{1:T}) = \frac{1}{T} \sum_{t=1}^T \tilde{Y}_{t}(a_{t}) \sim N \big( \lambda,\gamma^2/T).
\]
Under the prior distribution
\[
 \lambda|Y_{0} \sim N(\phi_0+\phi_1Y_{0},\underline{\Omega}),
\]
we obtain a posterior distribution of the form
\be
\label{eq.mc3.location.gibbs.lambda}
   \lambda|(a_{1:T},Y_{0:T})
   \sim N \big( \bar{\lambda}(a_{1:T}), \bar{\Omega} \big),
\ee
where
\begin{eqnarray*}
  \bar{\Omega}   &=& \big( \underline{\Omega}^{-1} + T/\gamma^2  \big)^{-1} \\	
  \bar{\lambda}(a_{1:T})  &=& \bar{\Omega} \big( (\phi_0+\phi_1Y_{0}) + (T/\gamma^2) \hat{\lambda}(a_{1:T}) \big).   
\end{eqnarray*}
The posterior probability of $a_{t}=1$ conditional on $(\lambda,Y_{0:T})$ is given by
\begin{eqnarray}
 \lefteqn{\mathbb{P}\big( a_t=1|\lambda,Y_{0:T}) } \label{eq.mc3.location.gibbs.at}\\
   &=& \frac{p_u \exp \left\{ -\frac{1}{2 \gamma^2}(Y_t-\rho Y_{t-1} -\lambda - \mu_+)^2\right\}} {p_u\exp \left\{ -\frac{1}{2 \gamma^2}(Y_t-\rho Y_{t-1} -\lambda - \mu_+)\right\} + (1-p_u)\exp \left\{ -\frac{1}{2 \gamma^2}(Y_t-\rho Y_{t-1} -\lambda + \mu_-)^2\right\}}. \nonumber
\end{eqnarray}
The posterior mean $\mathbb{E}[\lambda|Y_{0:T}]$ can be approximated with the following Gibbs sampler. Generate a sequence of draws $\{ \lambda^s,a_{1:T}^s\}_{s=1}^{N_{sim}}$
by iterating over the conditional distributions given in (\ref{eq.mc3.location.gibbs.lambda}) and (\ref{eq.mc3.location.gibbs.at}). Then,
\begin{eqnarray}
\widehat{\mathbb{E}}[\lambda|Y_{0:T}] 
&=& \frac{1}{N_{sim}} \sum_{s=1}^{N_{sim}} \bar{\lambda}(a^s_{1:T}), \\
\widehat{\mathbb{V}}[\lambda|Y_{0:T}] &=& \left( \bar{\Omega} + \frac{1}{N_{sim}} \sum_{s=1}^{N_{sim}} \bar{\lambda}^2(a_{1:T}^s) \right) - \left( \frac{1}{N_{sim}} \sum_{s=1}^{N_{sim}} \bar{\lambda}(a_{1:T}^s) \right)^2. \nonumber
\end{eqnarray}

\clearpage

\section{Data Set}

The construction of our data is based on \cite{CovasRumpZakrajsek2014}. We downloaded FR Y-9C BHC finanical statements for the years 2002 to 2014 using the web portal of the Federal Reserve Bank of Chicago. The financial statements are available at quarterly frequency. We define PPNR (relative to assets) as follows
\[
\mbox{PPNR} = 400 \big( \mbox{NII} + \mbox{ONII} - \mbox{ONIE} \big) / \mbox{ASSETS}, 
\]
where
\begin{center}
	\begin{tabular}{lll}
		NII & = Net Interest Income        & BHCK 4074 \\
		ONII &  = Total Non-Interest Income & BHCK 4079 \\
		ONIE & = Total Non-Interest Expenses & BHCK 4093 - C216 - C232 \\
		ASSETS  &= Consolidated Assets       & BHCK 3368 
	\end{tabular} 
\end{center}
Here net interest income is the difference between total interest income and expenses. It excludes provisions for loan and lease losses. Non-interest income includes various types of fees, trading revenue, as well as net gains on asset sales. Non-interest expenses include, for instance, salaries and employee benefits and expenses of premises and fixed assets. As in \cite{CovasRumpZakrajsek2014}, we exclude impairment losses (C216 and C232). We divide the net revenues by the amount of consolidated assets. This ratio is multiplied by 400 to annualize the flow variables and convert the ratio into percentages.

The raw data take the form of an unbalanced panel of BHCs. The appearance and disappearance of specific institutions in the data set is affected by entry and exit, mergers and acquisitions, as well as changes in reporting requirements for the FR Y-9C form. Because some of the quarter-over-quarter changes in the income and expense flows are a reflection of accounting practices rather than economic conditions of the institutions, we aggregate the quarterly data to annual data. However, prior to the temporal aggregation we eliminate certain types of outliers. Before describing our outlier removal procedure, we briefly discuss the structure of the rolling samples used for the forecast evaluation.

Our goal is to construct rolling samples that consist of T+2 observations, where $T$ is the size of the estimation sample and varies between $T=3$ and $T=11$. The additional two observations in each rolling sample are used, respectively, to initialize the lag in the first period of the estimation sample and to compute the error of the one-step-ahead forecast. We index each rolling sample by the forecast origin $t=\tau$. For instance, taking the time period $t$ to be a year, with data from 2002 to 2014 we can construct $M=9$ samples of size $T=3$ with forecast origins running from $\tau = 2005$ to $\tau = 2013$. Each rolling sample is indexed by the pair $(\tau,T)$.
The following adjustment procedure that eliminates BHCs with missing observations and outliers is applied to each rolling sample $(\tau,T)$ separately:

\begin{enumerate}
	\item Eliminate BCHs for which total assets are missing for all time periods in the sample.
	\item Compute average non-missing total assets and eliminate BCHs with average assets below 500 million dollars.
	\item Eliminate BCHs for which one or more PPNR components are missing for at least one period of the sample.
	\item Eliminate BCHs for which the absolute difference between the temporal mean and the temporal median exceeds 10.
	\item Define deviations from temporal means as $\delta_{it} = y_{it} - \bar{y}_i$. Pooling the $\delta_{it}$'s across institutions and time periods, compute the median $q_{0.5}$ and the 0.025 and 0.975 quantiles, $q_{0.025}$ and $q_{0.975}$. We delete institutions for which at least one $\delta_{it}$ falls outside of the range $q_{0.5} \pm (q_{0.975} - q_{0.025})$.  
\end{enumerate}	

The adjustment procedure is applied to quarterly observations. After the sample adjustments we aggregate from quarterly to annual frequency by averaging the PPNR ratios over the four quarters of the calendar year. The effect of the sample-adjustment procedure on the size of the rolling samples is summarized in Table~\ref{t_samplesizes}. Here we are focusing on the extreme cases $T=3$ (short sample) and $T=11$ (long sample). The column labeled $N_0$ provides the number of raw data for each sample. In columns $N_j$, $j=1,\ldots,4$, we report the observations remaining after adjustment $j$. Finally, $N$ is the number of observations after the fifth adjustment. This is the relevant sample size for the subsequent empirical analysis. For many BCHs we do not have information on the consolidated assets, which leads to reduction of the sample size by 60\% to 80\%. Once we restrict average consolidated assets to be above 500 million dollars, the sample size shrinks to approximately 900 to 1,400 institutions. Roughly 35\% to 65\% of these institutions have missing observations for PPNR components, which leads to $N_3$. The outlier elimination in Steps~4. and~5. have a relatively small effect on the sample size.

\begin{table}[t!]
	\caption{Size of Adjusted Rolling Samples}
	\label{t_samplesizes}
	\begin{center}
		\scalebox{0.9}{
		\begin{tabular}{llrrrrrr} \hline \hline	 
			\multicolumn{2}{c}{Sample} & \multicolumn{6}{c}{Adjustment Step} \\ 
			$T$ & $\tau$ & $N_0$ & $N_1$ & $N_2$ & $N_3$ & $N_4$ & $N$ \\ \hline
			3	&	2005	&	6,731	&	2,629	&	882	    &	580	&	580	&	551	\\
			3	&	2006	&	6,673	&	2,591	&	959	    &	650	&	650	&	615	\\
			3	&	2007	&	6,619	&	2,537	&	1,024	&	693	&	693	&	655	\\
			3	&	2008	&	6,519	&	2,456	&	1,074	&	716	&	716	&	670	\\
			3	&	2009	&	6,399	&	1,281	&	1,139	&	693	&	693	&	653	\\
			3	&	2010	&	6,223	&	1,287	&	1,157	&	683	&	683	&	639	\\
			3	&	2011	&	6,518	&	1,396	&	1,273	&	704	&	704	&	656	\\
			3	&	2012	&	6,343	&	1,413	&	1,301	&	755	&	755	&	710	\\
			3	&	2013	&	6,154	&	1,407	&	1,291	&	772	&	771	&	725	\\
			11	&	2013	&	8,011	&	2,957	&	1,431	&	497	&	496	&	461	\\ \hline
		\end{tabular}
	}
	\end{center}
\end{table}

\begin{table}[t!]
	\caption{Descriptive Statistics for Rolling Samples}
	\label{t_samplestatistics}
	\begin{center}
		\scalebox{0.9}{
		\begin{tabular}{llrrrrrrr} \hline \hline
			\multicolumn{2}{c}{Sample} & \multicolumn{7}{c}{Statistics} \\ 
			$T$	&	$\tau$	&	Min	    &	Mean	&	Median	&	Max 	&	StdD	&	Skew	&	Kurt	\\ \hline
			3	&	2005	&	-8.81	&	1.48	&	1.65	&	8.46	&	2.07	&	-0.80	&	5.36	\\
			3	&	2006	&	-7.61	&	1.50	&	1.54	&	8.46	&	1.95	&	-0.43	&	4.90	\\
			3	&	2007	&	-9.55	&	1.36	&	1.42	&	7.75	&	1.94	&	-0.61	&	5.51	\\
			3	&	2008	&	-9.55	&	1.12	&	1.22	&	7.75	&	1.93	&	-0.72	&	5.62	\\
			3	&	2009	&	-10.44	&	0.98	&	1.08	&	7.00	&	1.84	&	-0.82	&	6.01	\\
			3	&	2010	&	-7.46	&	0.87	&	0.96	&	6.60	&	1.74	&	-0.63	&	4.76	\\
			3	&	2011	&	-8.87	&	0.84	&	0.96	&	7.17	&	1.77	&	-0.70	&	5.04	\\
			3	&	2012	&	-7.65	&	0.79	&	0.90	&	7.81	&	1.86	&	-0.46	&	4.41	\\
			3	&	2013	&	-8.11	&	0.82	&	0.95	&	7.73	&	1.87	&	-0.53	&	4.62	\\
			11	&	2013	&	-8.89	&	1.15	&	1.23	&	7.00	&	1.82	&	-0.65	&	5.02	\\ \hline
		\end{tabular}
	}
	\end{center}
	{\footnotesize {\em Notes:} The descriptive statistics are computed for samples in which we pool observations across institutions and time periods. We did not weight the statistics by size of the institution.}\setlength{\baselineskip}{4mm}
\end{table}

Descriptive statistics for the $T=3$ and $T=11$ rolling samples are reported in Table~\ref{t_samplesizes}.
For each rolling sample we pool observations across institutions and time periods. We do not weight the observations by the size of the institution. Focusing on the $T=3$ samples, notice that the mean PPNR falls from about 1.5\% for the 2005 and 2006 samples to 0.80\% for the 2012 sample, which includes observations starting in 2009. In the 2013 sample the mean increased again to 1.15\%. The means are generally smaller than the medians, suggesting that the samples are left-skewed, which is confirmed by the skewness measures reported in the second to last column. The samples also exhibit fat tails. The kurtosis statistics range from 4.4 to 6.0.

\clearpage

\section{Additional Empirical Results}

\begin{table}[h!]
	\caption{Parameter Estimates: $\hat{\theta}_{QMLE}$, Parametric Tweedie Correction}
	\label{t_parameters_simple_T5}
	\begin{center}
		\begin{tabular}{lccccccccc} \hline \hline
			& & & \multicolumn{3}{c}{Intercept} & \multicolumn{3}{c}{Unemployment} & \\
			$\tau$ & $\hat{\rho}$ & $\hat{\sigma}^2$ & $\hat{\phi}_{10}$ & $\hat{\phi}_{11}$ & $\hat{\underline{\omega}}_1^2$ &
			$\hat{\phi}_{20}$ & $\hat{\phi}_{21}$ & $\hat{\underline{\omega}}_2^2$ &			
			N \\ \hline
			2007 & 0.91 & 1.10 & -0.99 & 0.08  & 4E-7 & 0.18 & -0.01 & 9E-9 & 537 \\ 
			2008 & 0.86	& 1.09 & -1.25 & -0.05 & 3E-6 & 0.28 &  0.02 & 1E-7 & 598 \\
			2009 & 0.86	& 1.14 & -0.27 & -0.06 & 1E-7 & 0.05 &  0.02 & 5E-9 & 613 \\
			2010 & 0.86 & 1.14 & -0.38 & -0.03 & 2E-8 & 0.07 &  0.01 & 1E-9 & 606 \\
			2011 & 0.94	& 1.12 & -0.22 & -0.17 & 2E-7 & 0.03 &  0.02 & 3E-9 & 582 \\
			2012 & 0.94	& 1.12 &  0.01 & -0.30 & 2E-8 & 0.00 &  0.03 & 1E-9 & 587 \\
			2013 & 0.93	& 1.12 & -0.47 & -0.30 & 3E-7 & 0.05 &  0.04 & 2E-9 & 608 \\ \hline
		\end{tabular}
	\end{center}
	{\footnotesize {\em Notes:} Point estimates for the model $Y_{it+1} = \lambda_{1i} + \lambda_{2i} UR_t + \rho Y_{it} + U_{it+1}$, $U_{it+1} \sim N(0,\sigma^2)$, $\lambda_{ji}|Y_{i0} \sim N(\phi_{j0} + \phi_{j1} Y_{i0}, \underline{\omega}_j^2)$ for $j=1,2$. The time-series dimension of the estimation sample is $T=5$.}\setlength{\baselineskip}{4mm}
\end{table}

\end{document}